\documentclass[10pt,a4paper]{article}
\pdfoutput=1
\usepackage[T1]{fontenc}
\usepackage[utf8]{inputenc}
\usepackage[ngerman,english]{babel}
\usepackage{cite}
\usepackage{amsmath}
\usepackage{amsfonts}
\usepackage{amssymb}
\usepackage{amsthm}
\usepackage{enumitem}
\usepackage{dcolumn}
\usepackage{bm}
\usepackage{bbm}
\usepackage{cancel}
\usepackage[pdftex]{graphicx}
\usepackage{tikz}
\usetikzlibrary{calc,arrows,positioning}
\usepackage{tabularx}
\usepackage{booktabs}
\usepackage{tensor}
\usepackage{scalerel}
\usepackage[small]{caption}
\usepackage{xcolor}
\usepackage{mathtools}
\usepackage{float}
\usepackage{xfrac}
\usepackage[pdfborder={0 0 0}]{hyperref}
\definecolor{darkblue}{rgb}{0,0,.5}
\hypersetup{colorlinks=true, breaklinks=true, linkcolor=darkblue, menucolor=darkblue, urlcolor=darkblue, linktocpage=true}
\usepackage{times}
\usepackage{textpos}
\usepackage{geometry}
\makeatletter 
\newsavebox\myboxA 
\newsavebox\myboxB 
\newlength\mylenA 
 
\newcommand*\xoverline[2][0.75]{%
    \sbox{\myboxA}{$\m@th#2$}%
    \setbox\myboxB\null
    \ht\myboxB=\ht\myboxA%
    \dp\myboxB=\dp\myboxA%
    \wd\myboxB=#1\wd\myboxA
    \sbox\myboxB{$\m@th\overline{\copy\myboxB}$}
    \setlength\mylenA{\the\wd\myboxA}
    \addtolength\mylenA{-\the\wd\myboxB}%
    \ifdim\wd\myboxB<\wd\myboxA%
       \rlap{\hskip 0.5\mylenA\usebox\myboxB}{\usebox\myboxA}%
    \else 
        \hskip -0.5\mylenA\rlap{\usebox\myboxA}{\hskip 0.5\mylenA\usebox\myboxB}%
    \fi}
\makeatother
\numberwithin{equation}{section}

\let\originalleft\left
\let\originalright\right
\renewcommand{\left}{\mathopen{}\mathclose\bgroup\originalleft}
\renewcommand{\right}{\aftergroup\egroup\originalright}
\newcommand{\e}{\operatorname{e}}
\newcommand{\SU}[1]{\operatorname{SU}\left(#1\right)}
\newcommand{\Un}[1]{\operatorname{U}\left(#1\right)}

\newcommand{\of}[1]{\left(#1\right)}
\newcommand{\bof}[1]{\biggl(#1\biggr)}
\newcommand{\sof}[1]{\bigl(#1\bigr)}
\newcommand{\ssof}[1]{(#1)}
\newcommand{\fof}[1]{\left[#1\right]}
\newcommand{\bfof}[1]{\biggl[#1\biggr]}

\newcommand{\cof}[1]{\left\{#1\right\}}
\newcommand{\bcof}[1]{\biggl\{#1\biggr\}}

\newcommand{\trace}{\operatorname{tr}}
\renewcommand*{\det}{\mathop{\mathrm{det}}\nolimits}
\newcommand{\Trace}[1]{\operatorname{Tr}\left(#1\right)}
\newcommand{\Det}[1]{\operatorname{Det}\left(#1\right)}

\newcommand{\avof}[1]{\left\langle #1\right\rangle}

\newcommand{\Repart}[1]{\operatorname{Re}\left(#1\right)}

\newcommand{\cRepart}[1]{\operatorname{Re}\left\{#1\right\}}

\newcommand{\repart}{\operatorname{Re}}

\newcommand{\ii}{\mathrm{i}}
\newcommand{\idd}[2]{\mathrm{d}^{#2}\,#1}
\newcommand{\dd}{\mathrm{d}}
\newcommand{\DD}[1]{\mathcal{D}\left[#1\right]}

\newcommand{\partd}[2]{\frac{\partial #1}{\partial #2}}

\newcommand{\order}[1]{\mathcal{O}\big(#1\big)}

\newcommand{\obs}{\mathcal{O}}

\newcommand{\id}{\mathbbm{1}}
\newcommand{\abs}[1]{\left|#1\right|}
\newcommand{\op}[1]{\operatorname{#1}}
\newcommand{\umod}{\operatorname{mod}}

\newcommand*\colvec[3][]{
    \begin{pmatrix}\ifx\relax#1\relax\else#1\\\fi#2\\#3\end{pmatrix}
}
\renewcommand*\[{\begin{equation}}
\renewcommand*\]{\end{equation}}
\renewcommand*\bar[1]{\ThisStyle{\xoverline{\SavedStyle #1}}}
\newcommand{\rbar}[1]{\xoverline{#1}}
\renewcommand*\hat[1]{\ThisStyle{\widehat{\SavedStyle #1}}}
\let\oldstackrel\stackrel
\renewcommand*\stackrel[2]{{\scriptstyle\oldstackrel{#1}{#2}}}

\definecolor{emphcol}{RGB}{0,0,0}
\let\oldemph\emph
\renewcommand*\emph[1]{\oldemph{\textcolor{emphcol}{#1}}}

\newlength{\hatchspread}
\newlength{\hatchthickness}
\newlength{\hatchshift}
\newcommand{\hatchcolor}{}
\tikzset{hatchspread/.code={\setlength{\hatchspread}{#1}},
         hatchthickness/.code={\setlength{\hatchthickness}{#1}},
         hatchshift/.code={\setlength{\hatchshift}{#1}},
         hatchcolor/.code={\renewcommand{\hatchcolor}{#1}}}
\tikzset{hatchspread=3pt,
         hatchthickness=0.4pt,
         hatchshift=0pt,
         hatchcolor=black}
\pgfdeclarepatternformonly[\hatchspread,\hatchthickness,\hatchshift,\hatchcolor]
   {custom north west lines}
   {\pgfqpoint{\dimexpr-2\hatchthickness}{\dimexpr-2\hatchthickness}}
   {\pgfqpoint{\dimexpr\hatchspread+2\hatchthickness}{\dimexpr\hatchspread+2\hatchthickness}}
   {\pgfqpoint{\dimexpr\hatchspread}{\dimexpr\hatchspread}}
   {
    \pgfsetlinewidth{\hatchthickness}
    \pgfpathmoveto{\pgfqpoint{\dimexpr0pt}{\dimexpr\hatchspread+\hatchshift}}
    \pgfpathlineto{\pgfqpoint{\dimexpr\hatchspread+0.15pt}{\dimexpr\hatchshift-0.15pt}}
    \ifdim \hatchshift > 0pt
      \pgfpathmoveto{\pgfqpoint{\dimexpr-0.15pt}{\dimexpr\hatchshift+0.15pt}}
      \pgfpathlineto{\pgfqpoint{\dimexpr\hatchshift+0.15pt}{\dimexpr-0.15pt}}
    \fi
    \pgfsetstrokecolor{\hatchcolor}
    \pgfusepath{stroke}
   }

\pgfdeclarepatternformonly[\hatchspread,\hatchthickness,\hatchshift,\hatchcolor]
   {custom north east lines}
   {\pgfqpoint{\dimexpr-2\hatchthickness}{\dimexpr-2\hatchthickness}}
   {\pgfqpoint{\dimexpr\hatchspread+2\hatchthickness}{\dimexpr\hatchspread+2\hatchthickness}}
   {\pgfqpoint{\dimexpr\hatchspread}{\dimexpr\hatchspread}}
   {
    \pgfsetlinewidth{\hatchthickness}
    \pgfpathmoveto{\pgfqpoint{0pt}{\dimexpr\hatchshift}}
    \pgfpathlineto{\pgfqpoint{\dimexpr\hatchspread+0.15pt}{\dimexpr\hatchspread+0.15pt+\hatchshift}}
    \ifdim \hatchshift > 0pt
      \pgfpathmoveto{\pgfqpoint{\dimexpr\hatchspread-\hatchshift-0.15pt}{\dimexpr-0.15pt}}
      \pgfpathlineto{\pgfqpoint{\dimexpr\hatchspread+0.15pt}{\dimexpr\hatchshift+0.15pt}}
    \fi
    \pgfsetstrokecolor{\hatchcolor}
    \pgfusepath{stroke}
   }

\pgfdeclarepatternformonly[\hatchspread,\hatchthickness,\hatchshift,\hatchcolor]
   {custom vertical lines}
   {\pgfqpoint{\dimexpr-2\hatchthickness}{\dimexpr-2\hatchthickness}}
   {\pgfqpoint{\dimexpr\hatchspread+2\hatchthickness}{\dimexpr\hatchspread+2\hatchthickness}}
   {\pgfqpoint{\dimexpr\hatchspread}{\dimexpr\hatchspread}}
   {
    \pgfsetlinewidth{\hatchthickness}
    \pgfpathmoveto{\pgfqpoint{\dimexpr\hatchshift}{0pt}}
    \pgfpathlineto{\pgfqpoint{\dimexpr\hatchshift}{\dimexpr\hatchspread+0.15pt}}
    \pgfsetstrokecolor{\hatchcolor}
    \pgfusepath{stroke}
   }

\pgfdeclarepatternformonly[\hatchspread,\hatchthickness,\hatchshift,\hatchcolor]
   {custom horizontal lines}
   {\pgfqpoint{\dimexpr-2\hatchthickness}{\dimexpr-2\hatchthickness}}
   {\pgfqpoint{\dimexpr\hatchspread+2\hatchthickness}{\dimexpr\hatchspread+2\hatchthickness}}
   {\pgfqpoint{\dimexpr\hatchspread}{\dimexpr\hatchspread}}
   {
    \pgfsetlinewidth{\hatchthickness}
    \pgfpathmoveto{\pgfqpoint{0pt}{\dimexpr\hatchshift}}
    \pgfpathlineto{\pgfqpoint{\dimexpr\hatchspread+0.15pt}{\dimexpr\hatchshift}}
    \pgfsetstrokecolor{\hatchcolor}
    \pgfusepath{stroke}
   }

\begin{document}\selectlanguage{english}
\newcounter{romanPagenumber}
\newcounter{arabicPagenumber}
\pagenumbering{Roman}
\title{
\begin{textblock*}{100pt}(390pt,-116pt)
\textnormal{\small \texttt{CERN-PH-TH-2015-206}}
\end{textblock*}
\begin{textblock*}{100pt}(381pt,-90pt)
\textnormal{\small \texttt{NSF-KITP-15-122}}
\end{textblock*}{\Large Two-Flavor Lattice QCD with a Finite Density of Heavy Quarks:\\
\vspace{5pt}\Large Heavy-Dense Limit and "Particle-Hole" Symmetry}}

\author{Tobias Rindlisbacher \thanks{E-mail: rindlisbacher@itp.phys.ethz.ch}\\
ETH Z\"urich, Institute for Theoretical Physics,\\ Wolfgang-Pauli-Str. 27, CH - 8093 Z\"urich, Switzerland\\
\and\\
Philippe de Forcrand \thanks{E-mail: forcrand@itp.phys.ethz.ch}\\
ETH Z\"urich, Institute for Theoretical Physics,\\ Wolfgang-Pauli-Str. 27, CH - 8093 Z\"urich, Switzerland\\
and\\
CERN, Physics Department, TH Unit,\\ CH-1211 Gen\`eve 23, Switzerland}

\maketitle

\begin{abstract}
We investigate the properties of the half-filling point in lattice QCD (LQCD), in particular the disappearance of the sign problem and the emergence of an apparent particle-hole symmetry, and try to understand where these properties come from by studying the heavy-dense fermion determinant and the corresponding strong-coupling partition function (which can be integrated analytically).
We then add in a first step an effective Polyakov loop gauge action (which reproduces the leading terms in the character expansion of the Wilson gauge action) to the heavy-dense partition function and try to analyze how some of the properties of the half-filling point change when leaving the strong coupling limit. In a second step, we take also the leading nearest-neighbor fermion hopping terms into account (including gauge interactions in the fundamental representation) and mention how the method could be improved further to incorporate the full set of nearest-neighbor fermion hoppings.
Using our mean-field method, we also obtain an approximate $\of{\mu,T}$ phase diagram for heavy-dense LQCD at finite inverse gauge coupling $\beta$.
Finally, we propose a simple criterion to identify the chemical potential beyond which lattice artifacts become dominant.
\end{abstract}
        
\newpage
\setcounter{romanPagenumber}{\value{page}} 
\pagenumbering{arabic}

\section{Introduction}\label{sec:intro}
Lattice QCD at non-zero baryon density suffers from the so-called "sign problem" \cite{DeForcrand}. It also displays a new type of lattice artifact: because of the Pauli exclusion principle, the lattice cannot support more than one quark per site (and per spin, color and flavor). The main goal of this paper is to characterize the effects of this new discretization error, so as to better disentangle genuine physics from lattice artifacts.\\
We first briefly recall the foundations of lattice QCD in the presence of a chemical potential $\mu$ and some well-known features of the sign problem which appears when $\mu\neq 0$. We then focus on the half-filling point (Sec. \ref{sec:sathf}) and the associated fermion saturation. Using the heavy-dense limit (Sec. \ref{sec:hdanalysis}), we elucidate the properties of the half-filling regime and analyze, using a mean-field calculation which takes leading nearest-neighbor fermion hoppings and gauge interactions into account (appendix \ref{sec:kappasqerms}), how far some of these findings generalize to full QCD.  Finally, using our mean-field method, we obtain an approximate $\of{\mu,T}$ phase diagram for heavy-dense LQCD, which can be compared with current state of the art complex Langevin investigations \cite{Aarts}.
\subsection{QCD at finite Density}\label{ssec:finitedens}
In this paper, we adopt the Wilson discretization of the Dirac operator. For a single flavor at chemical potential $\mu$ it reads:
\begin{align}
D_{x I a,y J b}\,&=\,\delta_{x y}\delta_{I J}\delta_{a b}\nonumber\\
-\,&\kappa\underbrace{\sum\limits_{\nu=1}^{3}\of{\delta_{x+\hat{\nu},y}\of{\id-\gamma_{\nu}}_{a b}U_{\nu,I J}\ssof{x}\,+\,\delta_{x-\hat{\nu},y}\of{\id+\gamma_{\nu}}_{a b}U^{\dagger}_{\nu,I J}\ssof{x-\hat{\nu}}}}_{S_{x I a,y J b}}\nonumber\\
-\,&\kappa\underbrace{\of{\delta_{x+\hat{4},y}\of{\id-\gamma_{4}}_{a b}U_{4,I J}\ssof{x}\e^{\mu}\,+\,\delta_{x-\hat{4},y}\of{\id+\gamma_{4}}_{a b}U^{\dagger}_{4,I J}\ssof{x-\hat{4}}\e^{-\mu}}}_{T_{x I a,y J b}}\label{eq:wilsondiracop}\ ,
\end{align}
where $\of{x,y}$ are multi-indices labeling different lattice sites, $\of{I,J}$, $\of{a,b}$ are color and Dirac indices respectively and $U_{\nu,I J}\of{x}$ is the $\of{I,J}$ component of a link variable that points from site $x$ to $\of{x+\hat{\nu}}$. The hopping parameter $\kappa$ is related to the bare lattice quark mass $m$ by $\kappa=\of{2\,m+8}^{-1}$, and the boundary conditions are assumed to be periodic in spatial directions and anti-periodic in the time direction.\\
Using matrix notation, we write \eqref{eq:wilsondiracop} simply as $D\of{\mu}$, keeping its dependency on the gauge field implicit. From the definition \eqref{eq:wilsondiracop}, it follows that
\[
D\of{\mu}\,=\,\gamma_{5}\,D^{\dagger}\of{-\mu^{*}}\,\gamma_{5}\ ,
\]
where $\mu^{*}$ is the complex conjugate of $\mu$, which shows that
\[
\Det{D\of{\mu}}\,=\,\int\DD{\bar{\psi},\psi}\,\e^{-\bar{\psi}\,D\of{\mu}\,\psi}\label{eq:fermiondetint}
\]
is real if $\Repart{\mu}=0$, as 
\begin{multline}
\Det{D\of{\mu}}\,=\,\Det{\gamma_{5}\,D^{\dagger}\of{-\mu^{*}}\,\gamma_{5}}\,=\,\Det{D^{\dagger}\of{-\mu^{*}}}\\
=\,\Det{D\of{-\mu^{*}}}^{*}\,\overset{\of{\Repart{\mu}=0}}{\displaystyle =}\,\Det{D\of{\mu}}^{*}.
\end{multline}
Therefore, if $\Repart{\mu}=0$, the full fermion determinant for two mass degenerate flavors is manifestly non-negative,
\begin{multline}
\Det{
\begin{matrix}
D\of{\mu} & 0\\
0 & D\of{\mu}
\end{matrix}}\,=\,\Det{D\of{\mu}}\Det{D\of{\mu}}\\=\,\Det{D\of{\mu}}\Det{D\of{\mu}}^{*}\,=\,\Det{D\of{\mu}\,D^{\dagger}\of{\mu}}\ ,
\end{multline}
and gives rise to a well defined effective fermion action
\[
S_{F}^{eff}\fof{U}\,=\,-\Trace{\log\of{D\of{\mu}D^{\dagger}\of{\mu}}}\ ,
\]
which can be used, together with Wilson's lattice gauge action,
\[
S_{g}\fof{U}\,=\,\frac{\beta}{N_{c}}\sum\limits_{x}\sum\limits_{\nu<\mu}\repart\trace\of{\id-U_{\mu \nu}\of{x}}\quad,\quad U_{\mu \nu}\of{x}\,=\,U_{\mu}\of{x}\,U_{\nu}\of{x+\hat{\mu}}\,U^{\dagger}_{\mu}\of{x+\hat{\nu}}\,U^{\dagger}_{\nu}\of{x}\ ,\label{eq:wilsongaugeaction}
\]
to sample gauge configurations by Monte Carlo. This generalizes to the case of $N_{f}$ flavors as long as the degeneracy factor for each quark mass is even, i.e. as long as there is always an even number of flavors that have the same quark mass.

\subsection{The Sign Problem}\label{ssec:signproblem}
For $\Repart{\mu}\neq 0$, the fermion determinant \eqref{eq:fermiondetint} becomes in general complex and loses its probabilistic interpretation. In order to continue sampling the gauge field by Monte Carlo, one is then forced to invoke reweighting techniques. For example, if the theory contains an even number $N_{f}$ of mass degenerate flavors, one can use $\abs{\Det{D\of{\mu}}}^{N_{f}}\,=\,\Det{D\of{\mu}}^{\sfrac{N_{f}}{2}}\Det{D\of{-\mu}}^{\sfrac{N_{f}}{2}}$ instead of $\Det{D\of{\mu}}^{N_{f}}$ to sample the gauge field and treat the complex phase,
\[
\op{R}\fof{U}\,=\,\e^{\ii\,N_{f}\,\phi\fof{U}}\,=\,\frac{\Det{D\of{\mu}}^{N_{f}}}{\abs{\Det{D\of{\mu}}}^{N_{f}}}\ ,
\]
as part of the observables. This is called \emph{phase quenching}. For $N_{f}=2$, the phase quenched ensemble just corresponds to LQCD with an isospin chemical potential. The expectation value of an observable $\obs\fof{U}$ in the original system can then be obtained as follows:
\begin{multline}
\avof{\obs}\,=\,\frac{\int\DD{U}\,\obs\fof{U}\,\Det{D\of{\mu}}^{N_{f}}\,\e^{-S_{g}\fof{U}}}{\int\DD{U}\,\Det{D\of{\mu}}^{N_{f}}\,\e^{-S_{g}\fof{U}}}\\
=\,\frac{\int\DD{U}\,\obs\fof{U}\,\op{R}\fof{U}\,\abs{\Det{D\of{\mu}}}^{N_{f}}\,\e^{-S_{g}\fof{U}}}{\int\DD{U}\,\op{R}\fof{U}\,\abs{\Det{D\of{\mu}}}^{N_{f}}\,\e^{-S_{g}\fof{U}}}\,=\,\frac{\avof{\obs\,\op{R}}_{q}}{\avof{\op{R}}_{q}}\ ,\label{eq:reweighting}
\end{multline}
where $\avof{\obs'}_{q}$ is the expectation value of $\obs'\fof{U}$ with respect to the \emph{phase quenched} system, i.e.
\[
\avof{\obs'}_{q}\,=\,\frac{\int\DD{U}\,\obs'\fof{U}\,\abs{\Det{D\of{\mu}}}^{N_{f}}\,\e^{-S_{g}\fof{U}}}{\int\DD{U}\,\abs{\Det{D\of{\mu}}}^{N_{f}}\,\e^{-S_{g}\fof{U}}}\ .
\]

Although the \emph{reweighting} procedure \eqref{eq:reweighting} is in principle exact, its applicability is limited by the statistical fluctuations of the phase $R\fof{U}$. To see this, consider the expectation value of the phase,
\[
\avof{\op{R}}_{q}\,=\,\frac{Z}{Z_{q}}\,=\,\e^{-L^{3}\,N_{t}\,\Delta f}\ ,\label{eq:avphase}
\]
where $\Delta f\,=\,f\,-\,f_{q}$ is the difference in the free energy densities of the unquenched and the quenched theory and $L$ and $N_{t}$ are respectively the spatial and temporal system sizes. For sufficiently large systems, $\Delta f$ becomes approximately independent of the system size and \eqref{eq:avphase} shows that $\avof{\op{R}}_{q}$ then decays exponentially with increasing system size. As per definition the modulus of $\op{R}\fof{U}$ remains equal to $1$, this shows that the fluctuations in the reweighted observable, $\sfrac{\obs\fof{U}\op{R}\fof{U}}{\avof{\op{R}}_{q}}$, grow exponentially with the system size and so does the statistical error. The dependency of the latter on the number of measurements is known to be proportional to $\sfrac{1}{\sqrt{\text{\# meas.}}}$. This means that the statistics required to obtain equally accurate results for different system sizes must scale like
\[
\text{required statistics}\,\propto\,\e^{2\,L^{3}\,N_{t}\,\Delta f}\ .\label{eq:signproblem}
\]
The exponential growth of the required statistics as a function of the system size in \eqref{eq:signproblem} is a manifestation of the so-called \emph{sign problem} which is generic in the numerical treatment of strongly correlated fermion systems at finite density.

\section{Half-filling and Saturation}\label{sec:sathf}
As mentioned above, the sign problem is generic in the numerical study of strongly correlated fermion systems at finite density and appears therefore also in solid state physics when working with e.g. the fermionic Hubbard model. There, finite density simulations are often carried out at the so-called half-filling point, i.e. at the value of the chemical potential where the fermion density assumes half its maximum value. Due to the particle - hole symmetry that exists at this point, the sign problem disappears and it is possible to simulate much larger systems than at generic non-zero values of the chemical potential.\\

\begin{figure}[h]
\centering
\includegraphics[width=0.5\linewidth]{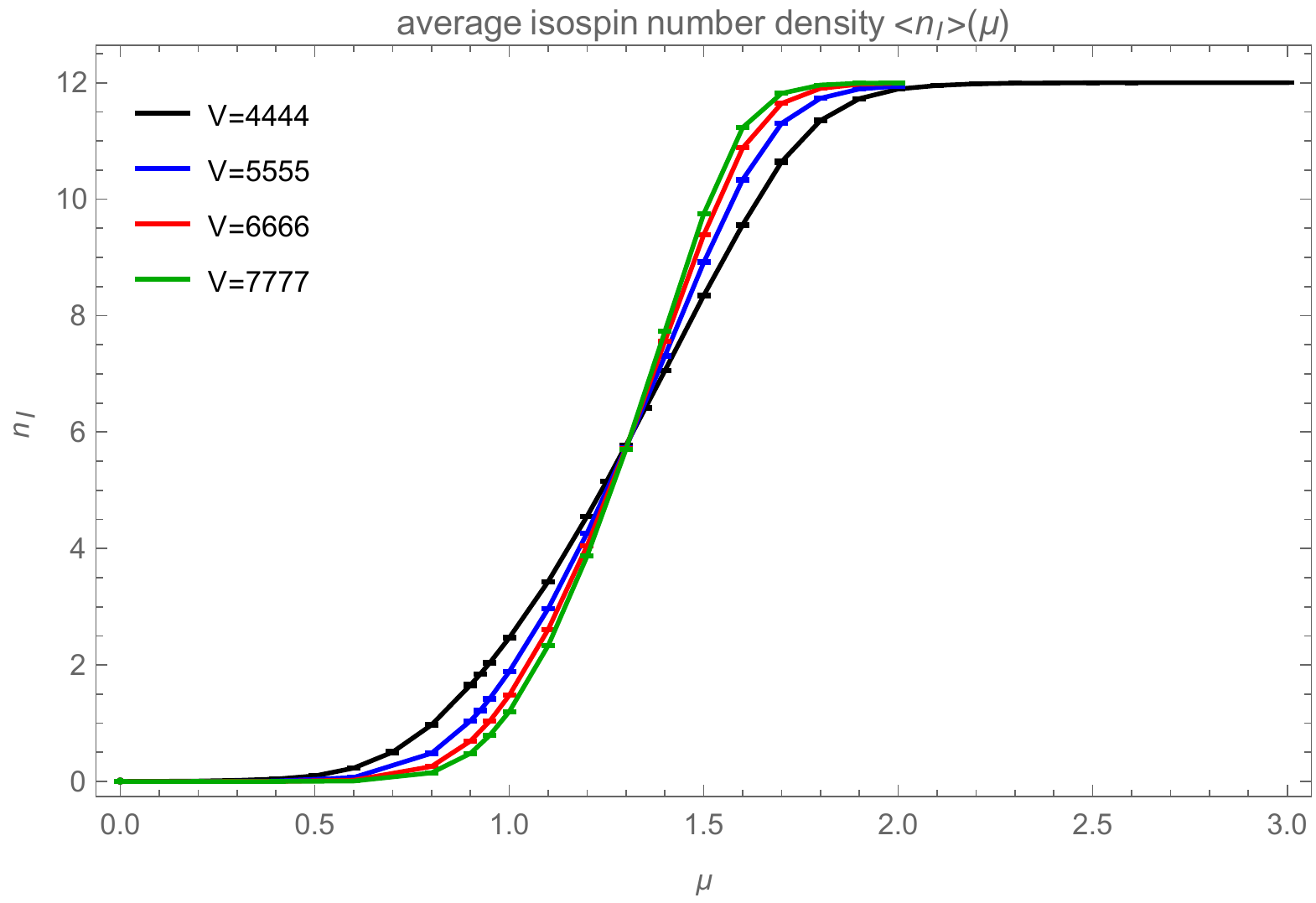}
\caption{Average isospin number density as a function of isospin $\mu$ for different system sizes, recorded at $\kappa=0.15$ and $\beta=5.0$. All curves cross in a single point, the half-filling point at $\mu\approx 1.3$, where the isospin number density assumes half its maximum value.}
  \label{fig:avisodensvarnt}
\end{figure}

\begin{figure}[h]
\centering
\begin{minipage}[t]{0.495\linewidth}
\centering
\includegraphics[width=\linewidth]{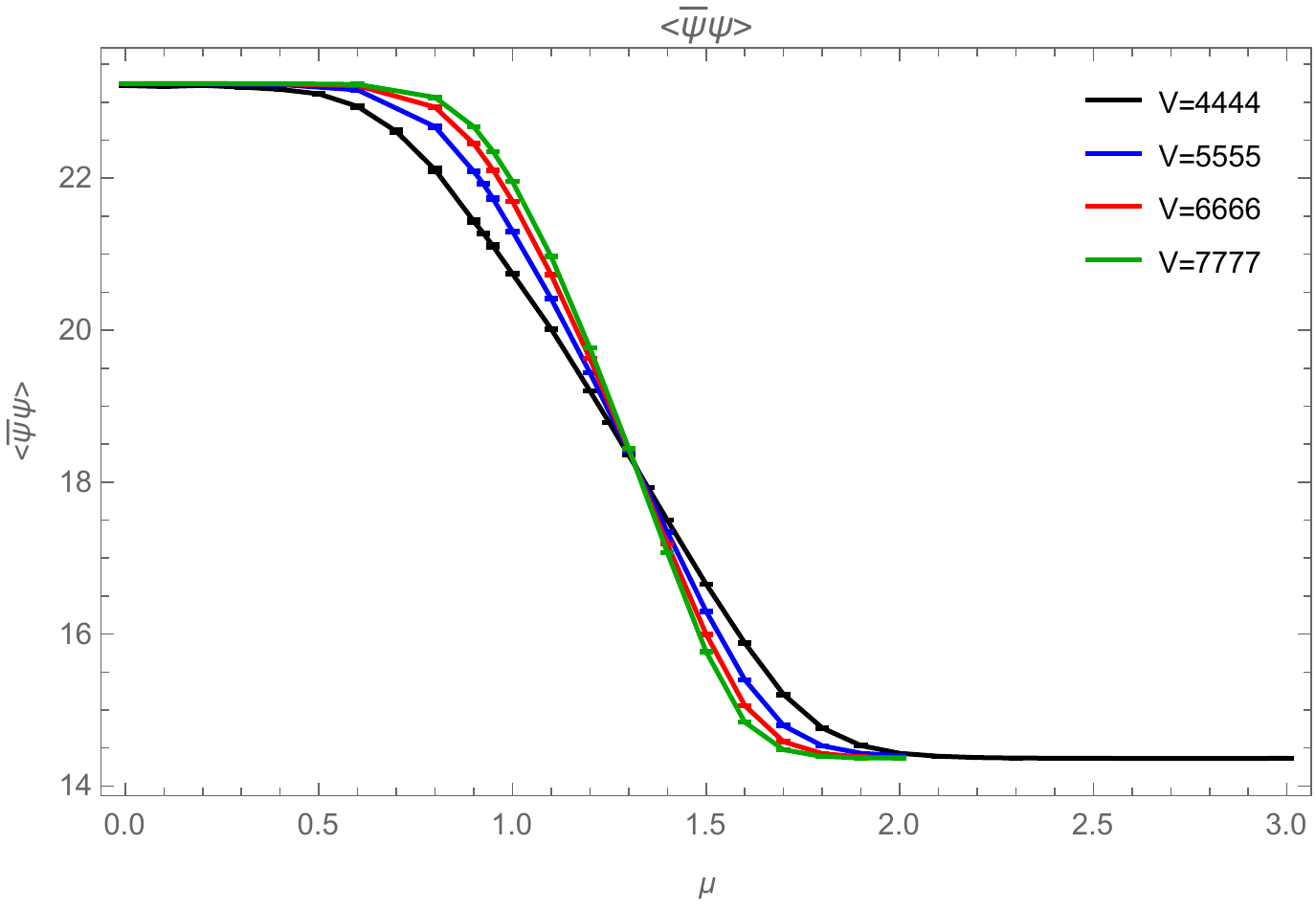}
\end{minipage}\hfill
\begin{minipage}[t]{0.495\linewidth}
\centering
\includegraphics[width=\linewidth]{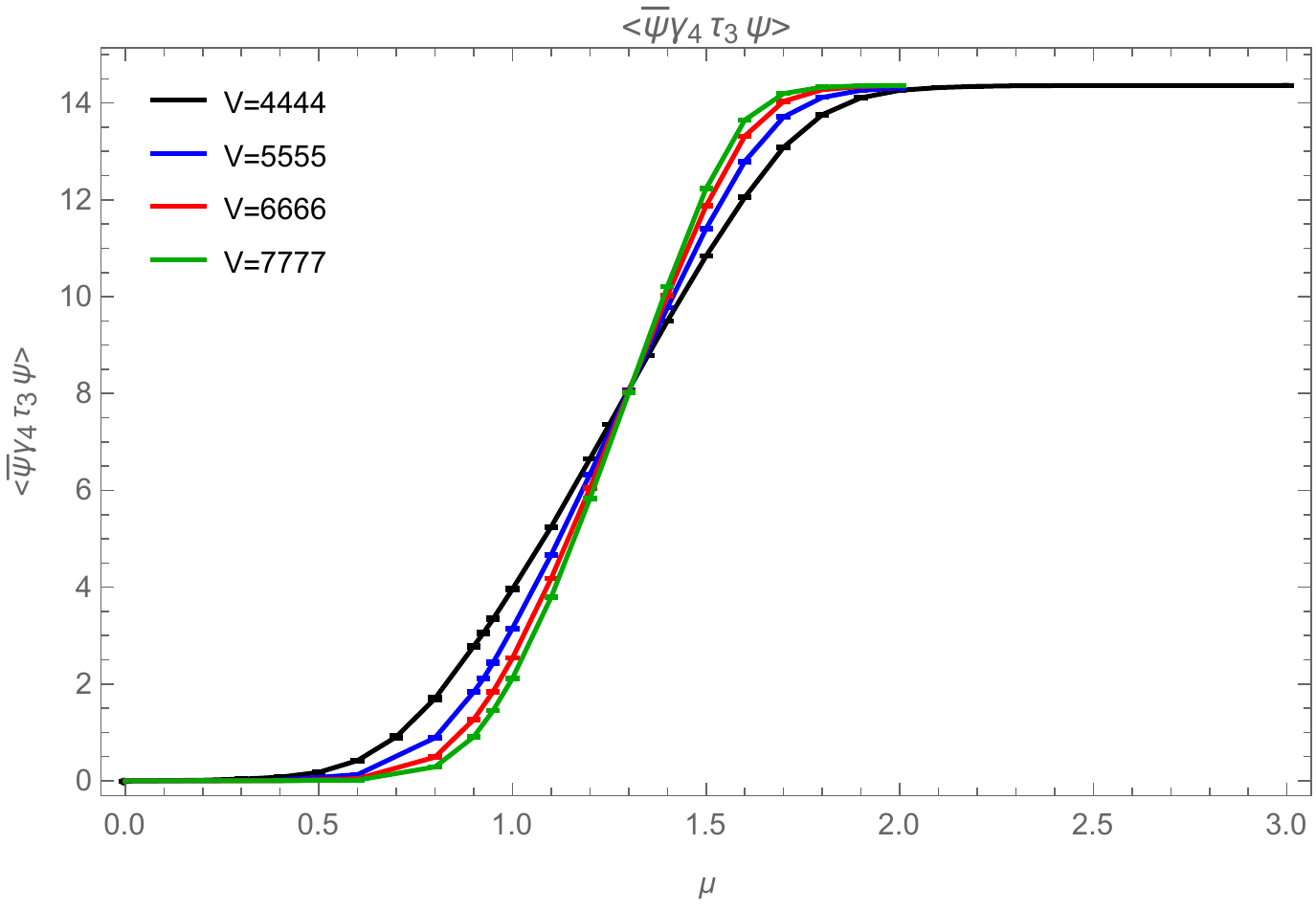}
\end{minipage}
\caption{Quark (left) and "isospin" (right) condensate as a function of isospin $\mu$ for different system sizes, recorded at $\kappa=0.15$ and $\beta=5.0$. All curves cross in a single point, the half-filling point at $\mu\approx 1.3$, but in contrast to the situation with the isospin density in Fig. \ref{fig:avisodensvarnt}, the values of the condensates at half-filling are not in the middle between their values at $\mu=0$ and $\mu\rightarrow\infty$.}
  \label{fig:avcondsvarnt}
\end{figure}

The half-filling point also exists in lattice QCD (LQCD). The reason for this is of course that the quarks, as they are fermions, are subject to the Pauli exclusion principle, which implies that on each site, each of the  $2\times N_{f}\times N_{c}$ fermionic states can be occupied only once. For two flavor QCD ($N_{f}=2$, $N_{c}=3$), the maximal number of fermions per site is $12$ and the half-filling point should therefore correspond to a fermion number density of 6. Figure \ref{fig:avisodensvarnt} illustrates this for the case of a tho flavor system with an isospin chemical potential (i.e. a phase quenched two flavor system): it shows the isospin number density
\[
n_{I}\of{\mu}\,=\,\frac{1}{L^{3}\,N_{t}}\partd{\log Z_{q}\of{\mu}}{\mu}\quad,\quad Z_{q}\of{\mu}\,=\,\int\DD{U}\,\Det{D\of{\mu}}\,\Det{D\of{-\mu}}\,\e^{-S_{g}\fof{U}}
\] 
as a function of the chemical potential $\mu$ for different system sizes. The figure also illustrates that the location of the half-filling point is independent of the system size as the different curves (corresponding to different system sizes) all cross in precisely the same point at $n_{I}=6$, $\mu\approx 1.3$. In figure \ref{fig:avcondsvarnt} we also show the condensates $\avof{\bar{\psi}\,\psi}$ and $\avof{\bar{\psi}\gamma_{4}\tau_{3}\psi}$ as a function of isospin-$\mu$ for the same set of system sizes. Also here we see volume independency at half-filling as the different curves cross in a single point at $\mu\approx 1.3$, but in contrast to the situation with the isospin density, the values of the condensates at half-filling are not half-way between their corresponding values at $\mu=0$ and at saturation $\mu\rightarrow\infty$.\\

\begin{figure}[h]
\centering
\begin{minipage}[t]{0.495\linewidth}
\centering
\includegraphics[width=\linewidth]{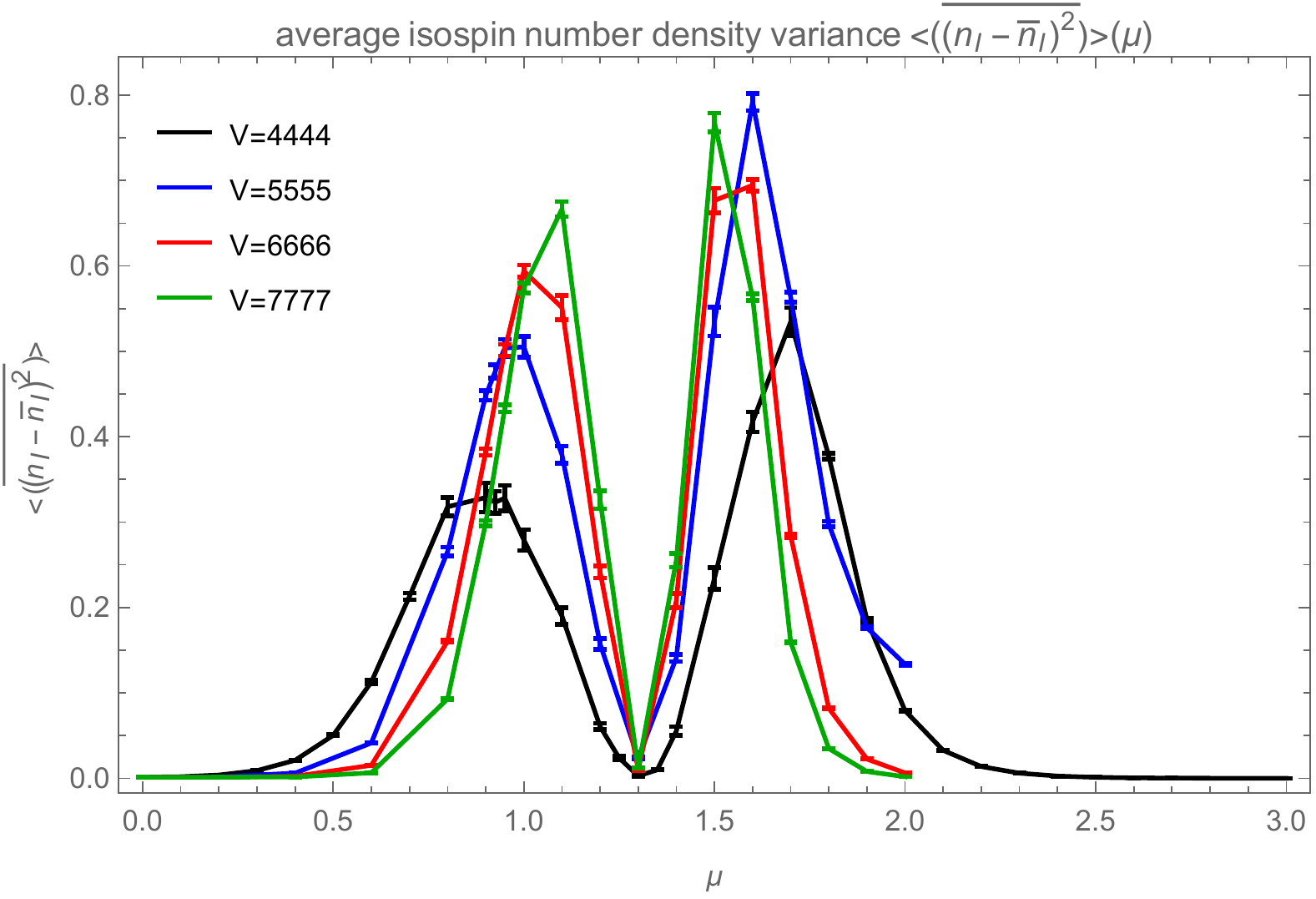}
\end{minipage}\hfill
\begin{minipage}[t]{0.485\linewidth}
\centering
\includegraphics[width=\linewidth]{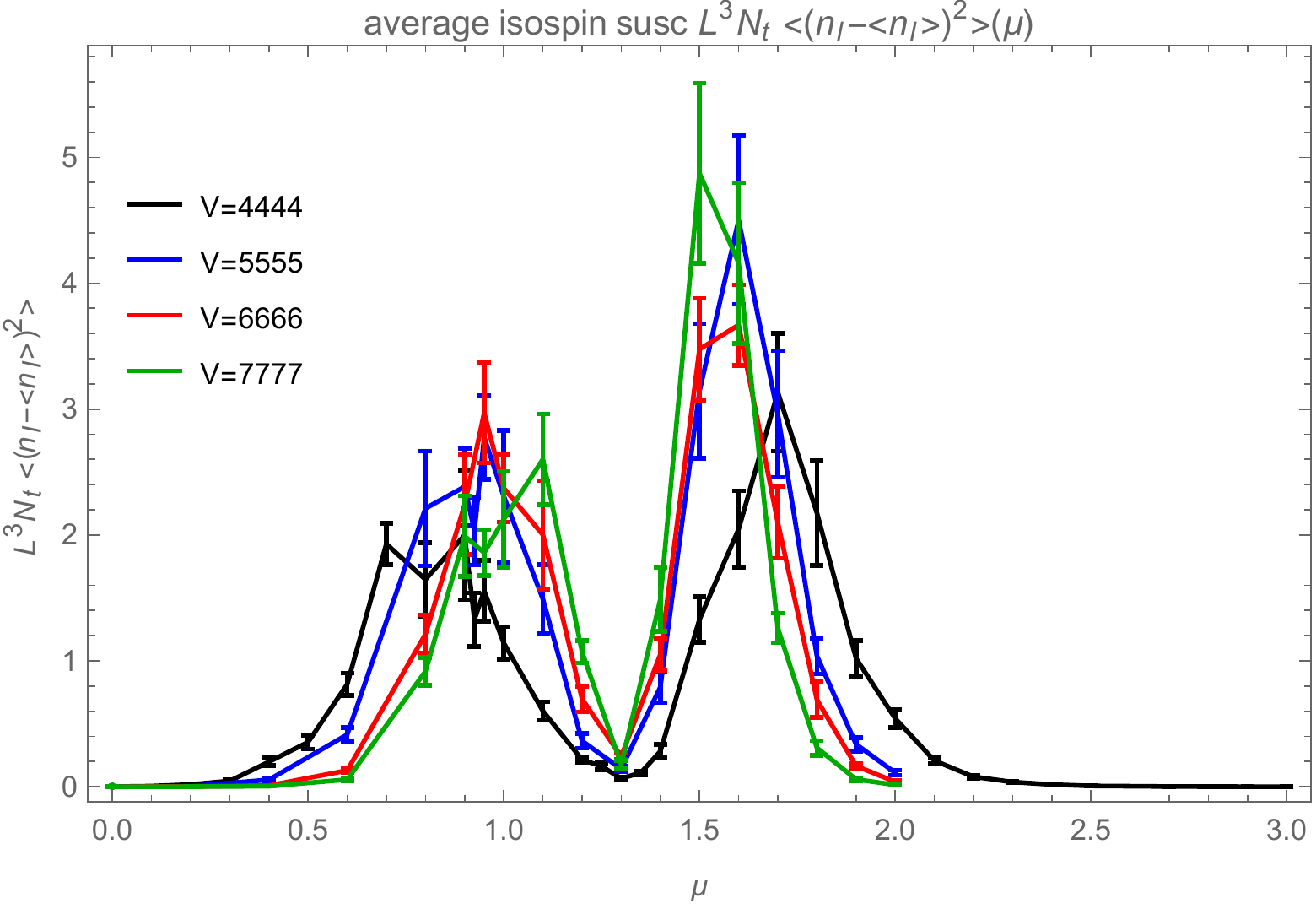}
\end{minipage}
\caption{Average isospin density space-time variance eq. \eqref{eq:isodensstvari} (left) and isospin density ensemble variance (right) as functions of $\mu$ for different system sizes, recorded at $\kappa=0.15$ and $\beta=5.0$. The figures show that at the half-filling point at $\mu\approx 1.3$, both the space-time variance and also the ensemble variance of the isospin density are essentially zero, which suggests that the isospin density becomes independent of the gauge fields for this particular value of $\mu$.}
  \label{fig:avisospatvarvarnt}
\end{figure}

To see where the volume independency comes from, we define the \emph{local isospin density},
\begin{multline}
n_{I}\of{x,\mu}\,=\,2\,\Repart{\trace_{c,d}\cof{D\ssof{x,x+\hat{4},\mu}\,D^{-1}\ssof{x+\hat{4},x,\mu}\right.\right.\\
\left.\left.-\,D\ssof{x+\hat{4},x,\mu}\,D^{-1}\ssof{x,x+\hat{4},\mu}}}\ ,\label{eq:localisodens}
\end{multline}
where $\trace_{c,d}$ is the trace with respect to color and Dirac indices, such that
\[
n_{I}\of{\mu}\,=\,\frac{1}{L^{3}\,N_{t}}\partd{\log Z_{q}\of{\mu}}{\mu}\,=\,\avof{\frac{1}{L^{3}\,N_{t}}\sum\limits_{x}\,n_{I}\of{x,\mu}}_{q}\ ,\label{eq:totisodens}
\]
and consider its average space-time variance, given by 
\[
\avof{\overline{\of{n_{I}\,-\,\bar{n}_{I}}^{2}}}_{q}\,=\,\avof{\frac{1}{L^{3}\,N_{t}}\sum\limits_{x}\of{n_{I}\of{x,\mu}\,-\,\frac{1}{L^{3}\,N_{t}}\sum\limits_{y}\,n_{I}\of{y,\mu}}^{2}}_{q}\ .\label{eq:isodensstvari}
\]
This quantity is shown on the left-hand side of figure \ref{fig:avisospatvarvarnt} for the same ensembles that were used in figure \ref{fig:avisodensvarnt}. As can be seen, \eqref{eq:isodensstvari} is essentially zero at the half-filling point, which means that the half-filling is realized very homogeneously. As, according to the right-hand side of figure \ref{fig:avisospatvarvarnt}, also the ensemble variance
\[
L^{3}\,N_{t}\,\avof{\sof{n_{I}\,-\,\avof{n_{I}}_{q}}^{2}}_{q}\,=\,L^{3}\,N_{t}\,\sof{\avof{n_{I}^{2}}_{q}\,-\,\avof{n_{I}}_{q}^{2}}
\]
of the total isospin density \eqref{eq:totisodens} is approximately zero at the half-filling point, we can conclude that the homogeneity of the local isospin density \eqref{eq:localisodens} at this point is due to the fact that its dependency on the gauge field is highly suppressed at half-filling, which explains the volume independency.\\

\begin{figure}[h]
\centering
\begin{minipage}[t]{0.485\linewidth}
\centering
\includegraphics[width=\linewidth]{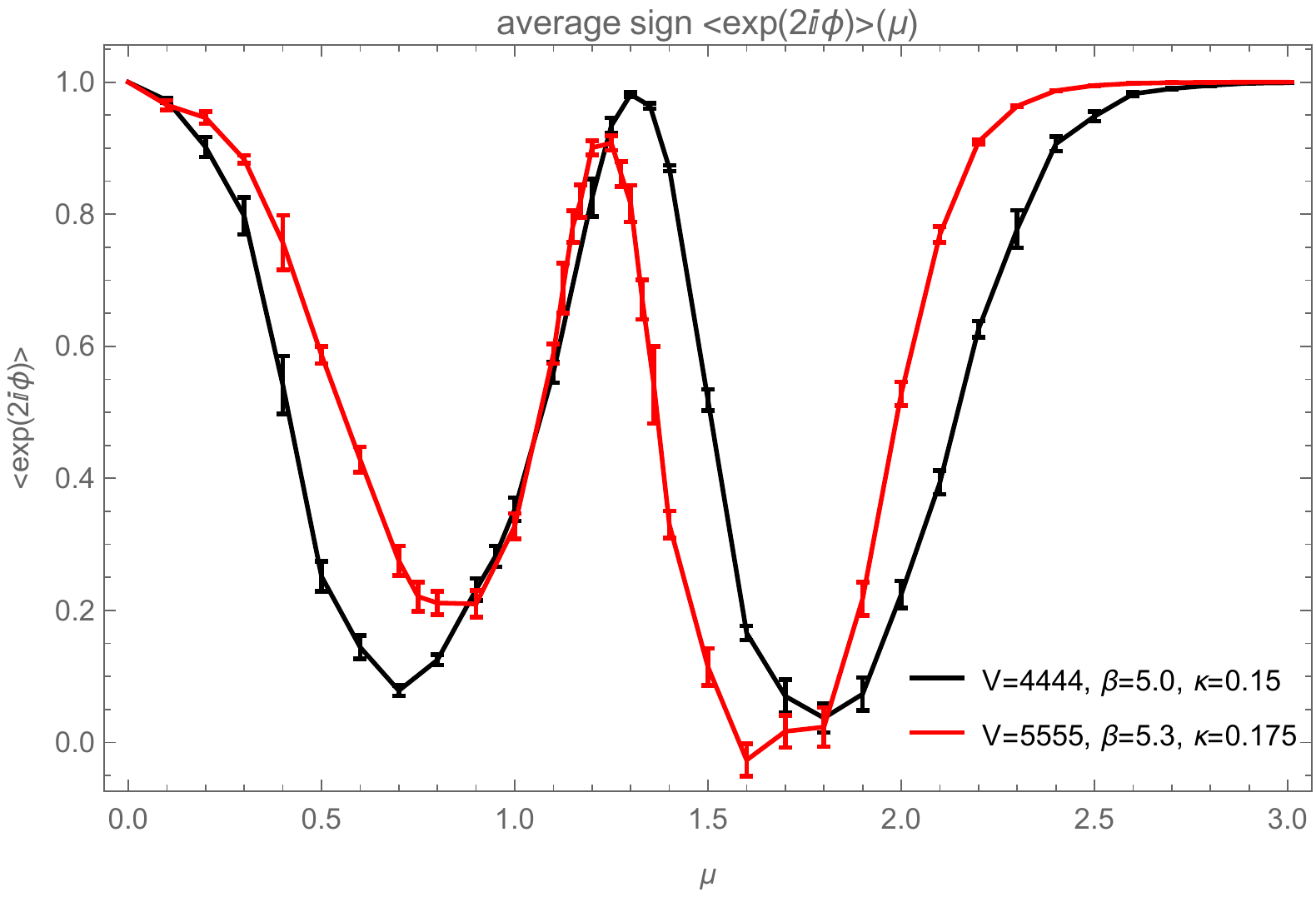}
\end{minipage}\hfill
\begin{minipage}[t]{0.485\linewidth}
\centering
\includegraphics[width=\linewidth]{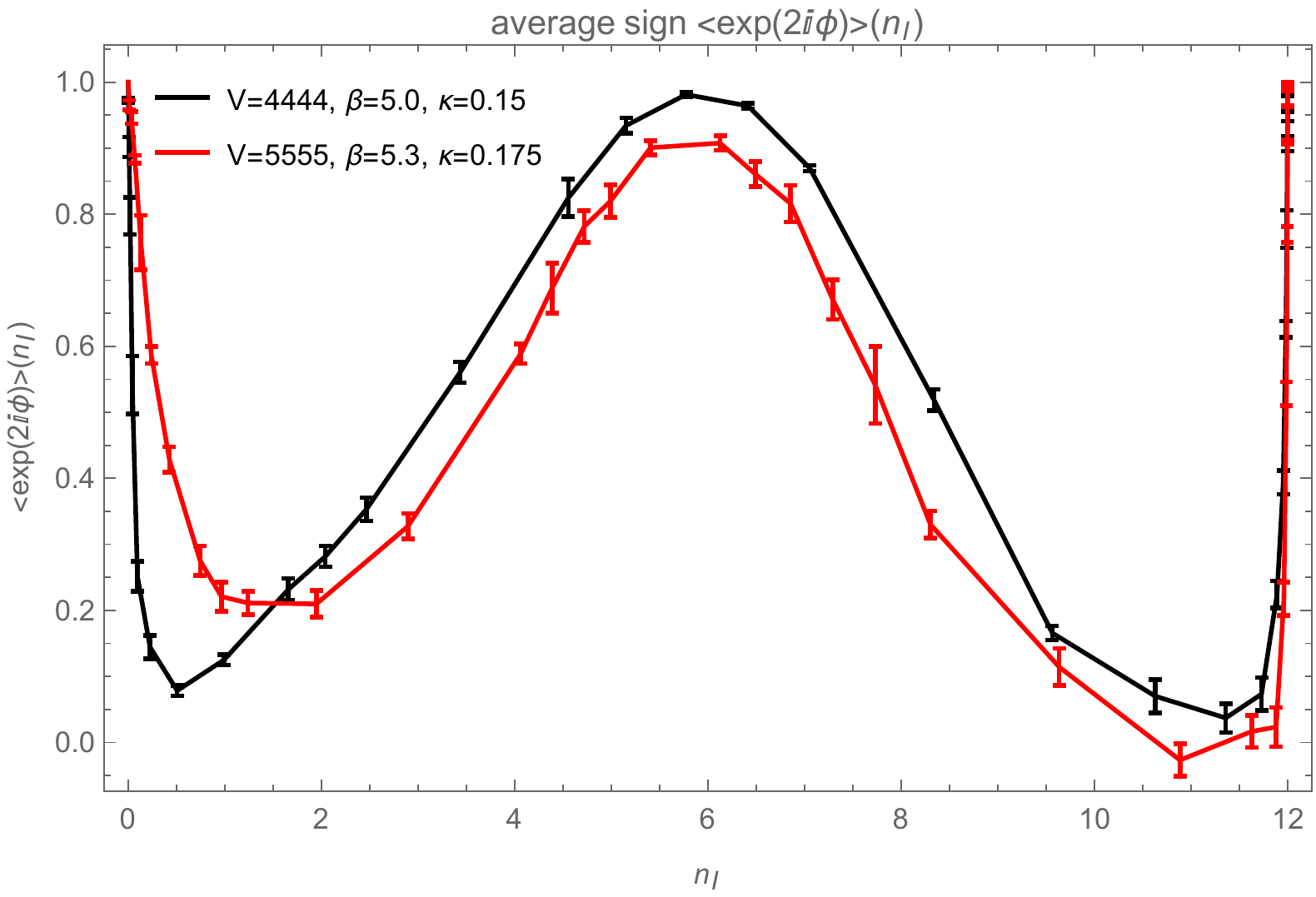}
\end{minipage}
\caption{Average sign as a function of $\mu$ (left) and as a function of the isospin number density (right) for two different sets of hopping parameter $\kappa$, inverse coupling $\beta$ and system size. The figures show that the average sign has a maximum at half-filling where its value is close to one. For the larger system there is however a clearly visible deviation from unity.}
  \label{fig:avsign}
\end{figure}

In figure \ref{fig:avsign} we also show the average sign (or average phase) \eqref{eq:avphase} for a two-flavor system as a function of the chemical potential (left) and as a function of the isospin number density (right), for two different sets of simulation parameters: once for a system with $L^3\,N_{t}\,=\,4^4$, inverse coupling $\beta\,=\,5.0$ and hopping parameter $\kappa\,=\,0.15$ and once for one with $L^3\,N_{t}\,=\,5^4$, $\beta\,=\,5.3$ and $\kappa\,=\,0.175$. As can be seen, for both systems the average sign has a local maximum at the half-filling point, where it almost reaches a value of one again. However, for the larger system a clear deviation from one is visible. The right-hand part of Fig. \ref{fig:avsign} also shows an approximate symmetry about the half-filling point.\\

\begin{figure}[h]
\centering
\begin{minipage}[t]{0.49\linewidth}
\includegraphics[width=\linewidth]{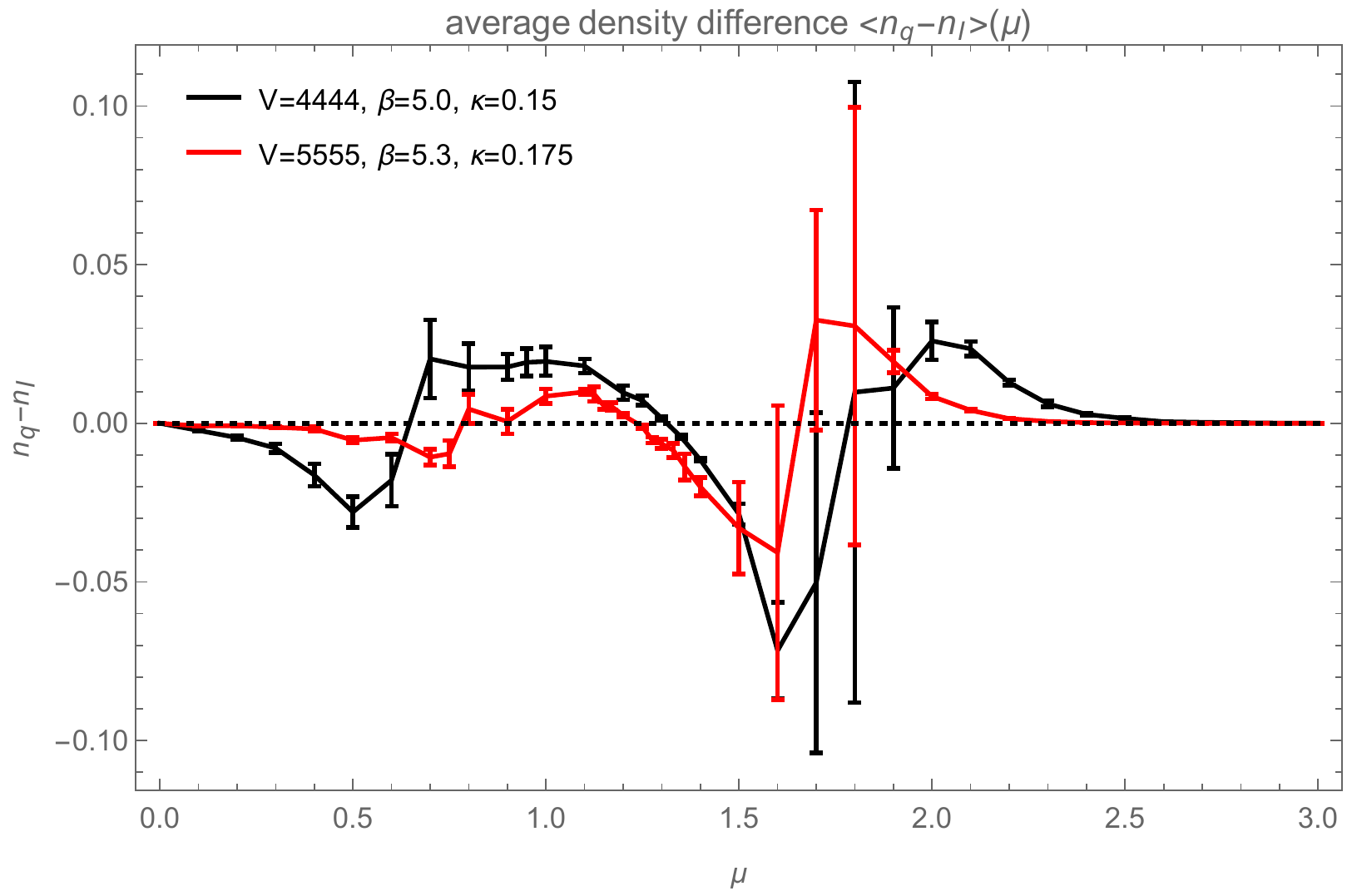}
\caption{Difference between average quark number density and average isospin number density as a function of $\mu$ for the same ensembles used in Fig. \ref{fig:avsign}. The quark number density was obtained by reweighting.}
  \label{fig:avdensdiff}
  \end{minipage}\hfill
  \begin{minipage}[t]{0.49\linewidth}
\centering
\includegraphics[width=\linewidth]{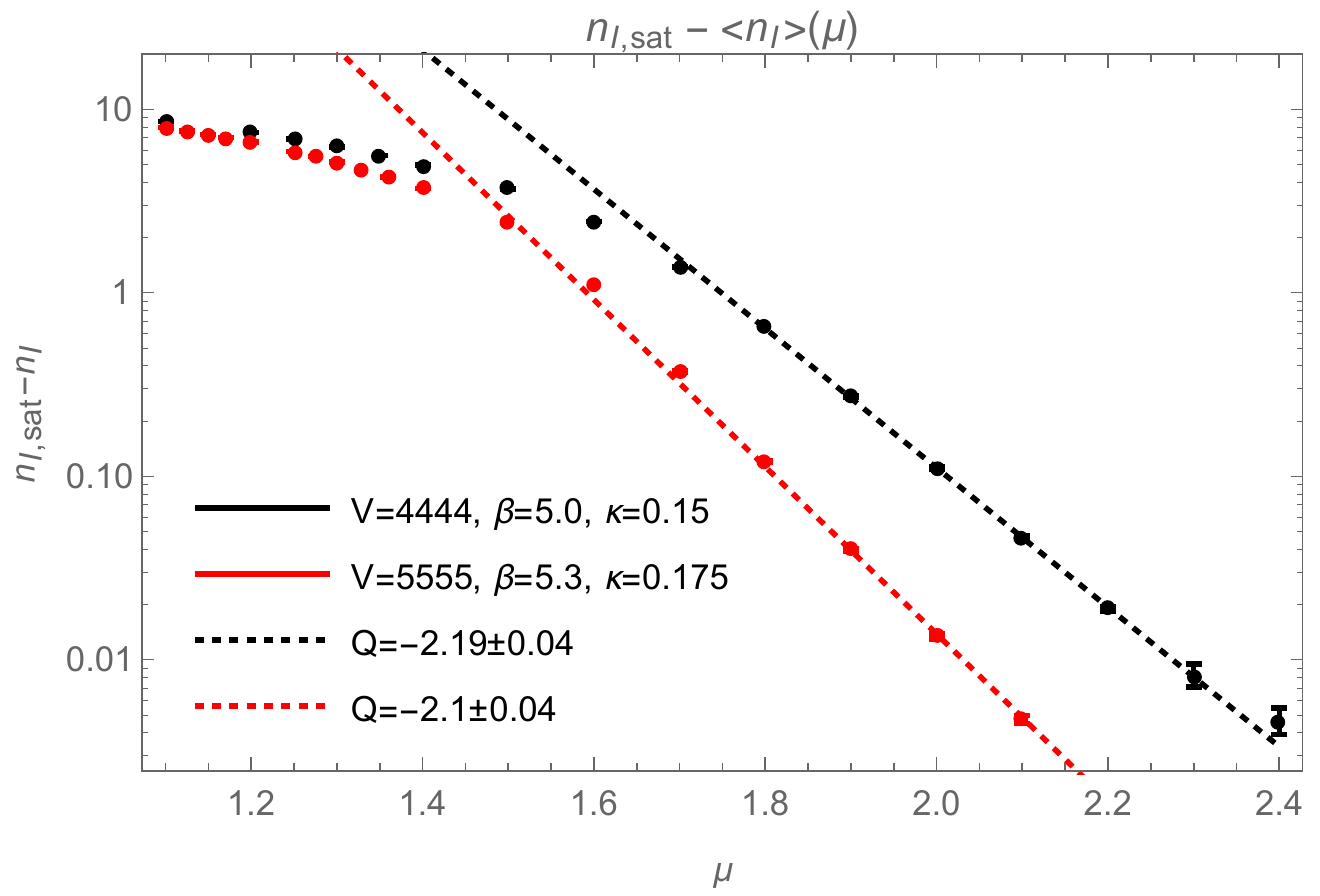}
\caption{The figure shows a log-plot of $n_{I,sat}-\avof{n_{I}}\of{\mu}$ for the two ensembles used in Fig. \ref{fig:avsign}. The dotted lines correspond to fits of the form $n_{I,sat}-\avof{n_{I}}\of{\mu}\,=\,a\,\e^{N_{t}\,Q\,\of{\mu}}$ to the data for $\mu>1.5\,\tilde{\mu}$, where $\tilde{\mu}=-\log\of{2\kappa}$ is the half-filling value of $\mu$ in each ensemble. As can be seen, both fits yield $Q\approx -2$, although the simulation parameters are quite different.}
  \label{fig:avisodensholecharge}
\end{minipage}
\end{figure}

Although the average sign is not exactly one at half-filling, the fact that it has a local maximum, implies that isospin number density and quark number density must be the same at this point, as we have
\[
\avof{n_{q}}\of{\mu}\,-\,\avof{n_{I}}\of{\mu}\,=\,\frac{1}{L^{3}\,N_{t}}\of{\partd{\log Z\of{\mu}}{\mu}\,-\,\partd{\log Z_{q}\of{\mu}}{\mu}}\,=\,\frac{1}{L^{3}\,N_{t}}\partd{\log \avof{\op{R}}_{q}\of{\mu}}{\mu}\ ,\label{eq:avdensdiff}
\]
which vanishes at an extremum of $\avof{\op{R}}_{q}\of{\mu}$. This is of course also true at the minima of $\avof{\op{R}}_{q}\of{\mu}$, which means, that although the sign problem is worst at these points, the quark number density could there be obtained with high accuracy. The problem is, however, to determine the location of these points, as they will not be volume independent as is the case for the half-filling point. In figure \ref{fig:avdensdiff}, we show \eqref{eq:avdensdiff} for the same data that was used in figure \ref{fig:avsign}. The density difference was obtained by reweighting the quark number density from the quenched to the un-quenched system and then subtracting the isospin number density.\\ 

We mentioned that, in the Hubbard model, the sign problem disappears at half-filling due to a particle-hole symmetry. We could now ask if also in LQCD such a particle-hole symmetry is the origin of the (almost) disappearing sign problem at half-filling. A first indication that the answer to this question could be yes, comes from Fig. \ref{fig:avisodensholecharge}, which shows, again for a two-flavor system, coupled to an isospin chemical potential $\mu$, the quantity $n_{I,sat}-\avof{n_{I}}\of{\mu}$ for two different ensembles, where $n_{I,sat}=\lim_{\mu\rightarrow\infty} \avof{n_{I}}\of{\mu}$ is the saturation value of the average isospin density $\avof{n_{I}}\of{\mu}$ at large $\mu$. We could think of $n_{I,sat}-\avof{n_{I}}\of{\mu}$ as being the density of "holes in the saturated state".\\
For the isospin density itself we would expect $\avof{n_{I}}\of{\mu}\propto \e^{N_{t}\,Q\,\mu}$ for $\mu\ll\tilde{\mu}$, where $\tilde{\mu}=-\log\ssof{2\kappa}$ is the half-filling value of the chemical potential and $Q=2$ is the charge of an elementary excitation (charged meson). The fits in Fig. \ref{fig:avisodensholecharge} now show that $n_{I,sat}-\avof{n_{I}}\of{\mu}$ behaves in precisely the same way for $\mu\gg\tilde{\mu}$, but with $Q=-2$,  which means that the "hole-excitations" from the saturating state at large chemical potential behave just like the anti-particles of the elementary excitations of the theory at small chemical potential. The strong interaction treats therefore fermion holes in the saturated state in a similar way as the fermions themselves in the vacuum, such that the holes also have to form gauge invariant combinations of the form of mesons and baryons.\\
Having this symmetry between particles at zero density and holes at saturation, makes it plausible that at half-filling, where particles and holes appear with equal densities, they indeed behave symmetrically. The same should be true for the non-phase-quenched system where the elementary excitations would be baryons with $Q=3$ at zero density and holes of the form of anti-baryons with $Q=-3$ at saturation. Again this suggests that at half-filling, where baryons and baryon-holes have equal density, we should have particle-hole symmetry.

\subsection{Dominance of Lattice Artifacts: a Simple Criterion}\label{ssec:doflatticeartifacts}

In contrast to the systems studied in solid state physics, where the lattice is physical and half-filling describes a physical state as in the fermionic Hubbard model, the lattice in LQCD is jut a regulator which has to be removed (e.g. in numerical works, by doing continuum extrapolation) in order to extract real physics. The half-filling state of a LQCD system is therefore just an unphysical lattice artifact and should be of no relevance for the continuum physics.\\
As long as we are running simulations sufficiently far away from the continuum limit (such that the maximal fermion density on the lattice (in physical units) is far below the density where the Pauli exclusion principle would become important in the continuum), the first appearance of a minimum in the average sign as a function of increasing chemical potential (or as a function of increasing isospin number density) as shown in figure \ref{fig:avsign}, indicates the point where lattice artifacts start to become dominant. This becomes clear by noting that the average sign is a measure for the overlap between a system with a quark chemical potential and the corresponding phase-quenched system (with an isospin instead of a quark chemical potential). As the right-hand side of figure \ref{fig:avsign} shows, the average sign drops dramatically as soon as the isospin density becomes non-zero, reflecting the fact that a pion condensate develops in the isospin system, leading to a ground state which is rather different from that of the corresponding system with a quark chemical potential. The fact that the overlap between the two systems starts to increase again with increasing $\mu$ (i.e. that the average sign starts to increase again) indicates, that the influence of the Pauli exclusion principle starts to become dominant on the lattice: the fact that the quark chemical potential favors the $u$ and $d$ quarks, while the isospin chemical potential favors $u$ and $\bar{d}$, starts to become less important as for example the "$d$-hole" in the system with the quark chemical potential starts to play the role of the $\bar{d}$ in the isospin system and vice versa. This happens at densities significantly below half-filling, and measurements of observables taken at larger densities should be considered with caution. 

\section{Heavy-Dense Analysis}\label{sec:hdanalysis}
In the following section we will analyze the half-filling point of LQCD in the heavy-dense limit and at strong coupling. These simplifications will allow us to understand the origin of the properties of the half-filling point.     

\subsection{The Heavy-Dense Approximation}\label{ssec:hdapprox}

The goal of this section is to obtain an expression for the Partition function in the so-called \emph{heavy-dense} approximation at strong coupling, which can be integrated analytically. The heavy-dense approximation usually consists of taking the limit $\kappa\rightarrow 0$, $\mu\rightarrow \infty$, keeping $\kappa \e^{\mu}$ finite while terms proportional to just $\kappa$ or $\kappa \e^{-\mu}$ vanish. In our case, it would be more appropriate to call it the \emph{static quark} limit, as we will just drop the spatial hopping terms from the Dirac operator \eqref{eq:wilsondiracop}, but not the one corresponding to backward hopping in time, which is proportional to $\kappa \e^{-\mu}$. The resulting fermion determinant is well-known as the leading term in the \emph{spatial hopping} expansion that was already used in \cite{Langelage} to derive an effective model for QCD or in \cite{Sexty} to obtain a holomorphic action for complex Langevin simulations of LQCD.\\

Let us start by expanding the determinant of the Dirac operator \eqref{eq:wilsondiracop} in terms of closed fermion loops $C_{l_{0}}$ of length $l_{0}$ \cite{Rothe}:
\begin{align}
\Det{D}\,&=\,\exp\sof{\Trace{\ln D}}\,=\,\exp\sof{\Trace{\ln\of{\id-\kappa\,H}}}\nonumber\\
&=\,\exp\bof{-\sum\limits_{k=1}^{\infty}\frac{\kappa^{k}}{k}\Trace{H^{k}}}\nonumber\\
&=\,\exp\bof{-\sum\limits_{l_{0}}\sum\limits_{\cof{C_{l_{0}}}}\sum_{n=1}^{\infty}\frac{\of{\kappa^{l_{0}}}^{n}}{n}\trace_{c,d}\sof{\of{M_{C_{l_{0}}}}^{n}}}\nonumber\\
&=\,\exp\bof{\sum\limits_{l_{0}}\sum\limits_{\cof{C_{l_{0}}}}\trace_{c,d}\sof{\ln\of{\id-\kappa^{l_{0}}M_{C_{l_{0}}}}}}\nonumber\\
&=\,\prod\limits_{l_{0}}\prod\limits_{\cof{C_{l_{0}}}}\det_{c,d}\sof{\id-\kappa^{l_{0}}M_{C_{l_{0}}}}\ ,\label{eq:fullfermiondet}
\end{align}
with $M_{C_{l_{0}}}\,=\,H_{x_{1} x_{2}}\cdots H_{x_{l_{0}} x_{1}}$ being the matrix product of hopping terms $H_{x y}\,=\,\of{S+T}_{x y}$ along the contour $C_{l_{0}}$ of length $l_{0}$, where $S$ and $T$ are the spatial and temporal hopping matrices defined in \eqref{eq:wilsondiracop}. The subscripts "c" and "d" of "$\det$" and "$\trace$" indicate that the operators act on color and Dirac space only. To get from the second to the third line in \eqref{eq:fullfermiondet}, we have used that $k$ can be written as $k=l_{0}\,n$, where $l_0$ is the length of the contour $C_{l_{0}}$, and $n$ the number of windings around $C_{l_{0}}$. The change in the denominator ($k\rightarrow n$ instead of $k\rightarrow l_{0}\,n$) comes from the fact that $\Trace{H^{l_{0}\,n}}$ contains $\trace_{c,d}\of{\of{M_{C_{l_0}}}^{n}}$ exactly $l_{0}$ times, once for every possible base point (or starting point) of a loop on $C_{l_0}$, which cancels the $l_{0}$ in the denominator.\\

As already mentioned, the static quark approximation now consists of dropping the spatial hopping matrix $S$ from \eqref{eq:wilsondiracop} and \eqref{eq:fullfermiondet}, which ultimately means that $M_{C_{l_{0}}}$ is non-zero only if $C_{l_{0}}$ wraps around the temporal direction, i.e. $l_{0}=n_{t}$ (we use from now on a lower-case $n_{t}$ to represent the temporal system size) and $M_{C_{n_{t}}}$ has to be one of the $n_{x}\cdot n_{y}\cdot n_{z}$ matrices
\[
-\e^{\mu\,n_{t}}\,P_{\bar{x}}\otimes\of{\id-\gamma_{4}}^{n_{t}}\,=\,T_{\bar{x}, \bar{x}+\hat{4}} \cdots T_{\bar{x}+\of{n_{t}-1}\hat{4}, \bar{x}}\label{eq:timeloop}
\]
or their time-reversed versions
\[
-\e^{-\mu\,n_{t}}P^{\dagger}_{\bar{\scriptstyle x}}\otimes\of{\id+\gamma_{4}}^{n_{t}}\,=\,T_{\bar{\scriptstyle x}, \bar{\scriptstyle x}+\of{n_{t}-1}\hat{\scriptstyle 4}} \cdots T_{\bar{\scriptstyle x}+\hat{\scriptstyle 4}, \bar{\scriptstyle x}}\label{eq:revtimeloop}
\]
where $P_{\bar{\scriptstyle x}}$ is the \emph{Polyakov loop} at spatial position $\bar{x}$. The minus signs on the left-hand sides of \eqref{eq:timeloop} and \eqref{eq:revtimeloop} come from the anti-periodic boundary conditions in temporal direction. With this approximation, the determinant \eqref{eq:fullfermiondet} reduces to a product of \emph{single site determinants}:
\begin{align}
\Det{D}\,=&\,\prod\limits_{\bar{\scriptstyle x}} \of{\det_{c,d}\sof{\id+\kappa^{n_{t}}\e^{\mu\,n_{t}}P_{\bar{\scriptstyle x}}\otimes\of{\id-\gamma_{4}}^{n_{t}}}\det_{c,d}\sof{\id+\kappa^{n_{t}}\e^{-\mu\,n_{t}}P^{\dagger}_{\bar{\scriptstyle x}}\otimes\of{\id+\gamma_{4}}^{n_{t}}}}\nonumber\\
=&\,\prod\limits_{\bar{\scriptstyle x}} \of{\det_{c,d}\sof{\id\,+\,\of{2\kappa\e^{\mu}}^{n_{t}}P_{\bar{\scriptstyle x}}\otimes\of{\begin{smallmatrix}0 & 0\\0 & \id_{2}\end{smallmatrix}}}\det_{c,d}\sof{\id+\of{2\kappa\e^{-\mu}}^{n_{t}}P^{\dagger}_{\bar{\scriptstyle x}}\otimes\of{\begin{smallmatrix}\id_{2} & 0\\0 & 0\end{smallmatrix}}}}\nonumber\\
=&\,\prod\limits_{\bar{\scriptstyle x}} \of{\det_{c}^{2}\sof{\id+\of{2\kappa\e^{\mu}}^{n_{t}}P_{\bar{\scriptstyle x}}}\det_{c}^{2}\sof{\id+\of{2\kappa\e^{-\mu}}^{n_{t}}P^{\dagger}_{\bar{\scriptstyle x}}}}\ ,\label{eq:fermiondetprod}
\end{align}
More precisely, the last line of \eqref{eq:fermiondetprod} shows that for each site, we get four factors, each in the form of a determinant in color space: the first two correspond to the two distinct states of a spin-1/2 particle while the second two are due to its anti-particle. These color-determinants can be re-written in terms of traces of powers of the Polyakov loop:
\begin{align}
\det_{c}\sof{\id+\of{2\kappa\e^{\mu}}^{n_{t}}&P_{\bar{\scriptstyle x}}}\,=\,\frac{1}{3!}\epsilon\indices{_{a_{1}}_{a_{2}}_{a_{3}}}\epsilon\indices{_{b_{1}}_{b_{2}}_{b_{3}}}\nonumber\\
&\,\cdot\sof{\id+\of{2\kappa\e^{\mu}}^{n_{t}}P_{\bar{\scriptstyle x}}}_{a_{1} b_{1}}\,\sof{\id+\of{2\kappa\e^{\mu}}^{n_{t}}P_{\bar{\scriptstyle x}}}_{a_{2} b_{2}} \sof{\id+\of{2\kappa\e^{\mu}}^{n_{t}}P_{\bar{\scriptstyle x}}}_{a_{3} b_{3}}\nonumber\\
=&\,1\,+\,\sum\limits_{k=1}^{3}\sof{2\kappa\e^{\mu}}^{n_{t}\,k}\delta_{a_{1}}^{[b_{1}}\cdots\delta_{a_{k}}^{b_{k}]} P_{\bar{\scriptstyle x},a_{1} b_{1}}\cdots P_{\bar{\scriptstyle x},a_{k} b_{k}}\nonumber\\
=&\,1\,+\,\of{2\kappa\e^{\mu}}^{n_{t}}\trace_{c}\sof{P_{\bar{\scriptstyle x}}}\,+\,\frac{\of{2\kappa\e^{\mu}}^{2\,n_{t}}}{2}\sof{\trace^{2}_{c}\sof{P_{\bar{\scriptstyle x}}}\,-\,\trace_{c}\sof{P_{\bar{\scriptstyle x}}^{2}}}\nonumber\\
&\hphantom{1\,}+\,\frac{\of{2\kappa\e^{\mu}}^{3\,n_{t}}}{6}\sof{\trace^{3}_{c}\sof{P_{\bar{\scriptstyle x}}}\,-\,3\,\trace_{c}\sof{P_{\bar{\scriptstyle x}}^{2}}\trace_{c}\sof{P_{\bar{\scriptstyle x}}}\,+\,2\,\trace_{c}\sof{P_{\bar{\scriptstyle x}}^{3}}}.\label{eq:singelsitedett}
\end{align}
By using that the characteristic equation for a matrix $X\,\in\,\SU{3}$ reads
\[
\chi_{{}_{\scriptstyle X}}\of{\lambda}\,=\,\det_{c}\of{\lambda\id-X}\,=\,\lambda^3-\trace_{c}\of{X}\lambda^2+\trace_{c}\of{X^{\dagger}}\lambda-1\,=\,0
\]
and, according to the Cayley-Hamilton theorem, $X$ satisfies its own characteristic equation, $\chi_{{}_{\scriptstyle X}}\of{X}\,=\,0$ (in the spectral sense), it follows that
\begin{align}
X^3\,=&\,\id\,+\trace_{c}\of{X}\,X^2\,-\,\trace_{c}\of{X^{\dagger}}\,X,\\
X^2\,=&\,X^3\,X^{\dagger}\,=\,\,X^{\dagger}\,+\trace_{c}\of{X}\,X\,-\,\trace_{c}\of{X^{\dagger}}\,\id,
\end{align}
and therefore
\begin{align}
\trace_{c}\of{X^3}\,=&\,3\,+\,\trace_{c}\of{X}\trace_{c}\of{X^2}\,-\,\trace_{c}\of{X^{\dagger}}\trace_{c}\of{X}\ ,\\
\trace_{c}\of{X^{\dagger}}\,=&\,\sof{\trace_{c}^{2}\of{X}\,-\,\trace_{c}\of{X^2}}/2,
\end{align}
which can be used to simplify \eqref{eq:singelsitedett} further:
\[
\det_{c}\sof{\id+\of{2\kappa\e^{\mu}}^{n_{t}}P_{\bar{\scriptstyle x}}}\,=\,1\,+\,\of{2\kappa\e^{\mu}}^{n_{t}}\trace_{c}\sof{P_{\bar{\scriptstyle x}}}\,+\,\of{2\kappa\e^{\mu}}^{2\,n_{t}}\trace_{c}\sof{P^{\dagger}_{\bar{\scriptstyle x}}}\,+\,\of{2\kappa\e^{\mu}}^{3\,n_{t}}.\label{eq:singelsitedet}
\]
Finally, expressing the traces in \eqref{eq:singelsitedet} in terms of the eigenvalues of the Polyakov loop $P_{\bar{\scriptstyle x}}$,
\[
\lambda_{1}\of{\bar{x}}\,=\,\e^{\ii\theta_{1}\of{\bar{\scriptstyle x}}},\,\lambda_{2}\of{\bar{x}}\,=\,\e^{\ii\theta_{2}\of{\bar{\scriptstyle x}}},\,\lambda_{3}\of{\bar{x}}=1/\of{\lambda_{1}\of{\bar{x}}\lambda_{2}\of{\bar{x}}}
\]
we arrive at
\begin{multline}
\det_{c}^{2}\sof{\id+\of{2\kappa\e^{\mu}}^{n_{t}}P_{\bar{\scriptstyle x}}}\,=\,\sof{\e^{\ii\of{\theta_{1}\of{\bar{\scriptstyle x}}+\theta_{2}\of{\bar{\scriptstyle x}}}}+\of{2\kappa\e^{\mu}}^{n_{t}}}^{2}\\
\cdot\sof{\e^{-\ii\theta_{1}\of{\bar{\scriptstyle x}}}+\of{2\kappa\e^{\mu}}^{n_{t}}}^{2}\sof{\e^{-\ii\theta_{2}\of{\bar{\scriptstyle x}}}+\of{2\kappa\e^{\mu}}^{n_{t}}}^{2}\ ,\label{eq:pfermiondet}
\end{multline}
and similarly,
\begin{multline}
\det_{c}^{2}\sof{\id+\of{2\kappa\e^{-\mu}}^{n_{t}}P^{\dagger}_{\bar{\scriptstyle x}}}\,=\,\sof{\e^{-\ii\of{\theta_{1}\of{\bar{\scriptstyle x}}+\theta_{2}\of{\bar{\scriptstyle x}}}}+\of{2\kappa\e^{-\mu}}^{n_{t}}}^{2}\\
\cdot\sof{\e^{\ii\theta_{1}\of{\bar{\scriptstyle x}}}+\of{2\kappa\e^{-\mu}}^{n_{t}}}^{2}\sof{\e^{\ii\theta_{2}\of{\bar{\scriptstyle x}}}+\of{2\kappa\e^{-\mu}}^{n_{t}}}^{2}.\label{eq:mfermiondet}
\end{multline}
The single site fermion determinants in \eqref{eq:fermiondetprod} reduce therefore to the simple form:
\begin{multline}
\Det{D\of{\theta_{1},\theta_{2};\mu,\kappa,n_{t}}}\,=\\
\sof{\e^{\ii\of{\theta_{1}+\theta_{2}}}+\of{2\kappa\e^{\mu}}^{n_{t}}}^{2}\sof{\e^{-\ii\theta_{1}}+\of{2\kappa\e^{\mu}}^{n_{t}}}^{2}\sof{\e^{-\ii\theta_{2}}+\of{2\kappa\e^{\mu}}^{n_{t}}}^{2}\\
\cdot\sof{\e^{-\ii\of{\theta_{1}+\theta_{2}}}+\of{2\kappa\e^{-\mu}}^{n_{t}}}^{2}\sof{\e^{\ii\theta_{1}}+\of{2\kappa\e^{-\mu}}^{n_{t}}}^{2}\sof{\e^{\ii\theta_{2}}+\of{2\kappa\e^{-\mu}}^{n_{t}}}^{2}.\label{eq:ssitedet}
\end{multline}
Using the Haar measure for the trace of the Polyakov loop,
\[
\dd P\,=\,H\of{\theta_{1},\theta_{2}}\,\dd\theta_{1}\dd\theta_{2}\ ,
\]
where
\[
H\of{\theta_{1},\theta_{2}}\,=\,-\frac{1}{24\,\pi^2}\prod\limits_{i<j}\abs{\lambda_{i}-\lambda_{j}}^{2}\,=\,\frac{\sof{\sin\of{\theta_{1}-\theta_{2}}-\sin\of{2\theta_{1}+\theta_{2}}+\sin\of{\theta_{1}+2\theta_{2}}}^{2}}{6\,\pi^{2}}\ ,
\] 
and neglecting the plaquette action (i.e. considering the strong coupling limit $\beta=0$), we can write down a heavy-dense partition function for QCD with two flavors, $u,d$:
\begin{multline}
Z\of{\mu_{u},\mu_{d},\kappa_{u},\kappa_{d},n_{t},\Lambda}\,=\,\prod\limits_{\bar{\scriptstyle x}\in \Lambda}\int\dd P_{\bar{\scriptstyle x}}\,\Det{D_{\bar{\scriptstyle x}}\of{\mu_{u},\kappa_{u},n_{t}}}\Det{D_{\bar{\scriptstyle x}}\of{\mu_{d},\kappa_{d},n_{t}}}\\
=\,\bof{\underbrace{\int\limits_{0}^{2\pi}\int\limits_{0}^{2\pi}\dd{\theta_{1}}\dd{\theta_{2}}\,H\of{\theta_{1},\theta_{2}}\Det{D\of{\theta_{1},\theta_{2};\mu_{u},\kappa_{u},n_{t}}}\Det{D\of{\theta_{1},\theta_{2};\mu_{d},\kappa_{d},n_{t}}}}_{Z_{s}\of{\mu_{u},\mu_{d},\kappa_{u},\kappa_{d},n_{t}}}}^{\abs{\Lambda}}\ ,\label{eq:qcd2fpartf}
\end{multline}
where $\kappa_{u/d}$ and $\mu_{u/d}$ are the hopping parameter and the chemical potential corresponding to the $u/d$ flavor, $\Lambda$ is the spatial lattice with $\abs{\Lambda}$ sites, and $Z_{s}\of{\mu_{u},\mu_{d},\kappa_{u},\kappa_{d},n_{t}}$ is the single site partition function.\\
The integral \eqref{eq:qcd2fpartf} can be evaluated explicitly but results in a rather lengthy expression. For the purpose of illustration its explicit form is shown in appendix \ref{sec:isospinssitepartf} for the case of degenerate quark masses, $\kappa_{u}=\kappa_{d}=\kappa$, once with an isospin ($\mu=\mu_{u}=-\mu_{d}$) and once with a quark ($\mu=\mu_{u}=\mu_{d}$) chemical potential.

\subsubsection{Heavy-Dense Staggered Fermions}\label{sssec:reltostaggeredfermions}
The \emph{Staggered fermion} operator is given by
\[
D_{st\,x,I;y,J}\,=\,m\of{\delta_{x y}\delta_{I J}+\frac{1}{2\,m}\sum\limits_{\mu=1}^{4}\,\eta_{\mu}\of{x}\of{U_{\mu}\of{x}\delta_{x+\hat{\mu},y}-U^{\dagger}_{\mu}\of{x-\hat{\mu}}\delta_{x-\hat{\mu},y}}}\ ,\label{eq:staggeredfop}
\]
with the \emph{Staggered phase} $\eta_{\mu}\of{x}=\of{-1}^{3\,x_{1}+2\,x_{2}+x_{3}}$ and the bare lattice fermion mass $m$. The full static quark fermion determinant for \eqref{eq:staggeredfop} is more difficult to obtain than the Wilson fermion analogue \eqref{eq:fermiondetprod}. The reason is that Staggered fermions allow for retracting fermion loops, such that even without spatial hoppings, several distinct temporal loops are possible. However, if we are just interested in the heavy-dense limit, where $\mu$ and $m$ are assumed to be large while $\e^{\mu-\log\of{2\,m}}$ is of order one, the leading term reads   
\[
\Det{D_{st}}\,\propto\,\prod\limits_{\bar{x}}\det_{c}\sof{\id+\e^{\of{\mu-\log\of{2 m}}n_{t}} P\of{\bar{x}}}\ .\label{eq:staggerdhddet}
\]
Comparing \eqref{eq:staggerdhddet} with \eqref{eq:fermiondetprod} reveals that the Staggered heavy-dense determinant \eqref{eq:staggerdhddet} can be obtained from the heavy-dense limit of \eqref{eq:fermiondetprod} by replacing $2\,\kappa$ by $1/\of{2\,m}$ and dropping the square of the color-determinant.   

\subsection{Analytic Expressions for some Observables}\label{ssec:analyticexpr}
In this section we derive some observables with respect to the system described by \eqref{eq:qcd2fpartf}.
\subsubsection{Average Sign}
In terms of \eqref{eq:qcd2fpartf}, with $\kappa_{u}=\kappa_{d}=\kappa$, the \emph{average sign} is computed as
\[
\avof{\e^{2\ii\,\phi}}\of{\mu,\kappa,n_{t},n_{x}\,n_{y}\,n_{z}}\,=\,\of{\frac{Z_{s}\of{\mu,\mu,\kappa,n_{t}}}{Z_{s}\of{\mu,-\mu,\kappa,n_{t}}}}^{n_{x}\,n_{y}\,n_{z}}\ ,\label{eq:avsign}
\] 
where $\phi$ is the complex phase of the single flavor determinant \eqref{eq:fermiondetprod}, and $n_{x}\,n_{y}\,n_{z}$ is the spatial system size.

\subsubsection{Isospin and Quark Number Density}
The isospin density is obtained from \eqref{eq:qcd2fpartf} by
\[
n_{I}\of{\mu,\kappa,n_{t}}\,=\,\frac{1}{n_{t}}\partd{\log Z_{s}\of{\mu,-\mu,\kappa,n_{t}}}{\mu}\ ,\label{eq:isodens}
\]
and the quark number density, analogously by
\[
n_{q}\of{\mu,\kappa,n_{t}}\,=\,\frac{1}{n_{t}}\partd{\log Z_{s}\of{\mu,\mu,\kappa,n_{t}}}{\mu}\ .\label{eq:quarkdens}
\]
For the difference $n_{q}\of{\mu,\kappa,n_{t}}\,-\,n_{I}\of{\mu,\kappa,n_{t}}$, we find
\[
n_{q}\of{\mu,\kappa,n_{t}}\,-\,n_{I}\of{\mu,\kappa,n_{t}}\,=\,\frac{1}{n_{x}\,n_{y}\,n_{z}\,n_{t}}\partd{\log \avof{\e^{2\ii\,\phi}}\of{\mu,\kappa,n_{t},n_{x}\,n_{y}\,n_{z}}}{\mu}\ ,
\]
as in full QCD.

\subsubsection{Heavy-Dense Quark Propagator}\label{sssec:hdquarkprop}
The heavy-dense quark propagator is most easily obtained by using that \cite{Langelage}
\[
\sof{\id-\kappa\,T\of{\mu}}^{-1}\,=\,\sof{\id-\kappa\,T^{+}\of{\mu}}^{-1}\,+\,\sof{\id-\kappa\,T^{-}\of{\mu}}^{-1}\,-\,\id\ ,\label{eq:temppropsplit}
\]
where
\begin{align}
T^{+}_{x,y}\of{\mu}\,&=\,\of{-1}^{\delta_{y,0}}\delta_{x+\hat{4},y}\,U_{4}\of{x}\otimes\of{\id-\gamma_{4}}\,\e^{\mu},\\
T^{-}_{x,y}\of{\mu}\,&=\,\of{-1}^{\delta_{x,0}}\delta_{x-\hat{4},y}\,U^{\dagger}_{4}\sof{x-\hat{4}}\otimes\of{\id+\gamma_{4}}\,\e^{-\mu}
\end{align}
and 
\[
T_{x,y}\,=\,T^{+}_{x,y}\,+\,T^{-}_{x,y}\ .
\]
We then obtain:
\begin{multline}
\sof{\id-\kappa\,T^{+}\of{\mu}}_{x,a,I;y,b,J}^{-1}\,=\,\bof{\sum\limits_{k=0}^{\infty}\of{\kappa\,T^{+}\of{\mu}}^{k}}_{x,a,I;y,b,J}\,=\,\delta_{\bar{x},\bar{y}}\,\of{2\,\kappa\e^{\mu}}^{\umod\of{y_{4}-x_{4},n_{t}}}\of{-1}^{\theta\of{x_{4}>y_{4}}}\\
\cdot\frac{\of{\id-\gamma_{4}}_{a,b}}{2}\,\bof{P_{\bar{y}}\of{x_{4},y_{4}}\,\sum\limits_{k=0}^{\infty}\of{-\of{2\,\kappa\e^{\mu}}^{n_{t}}P_{\bar{y}}\of{y_{4}}}^{k}}_{I,J}\,+\,\frac{\of{\id+\gamma_{4}}_{a,b}}{2}\delta_{x,y}\,\delta_{I,J}\\
=\,\delta_{\bar{x},\bar{y}}\,\of{2\,\kappa\e^{\mu}}^{\umod\of{y_{4}-x_{4},n_{t}}}\of{-1}^{\theta\of{x_{4}>y_{4}}}\,\frac{\of{\id-\gamma_{4}}_{a,b}}{2}\,\sof{P_{\bar{y}}\of{x_{4},y_{4}}\,\of{\id+\of{2\,\kappa\e^{\mu}}^{n_{t}}P_{\bar{y}}\of{y_{4}}}^{-1}}_{I,J}\\
+\,\frac{\of{\id+\gamma_{4}}_{a,b}}{2}\delta_{x,y}\,\delta_{I,J}\ ,\label{eq:hdquarkproppos}
\end{multline}
where $P_{\bar{y}}\of{y_{4}}$ is the Polyakov loop with base point $y$ and $P_{\bar{y}}\of{x_{4},y_{4}}$ is a Polyakov line at spatial position $\bar{y}$ that, for $x_{4}\neq y_{4}$, starts and ends at euclidean times $x_{4}$, $y_{4}$ respectively (by going in positive time direction), while $P_{\bar{y}}\of{x_{4},x_{4}}\,=\,\id$. The inverse of the matrix $\of{\id+\of{2\,\kappa\e^{\mu}}^{n_{t}}P_{\bar{y}}\of{y_{4}}}$ can easily be carried out (e.g. with Cayley-Hamilton and Leverrier-Faddeev), provided the matrix is not singular:
\begin{multline}
\of{\id+\of{2\,\kappa\e^{\mu}}^{n_{t}}P_{\bar{y}}\of{y_{4}}}^{-1}\,=\,\frac{1}{\det_{c}\sof{\id+\of{2\kappa\e^{\mu}}^{n_{t}}P_{\bar{y}}\of{y_{4}}}}\,\of{\id\of{1+\of{2\,\kappa\,\e^{\mu}}^{n_{t}}\,\trace_{c}\sof{P_{\bar{y}}\of{y_{4}}}}\right.\\
\left.-\,\of{2\,\kappa\,\e^{\mu}}^{n_{t}}\,P_{\bar{y}}\of{y_{4}}\,+\,\of{2\,\kappa\,\e^{\mu}}^{2\,n_{t}}\,P^{\dagger}_{\bar{y}}\of{y_{4}}}\ ,
\end{multline}
where the determinant in the denominator is given by \eqref{eq:singelsitedet}. Similarly we get:
\begin{multline}
\sof{\id-\kappa\,T^{-}\of{\mu}}_{x,a,I;y,b,J}^{-1}\,=\,\bof{\sum\limits_{k=0}^{\infty}\of{\kappa\,T^{-}\of{\mu}}^{k}}_{x,a,I;y,b,J}\,=\,\delta_{\bar{x},\bar{y}}\,\of{2\,\kappa\e^{-\mu}}^{\umod\of{x_{4}-y_{4},n_{t}}}\of{-1}^{\theta\of{y_{4}>x_{4}}}\\
\cdot\frac{\of{\id+\gamma_{4}}_{a,b}}{2}\,\bof{P^{\dagger}_{\bar{y}}\of{y_{4},x_{4}}\,\sum\limits_{k=0}^{\infty}\sof{-\of{2\,\kappa\e^{-\mu}}^{n_{t}}P^{\dagger}_{\bar{y}}\of{y_{4}}}^{k}}_{I,J}\,
+\,\frac{\of{\id-\gamma_{4}}_{a,b}}{2}\delta_{x,y}\,\delta_{I,J}\\
=\,\delta_{\bar{x},\bar{y}}\,\of{2\,\kappa\e^{-\mu}}^{\umod\of{x_{4}-y_{4},n_{t}}}\of{-1}^{\theta\of{y_{4}>x_{4}}}\,\frac{\of{\id+\gamma_{4}}_{a,b}}{2}\,\sof{P^{\dagger}_{\bar{y}}\of{y_{4},x_{4}}\,\sof{\id+\of{2\,\kappa\e^{-\mu}}^{n_{t}}P^{\dagger}_{\bar{y}}\of{y_{4}}}^{-1}}_{I,J}\\
+\,\frac{\of{\id-\gamma_{4}}_{a,b}}{2}\delta_{x,y}\,\delta_{I,J}\ ,
\end{multline}
and therefore with \eqref{eq:temppropsplit}:
\begin{multline}
\sof{\id-\kappa\,T\of{\mu}}_{x,a,I;y,b,J}^{-1}\\
=\,\frac{1}{2}\delta_{\bar{x},\bar{y}}\of{\of{2\,\kappa\e^{\mu}}^{\umod\of{y_{4}-x_{4},n_{t}}}\of{-1}^{\theta\of{x_{4}>y_{4}}}\,\of{\id-\gamma_{4}}_{a,b}\,\sof{P_{\bar{y}}\of{x_{4},y_{4}}\,\sof{\id+\of{2\,\kappa\e^{\mu}}^{n_{t}}P_{\bar{y}}\of{y_{4}}}^{-1}}_{I,J}\right.\\
\left.+\,\of{2\,\kappa\e^{-\mu}}^{\umod\of{x_{4}-y_{4},n_{t}}}\of{-1}^{\theta\of{y_{4}>x_{4}}}\,\of{\id+\gamma_{4}}_{a,b}\,\sof{P^{\dagger}_{\bar{y}}\of{y_{4},x_{4}}\,\sof{\id+\of{2\,\kappa\e^{-\mu}}^{n_{t}}P^{\dagger}_{\bar{y}}\of{y_{4}}}^{-1}}_{I,J}}\ .\label{eq:hdquarkprop}
\end{multline}
The heavy-dense quark propagator \eqref{eq:hdquarkprop} could be used to get analytic expressions for average meson and baryon propagators from which one could determine the corresponding masses. But due to the projection properties of \eqref{eq:hdquarkprop} in Dirac space, the mass spectrum will be highly degenerate. We will therefore use \eqref{eq:hdquarkprop} in the following just to facilitate the derivation of some other observables.

\subsubsection{Local Isospin Number Density}
The local isospin number density in the heavy-dense case can be obtained, by starting from equation \eqref{eq:localisodens} and replacing the full Dirac operators and propagators by their corresponding heavy-dense counterparts, i.e. just dropping the spatial hopping terms from \eqref{eq:wilsondiracop} and using \eqref{eq:hdquarkprop} as the corresponding propagator. We then obtain:
\begin{align}
n_{I}\of{x,\mu}\,=\,&2\,\Repart{\trace_{c,d}\cof{D\ssof{x,x+\hat{4},\mu}\,D^{-1}\ssof{x+\hat{4},x,\mu}\right.\right.\nonumber\\
&\qquad\left.\left.-\,D\ssof{x+\hat{4},x,\mu}\,D^{-1}\ssof{x,x+\hat{4},\mu}}}\nonumber\\
=\,&\Repart{\trace_{c,d}\cof{\of{2\,\kappa\e^{\mu}}^{n_{t}}\,\sof{P_{\bar{x}}\of{x_{4}}\,\sof{\id+\of{2\,\kappa\e^{\mu}}^{n_{t}}P_{\bar{x}}\of{x_{4}}}^{-1}}\otimes\of{\id-\gamma_{4}}\right.\right.\nonumber\\
&\qquad\left.\left.-\,\of{2\,\kappa\e^{-\mu}}^{n_{t}}\,\sof{P^{\dagger}_{\bar{x}}\of{x_{4}}\,\sof{\id+\of{2\,\kappa\e^{-\mu}}^{n_{t}}P^{\dagger}_{\bar{x}}\of{x_{4}}}^{-1}}\otimes\of{\id+\gamma_{4}}}}\nonumber\\
=\,&4\,\Repart{\trace_{c}\cof{\of{2\,\kappa\e^{\mu}}^{n_{t}}\,\sof{P_{\bar{x}}\of{x_{4}}\,\sof{\id+\of{2\,\kappa\e^{\mu}}^{n_{t}}P_{\bar{x}}\of{x_{4}}}^{-1}}\right.\right.\nonumber\\
&\qquad\left.\left.-\,\of{2\,\kappa\e^{-\mu}}^{n_{t}}\,\sof{P^{\dagger}_{\bar{x}}\of{x_{4}}\,\sof{\id+\of{2\,\kappa\e^{-\mu}}^{n_{t}}P^{\dagger}_{\bar{x}}\of{x_{4}}}^{-1}}}}\ ,\label{eq:hdlocalisodens}
\end{align}
where after the second equality sign it was used that $\of{\id\pm\gamma_{4}}$ are orthogonal projectors and that for example
\begin{multline}
\sof{-\kappa\e^{\mu}\,U_{4}\of{x}\otimes\of{\id-\gamma_{4}}}\cdot\sof{-\of{2\,\kappa\e^{\mu}}^{n_{t}-1}\,P_{\bar{x}}\of{x_{4}+1,x_{4}}\otimes\of{\id-\gamma_{4}}}\\
=\,\of{\of{2\,\kappa\e^{\mu}}^{n_{t}}\,P_{\bar{x}}\of{x_{4}}\otimes\of{\id-\gamma_{4}}}\ ,
\end{multline}
i.e. the link variable $U_{4}\of{x}$ coming from the term in the Dirac operator that is not projected out by the $\of{\id-\gamma_{4}}$ of the propagator, closes the Polyakov line $P_{\bar{x}}\of{x_{4}+1,x_{4}}$ to a Polyakov loop $P_{\bar{x}}\of{x_{4}}$, starting and ending at $x$. Analogously:
\begin{multline}
\sof{-\kappa\e^{-\mu}\,U^{\dagger}_{4}\ssof{x-\hat{4}}\otimes\of{\id+\gamma_{4}}}\cdot\sof{-\of{2\,\kappa\e^{-\mu}}^{n_{t}-1}\,P^{\dagger}_{\bar{x}}\of{x_{4},x_{4}+1}\otimes\of{\id+\gamma_{4}}}\\
=\,\sof{\of{2\,\kappa\e^{-\mu}}^{n_{t}}\,P^{\dagger}_{\bar{x}}\of{x_{4}}\otimes\of{\id+\gamma_{4}}}\ .
\end{multline}

\subsubsection{Condensates}\label{sssec:condensates}
The quark condensate for a two-flavor system,
\[
\avof{\bar{\psi}\,\psi}\,=\,\avof{\bar{u}\,u}\,+\,\avof{\bar{d}\,d}\ ,
\]
is in the static-quark limit obtained by taking the expectation value of
\begin{multline}
\frac{1}{n_{t}}\of{\trace_{c,d,t}\sof{\sof{\id-\kappa_{u}\,T\of{\mu_{u}}}_{\bar{x}}^{-1}}+\trace_{c,d,t}\sof{\sof{\id-\kappa_{d}\,T\of{\mu_{d}}}_{\bar{x}}^{-1}}}\\
=\,\frac{1}{2}\of{\trace_{c,d}\sof{\sof{\id+\sof{2\,\kappa_u\e^{\mu_u}}^{n_{t}}P_{\bar{x}}}^{-1}\otimes\of{\id-\gamma_{4}}}\,+\,\trace_{c,d}\sof{\sof{\id+\sof{2\,\kappa_u\e^{-\mu_u}}^{n_{t}}P^{\dagger}_{\bar{x}}}^{-1}\otimes\of{\id+\gamma_{4}}}}\\
+\,\frac{1}{2}\of{\trace_{c,d}\sof{\sof{\id+\sof{2\,\kappa_d\e^{\mu_d}}^{n_{t}}P_{\bar{x}}}^{-1}\otimes\of{\id-\gamma_{4}}}\,+\,\trace_{c,d}\sof{\sof{\id+\sof{2\,\kappa_d\e^{-\mu_d}}^{n_{t}}P^{\dagger}_{\bar{x}}}^{-1}\otimes\of{\id+\gamma_{4}}}}\\
=\,2\,\of{\trace_{c}\sof{\sof{\id+\sof{2\,\kappa_u\e^{\mu_u}}^{n_{t}}P_{\bar{x}}}^{-1}}\,+\,\trace_{c}\sof{\sof{\id+\sof{2\,\kappa_u\e^{-\mu_u}}^{n_{t}}P^{\dagger}_{\bar{x}}}^{-1}}}\\
+\,2\,\of{\trace_{c}\sof{\sof{\id+\sof{2\,\kappa_d\e^{\mu_d}}^{n_{t}}P_{\bar{x}}}^{-1}}\,+\,\trace_{c}\sof{\sof{\id+\sof{2\,\kappa_d\e^{-\mu_d}}^{n_{t}}P^{\dagger}_{\bar{x}}}^{-1}}}\ ,\label{eq:hdquarkcond}
\end{multline}
where $\sof{\id-\kappa\,T\of{\mu}}_{\bar{x}}^{-1}$ is the static-quark propagator \eqref{eq:hdquarkprop} for $\bar{y}=\bar{x}$ and $\kappa_{u}$, $\kappa_{d}$ and $\mu_{u}$, $\mu_{d}$ are the hopping parameters and chemical potentials for the two flavors.\\

Analogously the isospin condensate,
\[
\avof{\bar{\psi}\gamma_{4}\tau_{3}\psi}\,=\,\avof{\bar{u}\gamma_{4} u}\,-\,\avof{\bar{d}\gamma_{4} d}\ ,
\]
is obtained as the expectation value of
\begin{multline}
\frac{1}{n_{t}}\of{\trace_{c,d,t}\sof{\sof{\id_{c,t}\otimes\gamma_{4}}\sof{\id-\kappa_{u}\,T\of{\mu_{u}}}_{\bar{x}}^{-1}}+\trace_{c,d,t}\sof{\sof{\id_{c,t}\otimes\gamma_{4}}\sof{\id-\kappa_{d}\,T\of{\mu_{d}}}_{\bar{x}}^{-1}}}\\
=\,\frac{1}{2}\of{\trace_{c,d}\sof{\sof{\id+\sof{2\,\kappa_u\e^{\mu_u}}^{n_{t}}P_{\bar{x}}}^{-1}\otimes\of{\gamma_{4}-\id}}\,+\,\trace_{c,d}\sof{\sof{\id+\sof{2\,\kappa_u\e^{-\mu_u}}^{n_{t}}P^{\dagger}_{\bar{x}}}^{-1}\otimes\of{\gamma_{4}+\id}}}\\
-\,\frac{1}{2}\of{\trace_{c,d}\sof{\sof{\id+\sof{2\,\kappa_d\e^{\mu_d}}^{n_{t}}P_{\bar{x}}}^{-1}\otimes\of{\gamma_{4}-\id}}\,+\,\trace_{c,d}\sof{\sof{\id+\sof{2\,\kappa_d\e^{-\mu_d}}^{n_{t}}P^{\dagger}_{\bar{x}}}^{-1}\otimes\of{\gamma_{4}+\id}}}\\
=\,2\,\of{-\trace_{c}\sof{\sof{\id+\sof{2\,\kappa_u\e^{\mu_u}}^{n_{t}}P_{\bar{x}}}^{-1}}\,+\,\trace_{c}\sof{\sof{\id+\sof{2\,\kappa_u\e^{-\mu_u}}^{n_{t}}P^{\dagger}_{\bar{x}}}^{-1}}}\\
-\,2\,\of{-\trace_{c}\sof{\sof{\id+\sof{2\,\kappa_d\e^{\mu_d}}^{n_{t}}P_{\bar{x}}}^{-1}}\,+\,\trace_{c}\sof{\sof{\id+\sof{2\,\kappa_d\e^{-\mu_d}}^{n_{t}}P^{\dagger}_{\bar{x}}}^{-1}}}\ ,\label{eq:hdisocond}
\end{multline}
where in the first line $\id_{c,t}$ is the identity with respect to color and time indices.

\subsection{Half-filling Point}\label{ssec:hdhfpoint}
Having obtained the heavy-dense partition function for two-flavor LQCD in the strong coupling limit, and defined some basic observables, we now use these results to study the origin of the properties of the half-filling point in the heavy-dense/strong coupling limit, and check, how far our findings could also apply to full LQCD. 

\subsubsection{Location of the Half-filling Point}\label{sssec:hdhfpointlocation}
To find the location of the half-filling point, we consider the isospin density \eqref{eq:isodens} in its integral form:
\begin{align}
n_{I}\of{\mu,\kappa,n_{t}}\,&=\,\frac{1}{n_{t}}\partd{}{\mu}\log\of{\int\dd P\,\abs{\Det{D\of{\mu,\kappa,n_{t}}}}^{2}}\nonumber\\
&=\,\frac{1}{n_{t}\,Z_{s}\of{\mu,-\mu,\kappa,n_{t}}}\int\dd P\,\partd{\abs{\Det{D\of{\mu,\kappa,n_{t}}}}^{2}}{\mu}\nonumber\\
&=\,\frac{1}{n_{t}\,Z_{s}\of{\mu,-\mu,\kappa,n_{t}}}\int\dd P\,\partd{\abs{\det_{c}^{2}\sof{\id+\ssof{2\kappa\e^{\mu}}^{n_{t}}P}\det_{c}^{2}\sof{\id+\ssof{2\kappa\e^{-\mu}}^{n_{t}}P^{\dagger}}}^{2}}{\mu}\nonumber\\
&=\,\frac{4}{Z_{s}\of{\mu,-\mu,\kappa,n_{t}}}\int\dd P\,\cRepart{\ssof{2\kappa\e^{\mu}}^{n_{t}}\trace_{c}\sof{P\,\sof{\id+\ssof{2\kappa\e^{\mu}}^{n_{t}}P}^{-1}}\right.\nonumber\\
&\hspace{130pt}\left.-\ssof{2\kappa\e^{-\mu}}^{n_{t}}\trace_{c}\sof{P^{\dagger}\,\sof{\id+\ssof{2\kappa\e^{-\mu}}^{n_{t}}P^{\dagger}}^{-1}}}\nonumber\\
&\hspace{80pt}\times\abs{\det_{c}^{2}\sof{\id+\ssof{2\kappa\e^{\mu}}^{n_{t}}P}\det_{c}^{2}\sof{\id+\ssof{2\kappa\e^{-\mu}}^{n_{t}}P^{\dagger}}}^{2}\ .\label{eq:isodensintf}
\end{align}

\begin{figure}[H]
\centering
\begin{minipage}[t]{0.485\linewidth}
\centering
\includegraphics[width=\linewidth]{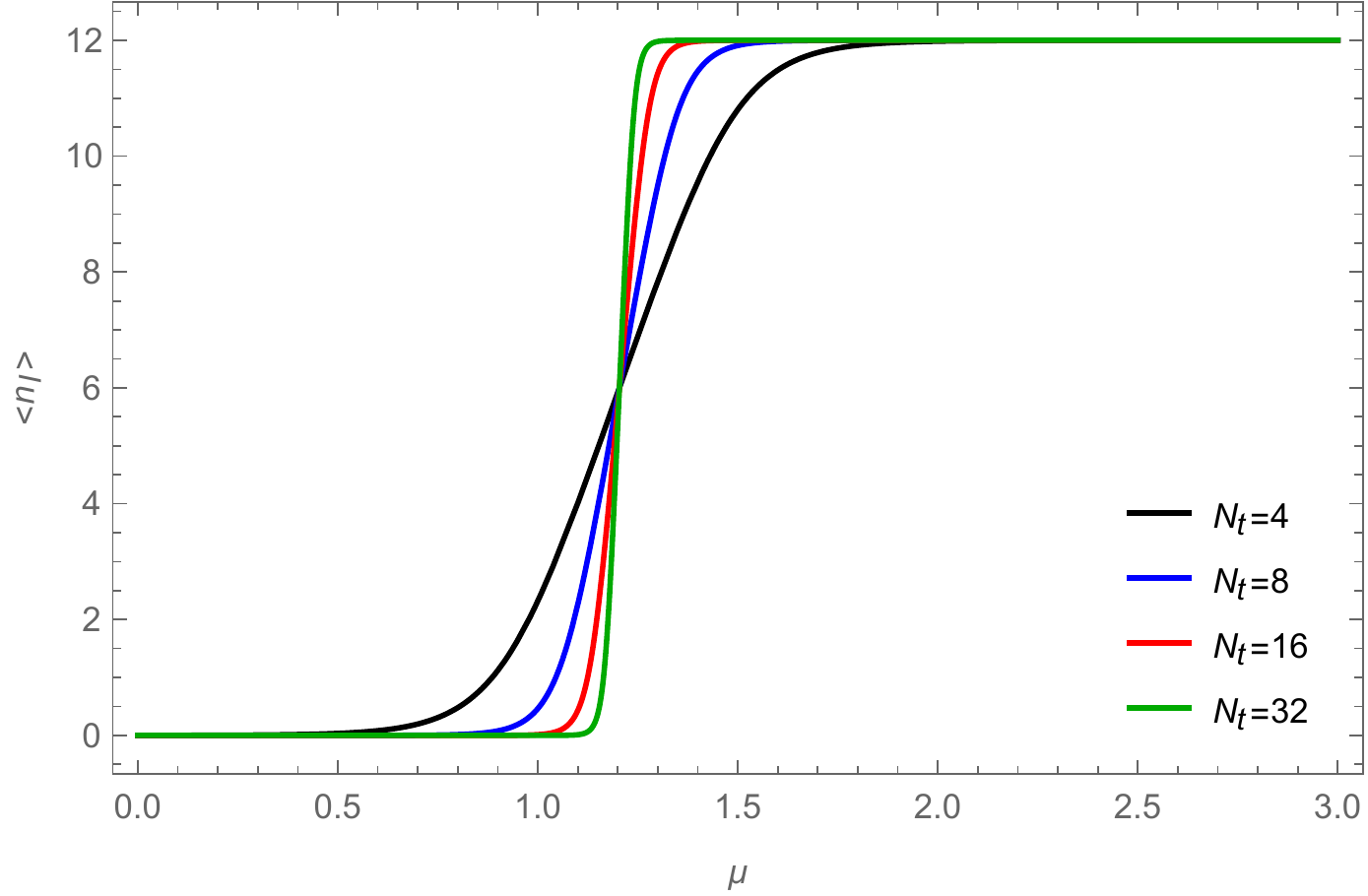}
\caption{Isospin density in the heavy-dense approximation as a function of $\mu$ for $n_{t}\in\cof{4,8,16,32}$ and $\kappa=0.15$. Compare with Fig. \ref{fig:avisodensvarnt}.}
  \label{fig:hdisodens}
\end{minipage}\hfill
\begin{minipage}[t]{0.495\linewidth}
\centering
\includegraphics[width=\linewidth]{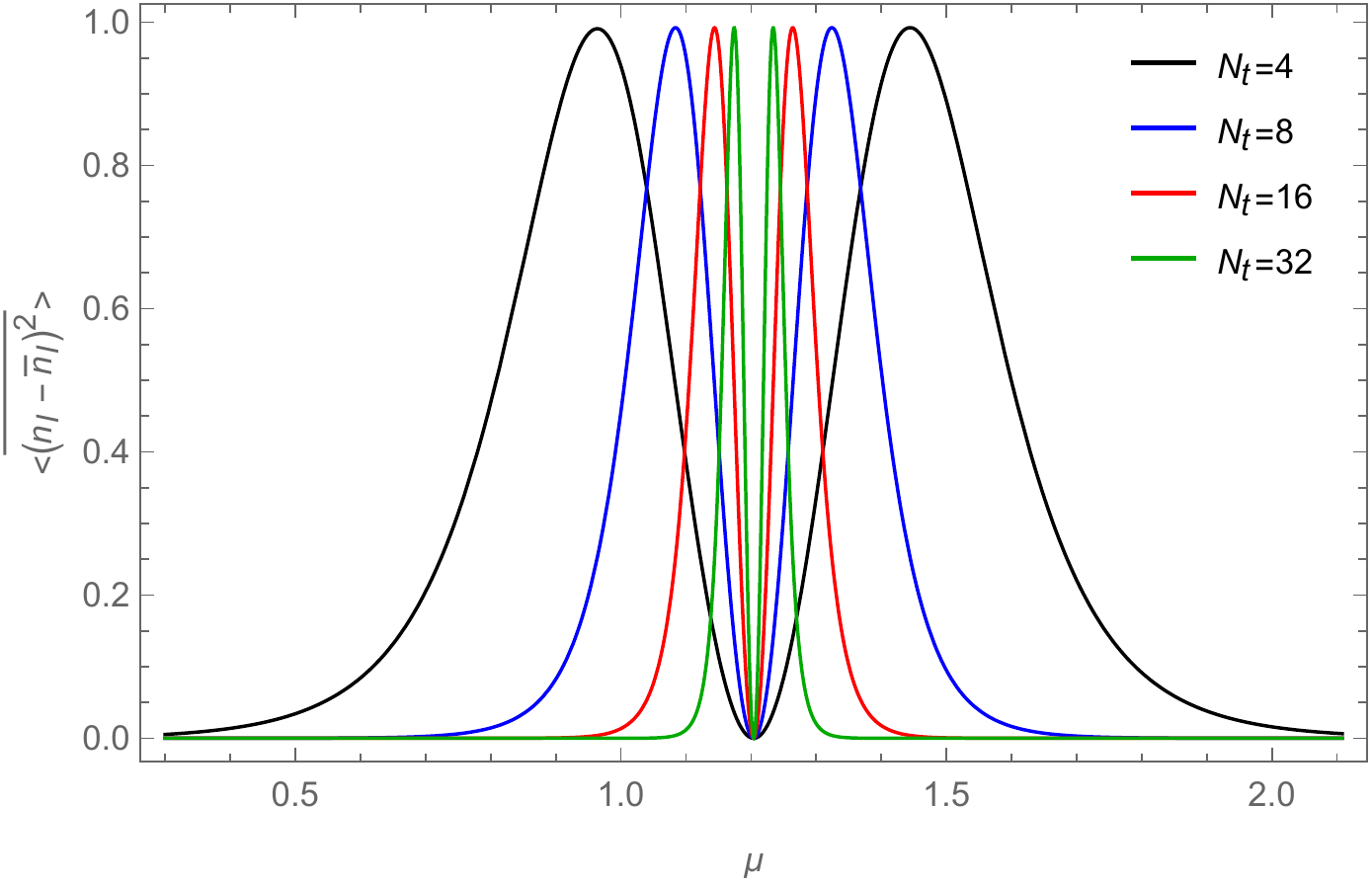}
\caption{Isospin density space-time variance in the heavy-dense approximation as a function of $\mu$ for $n_{t}\in\cof{4,8,16,32}$ and $\kappa=0.15$. Compare with Fig. \ref{fig:avisospatvarvarnt}.}
  \label{fig:hdisodensvari}
\end{minipage}
\end{figure}

For sufficiently large values of $\mu$, the second term within the curly brackets in \eqref{eq:isodensintf} can be neglected; setting $\mu\,=\,-\log\of{2\kappa}$, this term is of order $\order{\of{2\kappa}^{2 n_{t}}}$ while the first term is of order $1$ and becomes completely independent of $P$, as
\begin{align}
\cRepart{\trace_{c}\sof{P\,\sof{\id+P}^{-1}}}\,&=\,\frac{1}{2}\cof{\trace_{c}\sof{P\,\sof{\id+P}^{-1}}\,+\,\trace_{c}\sof{P^{\dagger}\,\sof{\id+P^{\dagger}}^{-1}}}\nonumber\\
&=\,\frac{1}{2}\cof{\trace_{c}\sof{P\,\sof{\id+P}^{-1}}\,+\,\trace_{c}\sof{\sof{P\,+\,\id}^{-1}}}\nonumber\\
&=\,\frac{1}{2}\trace_{c}\sof{\sof{\id+P}\sof{\id+P}^{-1}}\,=\,\frac{3}{2}\ ,\label{eq:isodensintfprt}
\end{align}
and we obtain for the isospin density \eqref{eq:isodensintf}:
\[
n_{I}\of{\mu=-\log\of{2\kappa},\kappa,n_{t}}\,=\,6\,+\,\order{\of{2\kappa}^{2 n_{t}}}\ ,\label{eq:hdhfdens}
\]
which means that $\mu=\tilde{\mu}\equiv-\log\of{2\kappa}$ corresponds to the half-filling point, independently of $n_{t}$ (up to $\order{\of{2\kappa}^{n_{t}}}$ corrections), as is visualized in figure \ref{fig:hdisodens}, where the isospin number density \eqref{eq:isodens} is shown as a function of $\mu$ for different system sizes and it can be seen that the curves, corresponding to different system sizes, all cross at the half-filling point, just as in the case of full LQCD shown above in figure \ref{fig:avisodensvarnt}. From \eqref{eq:isodensintf} and \eqref{eq:isodensintfprt} it also follows that the dependency of the \emph{local isospin density} \eqref{eq:hdlocalisodens} on the gauge field is highly suppressed at half-filling: gauge field dependent terms are suppressed by a factor of $\order{\of{2\kappa}^{2\,n_{t}}}$ such that the space time variance of \eqref{eq:hdlocalisodens} nearly vanishes at half-filling, again just as in the case of full LQCD shown above in figure \ref{fig:avisospatvarvarnt}.\\

Nevertheless, in full LQCD, the half-filling point is shifted towards a slightly larger value of $\mu$: while in the heavy-dense case, we have $\tilde{\mu}=-\log\of{2\,\kappa}\approx 1.2$ for $\kappa=0.15$, we find in full LQCD, for the same value of $\kappa$ and with $\beta = 5.0$, that the half-filling point is located at $\mu\approx 1.3$. This shift is due to spatial fermion hopping which causes corrections to the quark mass. Corrections coming from spatial hoppings and a finite value of $\beta$ will of course also appear in the expressions for observables and give rise to deviations that could be much larger than $\order{\ssof{2\,\kappa}^{2\,n_{t}}}$. These corrections are considered in Secs. \ref{ssec:meanfieldhdphasediagram} and \ref{ssec:fromhdtofulldet}.

\subsubsection{Particle-Hole Symmetry and Average Sign}\label{sssec:hdhfphsymmetry}
Consider the single flavor fermion determinant \eqref{eq:ssitedet} under the transformation $\mu\rightarrow 2\,\tilde{\mu}-\mu$ that corresponds to reflection about the half-filling point:
\begin{multline}
\Det{D\of{2\tilde{\mu}-\mu,\kappa,n_{t}}}\,=\,\det_{c}^{2}\sof{\id+\of{2\kappa\e^{\mu}}^{-n_{t}}P}\,\det_{c}^{2}\sof{\id+\ssof{\of{2\kappa}^{3}\e^{\mu}}^{n_{t}}P^{\dagger}}\\
=\,\of{2\kappa\e^{\mu}}^{-6\,n_{t}}\det_{c}^{2}\of{P}\det_{c}^{2}\sof{\id+\ssof{2\kappa\e^{\mu}}^{n_{t}}P^{\dagger}}\,\det_{c}^{2}\sof{\id+\ssof{\of{2\kappa}^{3}\e^{\mu}}^{n_{t}}P^{\dagger}}\\
=\,\of{2\kappa\e^{\mu}}^{-6\,n_{t}}\det_{c}^{2}\sof{\id+\ssof{2\kappa\e^{\mu}}^{n_{t}}P^{\dagger}}\,\of{1\,+\,\order{\ssof{2\kappa}^{n_{t}}}}\ ,\label{eq:ssitedettr}
\end{multline}
where the last line of \eqref{eq:ssitedettr} is true for $\mu\in\fof{0,2\,\tilde{\mu}}$. Comparing \eqref{eq:ssitedettr} with
\begin{multline}
\Det{D\of{\mu,\kappa,n_{t}}}^{*}\,=\,\Det{D\of{-\mu,\kappa,n_{t}}}\,=\,\det_{c}^{2}\sof{\id+\ssof{2\kappa\e^{-\mu}}^{n_{t}}P}\,\det_{c}^{2}\sof{\id+\ssof{2\kappa\e^{\mu}}^{n_{t}}P^{\dagger}}\\
=\,\det_{c}^{2}\sof{\id+\ssof{2\kappa\e^{\mu}}^{n_{t}}P^{\dagger}}\,\of{1\,+\,\order{\ssof{2\kappa}^{n_{t}}}}\ ,\label{eq:ssitedetm}
\end{multline}
we see that they agree up to an irrelevant gauge field independent pre-factor (which is unity at the half-filling point) and terms which are suppressed by at least a factor of $\of{2\kappa}^{n_{t}}$. As the complex conjugate of the fermion determinant can be interpreted as resulting from a charge conjugation of the original fermion fields, this shows that the half-filling state possesses an exact $T=0$ particle-hole symmetry also in the heavy-dense/strong coupling limit of LQCD.\\

We can also check the behavior of the heavy-dense quark propagator \eqref{eq:hdquarkprop},
\begin{multline}
\sof{\id-\kappa\,T\of{\mu}}_{x,a,I;y,b,J}^{-1}\\
=\,\frac{1}{2}\delta_{\bar{x},\bar{y}}\of{\of{2\,\kappa\e^{\mu}}^{\umod\of{y_{4}-x_{4},n_{t}}}\of{-1}^{\theta\of{x_{4}>y_{4}}}\,\of{\id-\gamma_{4}}_{a,b}\,\sof{P_{\bar{y}}\of{x_{4},y_{4}}\,\sof{\id+\of{2\,\kappa\e^{\mu}}^{n_{t}}P_{\bar{y}}\of{y_{4}}}^{-1}}_{I,J}\right.\\
\left.+\,\of{2\,\kappa\e^{-\mu}}^{\umod\of{x_{4}-y_{4},n_{t}}}\of{-1}^{\theta\of{y_{4}>x_{4}}}\,\of{\id+\gamma_{4}}_{a,b}\,\sof{P^{\dagger}_{\bar{y}}\of{y_{4},x_{4}}\,\sof{\id+\of{2\,\kappa\e^{-\mu}}^{n_{t}}P^{\dagger}_{\bar{y}}\of{y_{4}}}^{-1}}_{I,J}}\ ,
\end{multline}
under the transformation $\mu\rightarrow 2\,\tilde{\mu}-\mu$. For $y_{4}\neq x_{4}$ and $\mu\in\fof{0,2\,\tilde{\mu}}$ we find:
\begin{multline}
\sof{\id-\kappa\,T\of{\mu}}_{x,a,I;y,b,J}^{-1}\rightarrow -\sof{1+\order{\of{2\kappa}^{2}}}\cdot\frac{1}{2}\delta_{\bar{x},\bar{y}}\,\of{2\,\kappa\e^{\mu}}^{\umod\of{x_{4}-y_{4},n_{t}}}\of{-1}^{\theta\of{y_{4}>x_{4}}}\\
\of{\id-\gamma_{4}}_{a,b}\,\sof{P^{\dagger}_{\bar{y}}\of{y_{4},x_{4}}\,\sof{\id+\of{2\,\kappa\e^{\mu}}^{n_{t}}P^{\dagger}_{\bar{y}}\of{y_{4}}}^{-1}}_{I,J}
\end{multline}
and
\begin{multline}
\sof{\id-\kappa\,T\of{-\mu}}_{x,a,I;y,b,J}^{-1}\rightarrow -\sof{1+\order{\of{2\kappa}^{2}}}\cdot\frac{1}{2}\delta_{\bar{x},\bar{y}}\,\of{2\,\kappa\e^{\mu}}^{\umod\of{y_{4}-x_{4},n_{t}}}\of{-1}^{\theta\of{x_{4}>y_{4}}}\\
\of{\id+\gamma_{4}}_{a,b}\,\sof{P_{\bar{y}}\of{x_{4},y_{4}}\,\sof{\id+\of{2\,\kappa\e^{\mu}}^{n_{t}}P_{\bar{y}}\of{y_{4}}}^{-1}}_{I,J}\ ,
\end{multline}
which by noting that 
\[
\sof{P_{\bar{y}}\of{x_{4},y_{4}}\,\sof{\id+\of{2\,\kappa\e^{\mu}}^{n_{t}}P_{\bar{y}}\of{y_{4}}}^{-1}}\,=\,\sof{\sof{\id+\of{2\,\kappa\e^{\mu}}^{n_{t}}P_{\bar{y}}\of{x_{4}}}^{-1}\,P_{\bar{y}}\of{x_{4},y_{4}}}\ ,
\]
and remembering that
\[
\gamma_{5}\,\of{\id\pm\gamma_{4}}\,\gamma_{5}\,=\,\of{\id\mp\gamma_{4}}\ ,
\]
shows that:
\begin{multline}
\sof{\id-\kappa\,T\of{2\,\tilde{\mu}-\mu}}\,=\,-\gamma_{5}\,\sof{\id-\kappa\,T\of{-\mu}}\,\gamma_{5}\,\sof{1+\order{\of{2\kappa}^{2}}}\\
=\,-\sof{\id-\kappa\,T\of{\mu}}^{\dagger}\,\sof{1+\order{\of{2\kappa}^{2}}}\ ,
\end{multline}
and 
\begin{multline}
\sof{\id-\kappa\,T\of{-\of{2\,\tilde{\mu}-\mu}}}\,=\,-\gamma_{5}\,\sof{\id-\kappa\,T\of{\mu}}\,\gamma_{5}\,\sof{1+\order{\of{2\kappa}^{2}}}\\
=\,-\sof{\id-\kappa\,T\of{-\mu}}^{\dagger}\,\sof{1+\order{\of{2\kappa}^{2}}}\ ,
\end{multline}
i.e. up to a minus sign and terms of higher order in $\of{2\kappa}$, we find that for $y_{4}\neq x_{4}$, reflecting $\mu$ about $\tilde{\mu}$ turns the quark propagator into its Hermitian conjugate. For the equal time part, $y_{4}=x_{4}$ we find in a similar way:
\begin{multline}
\sof{\id-\kappa\,T\of{\mu}}_{x,a,I;x,b,J}^{-1}\rightarrow \frac{1}{2}\of{\of{\id-\gamma_{4}}_{a,b}\,\of{\delta_{I,J}-\sof{\id+\of{2\,\kappa\e^{\mu}}^{n_{t}}P^{\dagger}_{\bar{y}}\of{y_{4}}}^{-1}_{I,J}}\right.\\
\left.+\,\of{\id+\gamma_{4}}_{a,b}\,\sof{\delta_{I,J}\,+\,\order{\of{2\kappa}^{2\,n_{t}}\of{2\kappa\e^{\mu}}^{n_{t}}}}}\ ,
\end{multline}
and
\begin{multline}
\sof{\id-\kappa\,T\of{-\mu}}_{x,a,I;x,b,J}^{-1}\rightarrow \frac{1}{2}\of{\of{\id+\gamma_{4}}_{a,b}\,\of{\delta_{I,J}-\sof{\id+\of{2\,\kappa\e^{\mu}}^{n_{t}}P_{\bar{y}}\of{y_{4}}}^{-1}_{I,J}}\right.\\
\left.+\,\of{\id-\gamma_{4}}_{a,b}\,\sof{\delta_{I,J}\,+\,\order{\of{2\kappa}^{2\,n_{t}}\of{2\kappa\e^{\mu}}^{n_{t}}}}}\ .
\end{multline}

\begin{figure}[h]
\centering
\begin{minipage}[t]{0.495\linewidth}
\centering
\includegraphics[width=\linewidth]{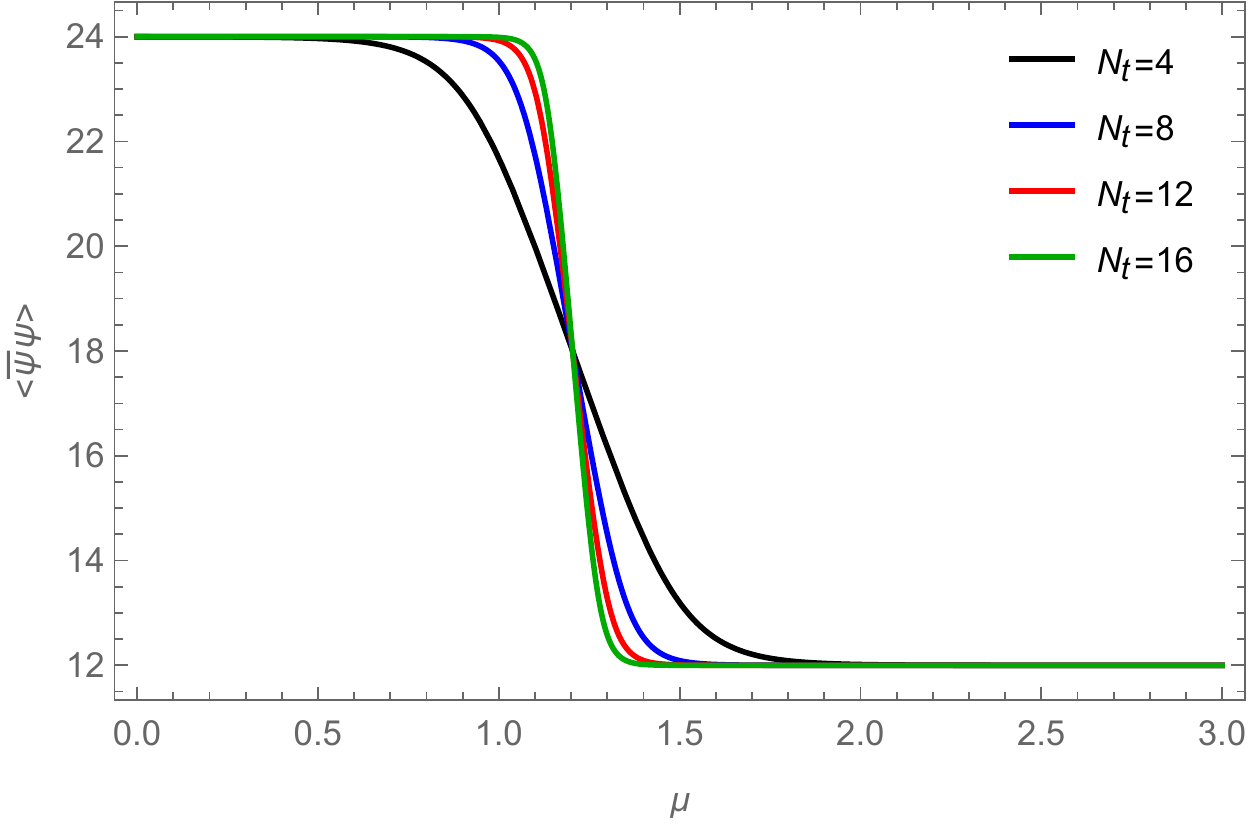}
\end{minipage}\hfill
\begin{minipage}[t]{0.495\linewidth}
\centering
\includegraphics[width=\linewidth]{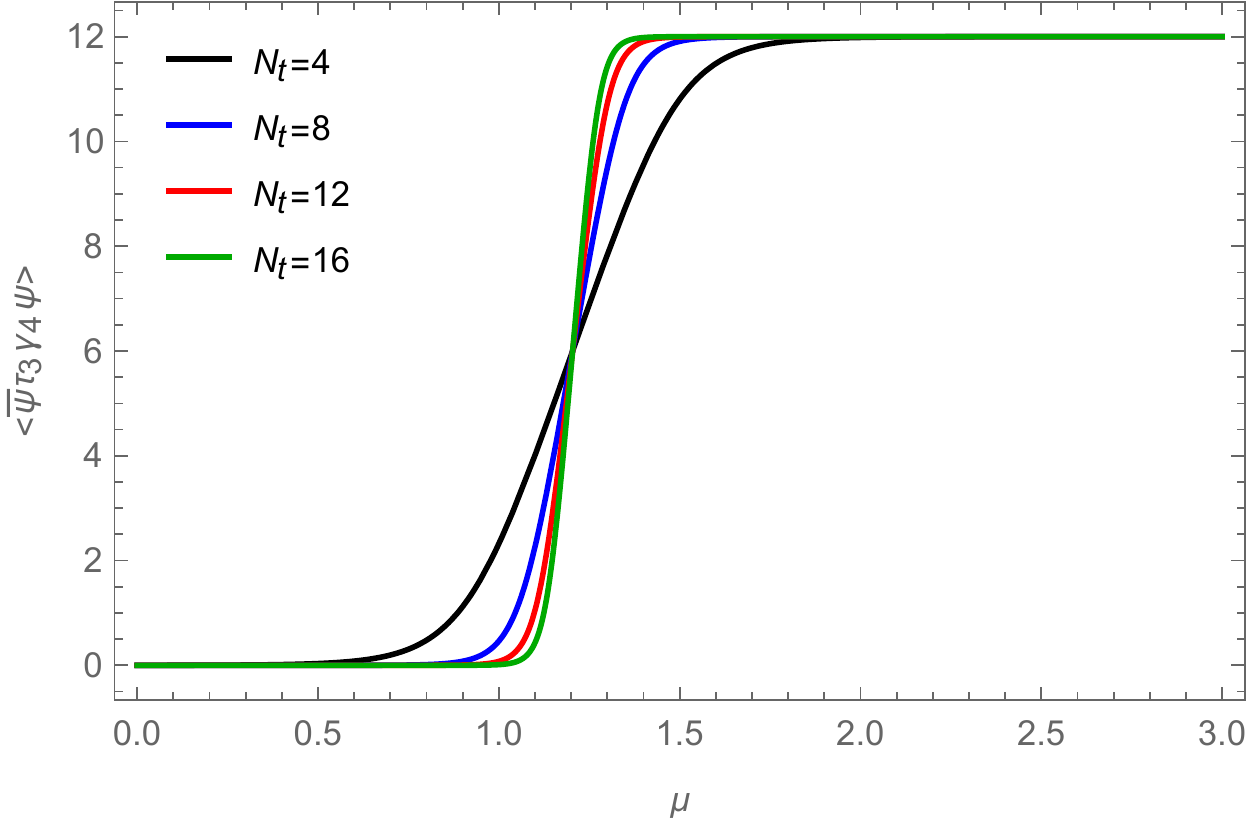}
\end{minipage}
\caption{Quark (left) and isospin condensate (right) in the static-quark limit, defined as the expectation values of \eqref{eq:hdquarkcond} and \eqref{eq:hdisocond} respectively, shown for $\kappa_{u}=\kappa_{d}=0.15$ as functions of an isospin chemical potential $\mu=\mu_{u}=-\mu_{d}$ for $n_{t}\in\cof{4,8,12,16}$. Compare with Fig. \ref{fig:avcondsvarnt}.}
  \label{fig:hdcondensates}
\end{figure}

For the two condensates introduced in Sec. \ref{sssec:condensates}, we then find for the phase quenched case:
\[
\avof{\bar{\psi}\,\psi}_{q}\of{2\,\tilde{\mu}-\mu}\,=\,12+\avof{\bar{\psi}\gamma_{4}\tau_{3}\psi}_{q}\of{\mu}\,+\,\order{\of{2\kappa}^{n_{t}}}\ ,
\]
and
\[
\avof{\bar{\psi}\gamma_{4}\tau_{3}\psi}_{q}\of{2\,\tilde{\mu}-\mu}\,=\,\avof{\bar{\psi}\,\psi}_{q}\of{\mu}\,-\,12\,+\,\order{\of{2\kappa}^{n_{t}}}\ ,
\] 
which is illustrated in Fig. \ref{fig:hdcondensates}. In contrast to the situation in full LQCD shown above in Fig. \ref{fig:avcondsvarnt}, in the static quark/strong coupling limit, the values of the condensates are at half-filling exactly in the middle between their corresponding values at $\mu=0$ and for $\mu\rightarrow\infty$.\\

In the un-quenched case, the relations would be:
\[
\avof{\bar{\psi}\,\psi}\of{2\,\tilde{\mu}-\mu}\,=\,12+\avof{\bar{\psi}\gamma_{4}\psi}\of{\mu}\,+\,\order{\of{2\kappa}^{n_{t}}}\ ,
\]
and
\[
\avof{\bar{\psi}\gamma_{4}\psi}\of{2\,\tilde{\mu}-\mu}\,=\,\avof{\bar{\psi}\,\psi}\of{\mu}\,-\,12\,+\,\order{\of{2\kappa}^{n_{t}}}\ .
\]

Returning to the single flavor fermion determinant \eqref{eq:ssitedet}, we can see from \eqref{eq:ssitedettr} and \eqref{eq:ssitedetm} that it is essentially real at $\mu=\tilde{\mu}$ and the average sign \eqref{eq:avsign} should therefore be $1+\order{\ssof{2\kappa}^{n_{t}}}$. This can also be verified more directly by using \eqref{eq:ssitedet} to show that for $\mu=\tilde{\mu}$, we have:
\begin{multline}
\det_{c}\sof{\id+\ssof{2\kappa\e^{\tilde\mu}}^{n_{t}}P}\,=\,1\,+\,\ssof{2\kappa\e^{\tilde{\mu}}}^{n_{t}}\trace_{c}\sof{P}\,+\,\ssof{2\kappa\e^{\tilde{\mu}}}^{2\,n_{t}}\trace_{c}\sof{P^{\dagger}}\,+\,\ssof{2\kappa\e^{\tilde{\mu}}}^{3\,n_{t}}\\
=\,2\,+\,\trace_{c}\ssof{P}\,+\,\trace_{c}\ssof{P^{\dagger}}\,=\,2\,\of{1\,+\,\Repart{\trace_{c}\of{P}}}\ ,
\end{multline}
i.e. we find again that the dominant part of the single flavor fermion determinant \eqref{eq:singelsitedett} at half-filling,
\[
\det_{c}^{2}\sof{\id+P}\,\sof{1\,+\,\order{\ssof{2\kappa}^{2\,n_{t}}}}
\]
is real and positive up to terms of order $\order{\ssof{2\kappa}^{2\,n_{t}}}$.\\

\begin{figure}[h]
\centering
\begin{minipage}[t]{0.485\linewidth}
\centering
\includegraphics[width=\linewidth]{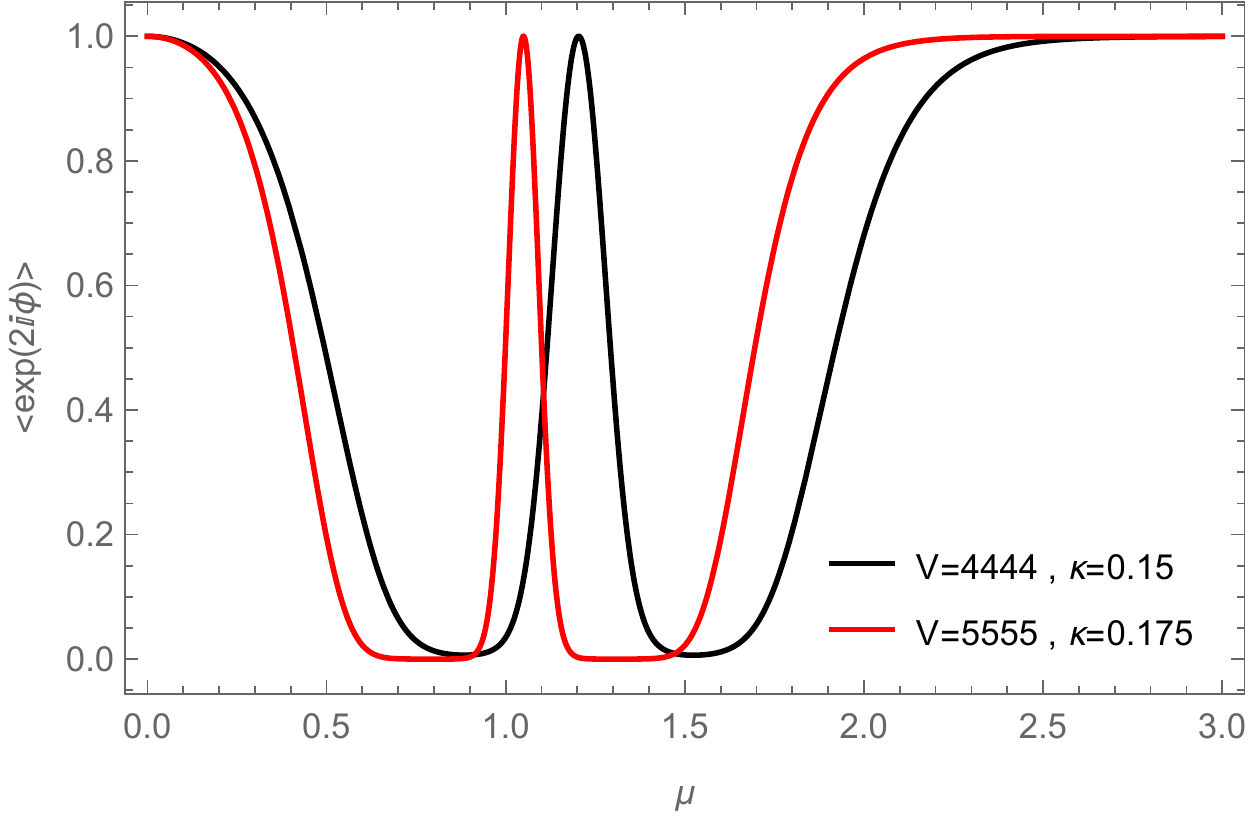}
\end{minipage}\hfill
\begin{minipage}[t]{0.485\linewidth}
\centering
\includegraphics[width=\linewidth]{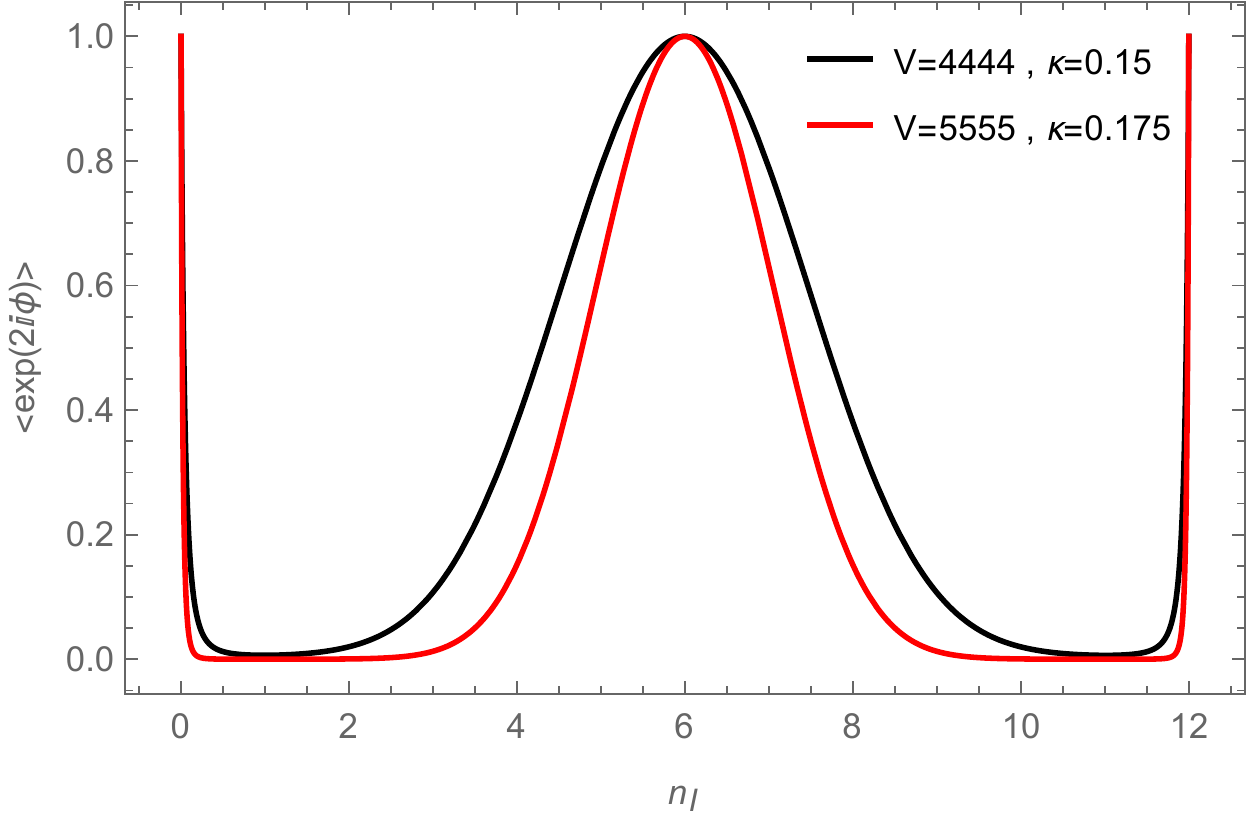}
\end{minipage}
\caption{Average sign \eqref{eq:avsign} in the heavy-dense/strong coupling approximation as a function of the chemical potential $\mu$ (left) and as a function of the isospin number density (right) for the same sets of hopping parameter and system size that were used in figure \ref{fig:avsign} above for full LQCD.}
  \label{fig:hdavsignvsmu}
\end{figure}

The average sign \eqref{eq:avsign} in the heavy-dense limit is shown in figure \ref{fig:hdavsignvsmu} as a function of the chemical potential (left) and as a function of the isospin number density (right) for the same system sizes and values of the hopping parameter $\kappa$ that were used to generate the corresponding plots in full LQCD shown above in figure \ref{fig:avsign}. By comparing the two figures, it can be seen, that in the heavy-dense case, the average sign is much more symmetric about the half-filling point and at the half-filling point itself, it deviates much less from being unity than in full LQCD, which is again due to the absence of spatial fermion hoppings, such that the deviation is really just of order $\order{\of{2\kappa}^{2\,n_{t}}}$.

\subsection{Mean-Field Method for Heavy-Dense QCD at Finite Gauge Coupling $\beta$}\label{ssec:mfmethod}
So far we have only considered heavy dense LQCD in the strong coupling limit, $\beta=0$. In this section, we would now like to get a glimpse of finite $\beta$ effects by introducing an effective nearest-neighbor Polyakov loop action, which is motivated by the leading terms of the character expanded gauge field Boltzmann factor, as derived for example in \cite{Langelage}, and then using a mean-field approximation. Our mean-field treatment is similar to one of the approaches considered in \cite{Fukushima} but differs from the mean-field calculation in \cite{Feo} as we are considering the Polyakov loops as our effective degrees of freedom while in \cite{Feo}, this role was taken by spatial links.\\

The nearest-neighbor Polyakov loop action is obtained as follows: we start with a character expansion of the $\SU{3}$ Yang-Mills Boltzmann factor (c.f. \cite{Montvay}, chapter 3.4),
\[
\e^{-S_{g}}\,=\,\prod\limits_{p}\bcof{\sum\limits_{r}\,d_{r}\,c_{r}\of{\beta}\,\chi_{r}\ssof{U_{p}}}\,=\,c_{0}\of{\beta}^{6\,\abs{\Lambda}\,n_{t}}\,\prod\limits_{p}\bcof{1\,+\,\sum\limits_{r\neq 0}\,d_{r}\,a_{r}\of{\beta}\,\chi_{r}\ssof{U_{p}}}\ ,\label{eq:gaugefieldboltzannfactorexpnd}
\]
where the product is over all plaquettes $p$, $\chi_{r}\of{U_{p}}$ is the character of the $\SU{3}$ group element $U_{p}$ (corresponding to the plaquette $p$) in the representation $r$, and $a_{r}\of{\beta}\,=\,c_{r}\of{\beta}/c_{0}\of{\beta}$ with $c_{0}\of{\beta}$ being the expansion coefficient of the trivial character $\chi_{0}\of{U_{p}}=1$. We now drop the summation over $r$ and keep just the terms corresponding to the fundamental representation. After integrating out the spatial links, we are then left with \cite{Langelage}:
\begin{multline}
\e^{-S_{g}}\,\approx\,c_{0}\of{\beta}^{6\,\abs{\Lambda}\,n_{t}}\,\prod\limits_{\avof{i,j}}\cof{1\,+\,a_{f}^{n_{t}}\of{\beta}\of{\chi_{f}\ssof{P_{i}}\,\chi_{f}\ssof{P^{\dagger}_{j}}+\chi_{f}\ssof{P^{\dagger}_{i}}\,\chi_{f}\ssof{P_{j}}}}\\
=\,c_{0}\of{\beta}^{6\,\abs{\Lambda}\,n_{t}}\,\prod\limits_{\avof{i,j}}\cof{1\,+\,a_{f}^{n_{t}}\of{\beta}\of{\trace_{c}\ssof{P_{i}}\,\trace_{c}\ssof{P^{\dagger}_{j}}+\trace_{c}\ssof{P^{\dagger}_{i}}\,\trace_{c}\ssof{P_{j}}}}\ ,\label{eq:strongcouplingexp}
\end{multline}
where the products run over all pairs of nearest-neighboring sites, $\avof{i,j}$. If we now require that the Boltzmann factor corresponding to the desired effective nearest-neighbor Polyakov loop action should, to lowest order, be proportional to \eqref{eq:strongcouplingexp}, we find that the effective action should be given by:
\[
-S_{eff}\,=\,a_{f}^{n_{t}}\of{\beta}\sum\limits_{\avof{i,j}}\of{L_{i}\,L^{*}_{j}+L^{*}_{i}\,L_{j}}\ ,\label{eq:effplaction}
\]
where $L_{i}\,=\,\trace_{c}\of{P_{i}}$. At this point it should be mentioned that in contrast to the situation with the $\Un{1}$ or $\SU{2}$ Yang-Mills action, where the coefficients of the character expansion can be written in closed form in terms of Bessel functions, no closed form is known for these coefficients in the case of $\SU{3}$. To compute $a_{f}\of{\beta}$, we made use of the power series representations for $c_{0}\of{\beta}$ and $c_{f}\of{\beta}$, given in \cite{Montvay} up to order $\order{\beta^{14}}$, i.e.:
\begin{multline}
c_{0}\of{\beta}\,=\,1+\of{\beta/6}^2+\frac{\of{\beta/6}^3}{3}+\frac{\of{\beta/6}^4}{2}+\frac{\of{\beta/6}^5}{4}+\frac{13 \of{\beta/6}^6}{72}+\frac{11 \of{\beta/6}^7}{120}+\frac{139 \of{\beta/6}^8}{2880}+\frac{19 \of{\beta/6}^9}{864}\\
+\frac{23 \of{\beta/6}^{10}}{2400}+\frac{29 \of{\beta/6}^{11}}{7560}+\frac{2629 \of{\beta/6}^{12}}{1814400}+\frac{1241 \of{\beta/6}^{13}}{2419200}+\frac{17449 \of{\beta/6}^{14}}{101606400}+\ldots
\end{multline}
and
\begin{multline}
c_{f}\of{\beta}\,=\,\of{\beta/6}+\frac{\of{\beta/6}^2}{2}+\of{\beta/6}^3+\frac{5 \of{\beta/6}^4}{8}+\frac{13 \of{\beta/6}^5}{24}+\frac{77 \of{\beta/6}^6}{240}+\frac{139 \of{\beta/6}^7}{720}+\frac{19 \of{\beta/6}^8}{192}+\frac{23 \of{\beta/6}^9}{480}\\+\frac{319 \of{\beta/6}^{10}}{15120}+\frac{2629 \of{\beta/6}^{11}}{302400}+\frac{16133 \of{\beta/6}^{12}}{4838400}+\frac{17449 \of{\beta/6}^{13}}{14515200}+\frac{35531 \of{\beta/6}^{14}}{87091200}+\ldots\ .
\end{multline}

We now proceed by applying the mean-field approximation to \eqref{eq:effplaction}, i.e. we write $L_{i}\,=\,\bar{L}+\delta L_{i}$, keep only terms of order $\order{\delta L}$, and then write again $\delta L_{i}\,=\,L_{i}-\bar{L}$ (and proceed analogously for $L_{i}^{*}$), which, after dropping constant terms, leads to the following single site mean-field action:
\[
-S_{mf,i}\,=\,6\,a_{f}^{n_{t}}\of{\beta}\,\of{L^{*}_{i}\,\bar{L}\,+\,\bar{L}^{*}\,L_{i}}\ ,\label{eq:mfgaugeaction}
\]
which we use to define an additional probability weight,
\[
w\of{\bar{L},\bar{L}^{*},L,\beta,n_{t}}\,=\,\e^{6\,a_{f}^{n_{t}}\of{\beta}\,\of{L^{*}\,\bar{L}\,+\,\bar{L}^{*}\,L}}\ ,\label{eq:mfweight}
\]
that has to be included in the single site partition function defined in \eqref{eq:qcd2fpartf}, i.e.:
\begin{multline}
Z_{s}\sof{\mu_{u},\mu_{d},\kappa_{u},\kappa_{d},n_{t},\bar{L},\bar{L}^{*}}\,=\,\int\limits_{0}^{2\pi}\int\limits_{0}^{2\pi}\dd{\theta_{1}}\dd{\theta_{2}}\bcof{H\of{\theta_{1},\theta_{2}}\Det{D\of{\theta_{1},\theta_{2};\mu_{u},\kappa_{u},n_{t}}}\bigg.\\
\bigg.\Det{D\of{\theta_{1},\theta_{2};\mu_{d},\kappa_{d},n_{t}}}\,w\sof{\bar{L},\bar{L}^{*},L\of{\theta_{1},\theta_{2}},\beta,n_{t}}}\ .\label{eq:qcd2fpartfmf}
\end{multline}
For general values of $\mu_{u}$, $\mu_{d}$, $\kappa_{u}$ and $\kappa_{d}$, the product of the two fermion determinants in \eqref{eq:qcd2fpartfmf} is usually complex. As a consequence $\avof{L}$ and $\avof{L^{*}}$ differ and it is a subtle issue how to proceed in a mean-field treatment \cite{Fukushima}. To bypass the subtleties associated with $\bar{L}^{*}\neq\bar{L}$ when setting $\bar{L}=\avof{L}$ and $\bar{L}^{*}=\avof{L^{*}}$, we define $\bar{L}$ to be the mean-field value of $L$ with respect to the phase quenched system, where $\avof{L^{*}}_{q}=\avof{L}_{q}$ holds and the identification $\bar{L}=\avof{L}_{q}$, $\bar{L}^{*}=\avof{L^{*}}_{q}$ therefore leads to $\bar{L}^{*}=\bar{L}$. The mean-field $\bar{L}$ is therefore determined by solving the following self-consistency equation:
\begin{multline}
\bar{L}\,=\,\avof{L}_{q}\sof{\mu_{u},\mu_{d},\kappa_{u},\kappa_{d},n_{t},\bar{L}}\\
=\,\frac{1}{Z_{s,q}\of{\mu_{u},\mu_{d},\kappa_{u},\kappa_{d},n_{t},\bar{L}}}\,\int\limits_{0}^{2\pi}\int\limits_{0}^{2\pi}\dd{\theta_{1}}\dd{\theta_{2}}\bcof{L\of{\theta_{1},\theta_{2}}\,H\of{\theta_{1},\theta_{2}}\bigg.\\
\bigg.\abs{\Det{D\of{\theta_{1},\theta_{2};\mu_{u},\kappa_{u},n_{t}}}\,\Det{D\of{\theta_{1},\theta_{2};\mu_{d},\kappa_{d},n_{t}}}}\,w\sof{\bar{L},L\of{\theta_{1},\theta_{2}},\beta,n_{t}}}\ ,\label{eq:mfselfconseq}
\end{multline}
where $w\sof{\bar{L},L\of{\theta_{1},\theta_{2}},\beta,n_{t}}\equiv w\sof{\bar{L},\bar{L},L\of{\theta_{1},\theta_{2}},\beta,n_{t}}$ and where
\begin{multline}
Z_{s,q}\sof{\mu_{u},\mu_{d},\kappa_{u},\kappa_{d},n_{t},\bar{L}}\,=\,\int\limits_{0}^{2\pi}\int\limits_{0}^{2\pi}\dd{\theta_{1}}\dd{\theta_{2}}\bcof{H\of{\theta_{1},\theta_{2}}\abs{\Det{D\of{\theta_{1},\theta_{2};\mu_{u},\kappa_{u},n_{t}}}\right.\bigg.\\
\bigg.\left.\Det{D\of{\theta_{1},\theta_{2};\mu_{d},\kappa_{d},n_{t}}}}\,w\sof{\bar{L},L\of{\theta_{1},\theta_{2}},\beta,n_{t}}}\label{eq:qcd2fpartfmfq}
\end{multline}
is the phase-quenched partition function. After one has determined the stationary $\bar{L}$ with respect to \eqref{eq:mfselfconseq} for a particular set $\of{\mu_{u},\mu_{d},\kappa_{u},\kappa_{d},n_{t}}$ of parameters, one can reweight to the non-phase-quenched system by using $\bar{L}\sof{\mu_{u},\mu_{d},\kappa_{u},\kappa_{d},n_{t}}$ in \eqref{eq:qcd2fpartfmf} when computing expectation values of observables.\\

\begin{figure}[h]
\centering
\includegraphics[width=0.5\linewidth]{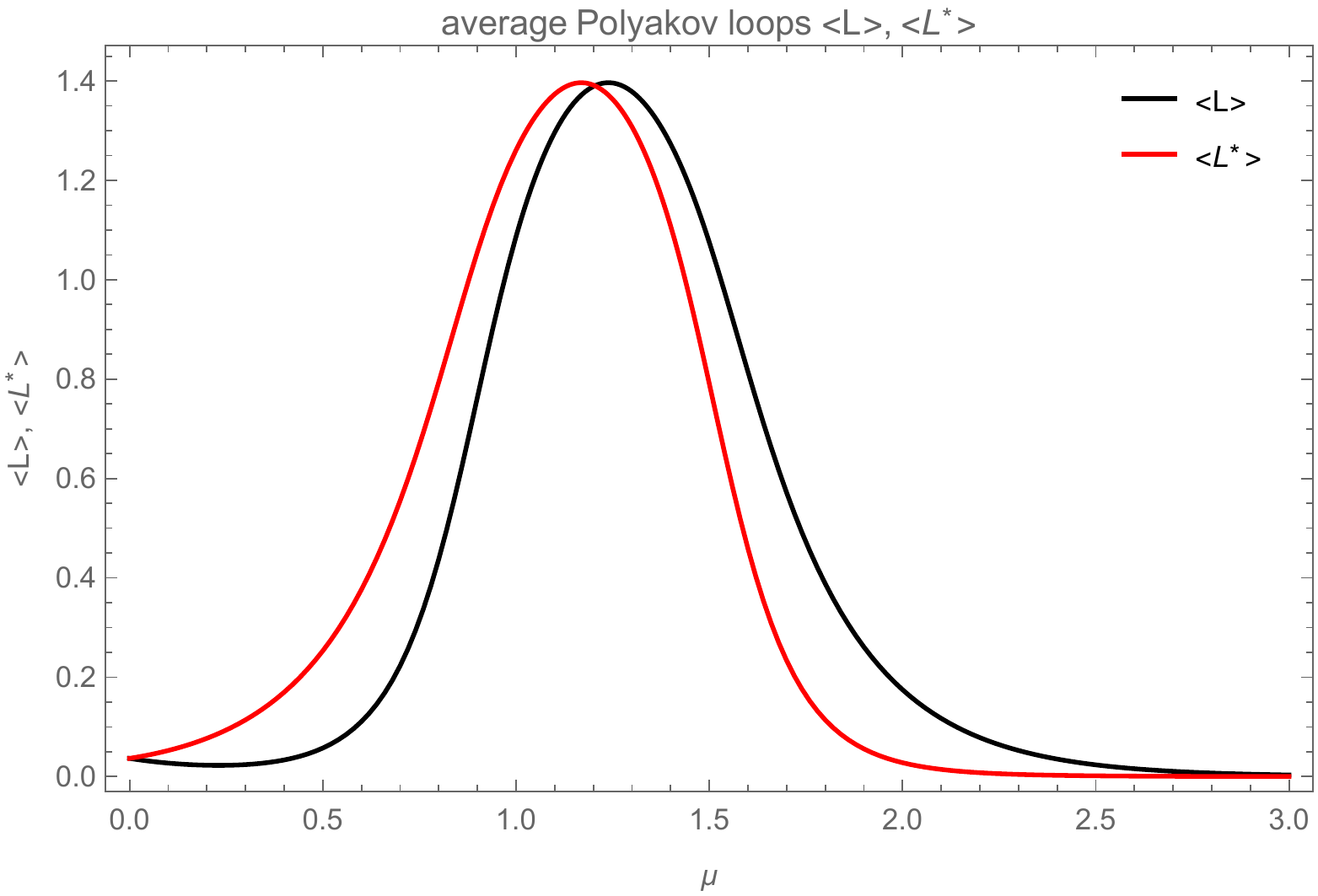}
\caption{Comparison of the average Polyakov loop and average complex conjugate Polyakov loop for a system with $n_{t}=4$, $\kappa_{u}=\kappa_{d}=0.15$ and $\beta=5.0$ as a function of a quark chemical potential $\mu=\mu_{u}=\mu_{d}$, obtained according to \eqref{eq:mfpolyakovl} and \eqref{eq:mfinvpolyakovl}. The average Polyakov loop (black curve) initially decreases with increasing chemical potential. Such a behavior has also been found in \cite{Pisarski} where in their figure 2, the expectation values for Polyakov loop and anti-Polyakov loop are shown for a $\SU{3}$ matrix model with a (quark) chemical potential.}
\label{fig:hdpolyakovlvspolyakovlc}
\end{figure}

Of particular interest among the observables that can be measured for the phase-unquenched system, are the average Polyakov loop,
\begin{multline}
\avof{L}\sof{\mu_{u},\mu_{d},\kappa_{u},\kappa_{d},n_{t}}\\
=\,\frac{1}{Z_{s}\of{\mu_{u},\mu_{d},\kappa_{u},\kappa_{d},n_{t},\bar{L}\sof{\mu_{u},\mu_{d},\kappa_{u},\kappa_{d},n_{t}}}}\,\int\limits_{0}^{2\pi}\int\limits_{0}^{2\pi}\dd{\theta_{1}}\dd{\theta_{2}}\bcof{L\of{\theta_{1},\theta_{2}}\,H\of{\theta_{1},\theta_{2}}\bigg.\\
\bigg.\Det{D\of{\theta_{1},\theta_{2};\mu_{u},\kappa_{u},n_{t}}}\,\Det{D\of{\theta_{1},\theta_{2};\mu_{d},\kappa_{d},n_{t}}}\,w\sof{\bar{L}\sof{\mu_{u},\mu_{d},\kappa_{u},\kappa_{d},n_{t}},L\of{\theta_{1},\theta_{2}},\beta,n_{t}}}\ ,\label{eq:mfpolyakovl}
\end{multline}
and the average complex conjugate Polyakov loop,
\begin{multline}
\avof{L^{*}}\sof{\mu_{u},\mu_{d},\kappa_{u},\kappa_{d},n_{t}}\\
=\,\frac{1}{Z_{s}\of{\mu_{u},\mu_{d},\kappa_{u},\kappa_{d},n_{t},\bar{L}\sof{\mu_{u},\mu_{d},\kappa_{u},\kappa_{d},n_{t}}}}\,\int\limits_{0}^{2\pi}\int\limits_{0}^{2\pi}\dd{\theta_{1}}\dd{\theta_{2}}\bcof{L^{*}\of{\theta_{1},\theta_{2}}\,H\of{\theta_{1},\theta_{2}}\bigg.\\
\bigg.\Det{D\of{\theta_{1},\theta_{2};\mu_{u},\kappa_{u},n_{t}}}\,\Det{D\of{\theta_{1},\theta_{2};\mu_{d},\kappa_{d},n_{t}}}\,w\sof{\bar{L}\sof{\mu_{u},\mu_{d},\kappa_{u},\kappa_{d},n_{t}},L\of{\theta_{1},\theta_{2}},\beta,n_{t}}}\ ,\label{eq:mfinvpolyakovl}
\end{multline}
for which we now have $\avof{L^{*}}\sof{\mu_{u},\mu_{d},\kappa_{u},\kappa_{d},n_{t}}\neq\avof{L}^{*}\sof{\mu_{u},\mu_{d},\kappa_{u},\kappa_{d},n_{t}}=\avof{L}\sof{\mu_{u},\mu_{d},\kappa_{u},\kappa_{d},n_{t}}$ due to the in general complex product of fermion determinants in \eqref{eq:mfpolyakovl} and \eqref{eq:mfinvpolyakovl}. This difference between $\avof{L^{*}}\sof{\mu_{u},\mu_{d},\kappa_{u},\kappa_{d},n_{t}}$ and $\avof{L}\sof{\mu_{u},\mu_{d},\kappa_{u},\kappa_{d},n_{t}}$ can clearly be seen in Fig. \ref{fig:hdpolyakovlvspolyakovlc} where we show the two quantities for the mass degenerate case (i.e. $\kappa_{u}=\kappa_{d}$) as a function of a quark chemical potential $\mu=\mu_{u}=\mu_{d}$.\\

It has been mentioned in \cite{Fukushima} that in a mean-field treatment, the reweighting method can be no more than an approximation scheme: while in a full Monte Carlo simulation, reweighting is exact in the limit of infinitely many sampled configurations, a mean-field treatment relies only on the most dominant configuration. In our opinion this argument does not hold as in a mean-field treatment, the expectation value of an observable is determined on the \emph{active site}, usually by integrating exactly over all possible values of its configuration variables, and reweighting should therefore work perfectly well!\\

Setting $\bar{L}^{*}=\bar{L}$, the integrals in \eqref{eq:qcd2fpartfmf} and \eqref{eq:qcd2fpartfmfq} can be carried out exactly, leading to solutions in terms of infinite sums of modified Bessel functions of the first kind:
\begin{multline}
Z_{s}\sof{\mu_{u},\mu_{d},\kappa_{u},\kappa_{d},n_{t},\bar{L}}\,=\\
\sum\limits_{l_1,l_2}\,C_{l_1,l_2}\sof{\mu_{u},\mu_{d},\kappa_{u},\kappa_{d},n_{t}}\,\sum\limits_{k=0}^{\infty}\,\frac{\sof{6\,a_{f}^{n_{t}}\of{\beta}\,\bar{L}}^{2 k-l_1}}{k!\,\of{k-l_1}!}\,\sum\limits_{m=0}^{2 k-l_1}\,\binom{2 k-l_1}{m}\,I_{\of{k-l_2-m}}\sof{12\,a_{f}^{n_{t}}\of{\beta}\,\bar{L}}\label{eq:exactmfpartf}
\end{multline}
and
\begin{multline}
Z_{s,q}\sof{\mu_{u},\mu_{d},\kappa_{u},\kappa_{d},n_{t},\bar{L}}\,=\\
\sum\limits_{l_1,l_2}\,\abs{C_{l_1,l_2}\sof{\mu_{u},\mu_{d},\kappa_{u},\kappa_{d},n_{t}}}\,\sum\limits_{k=0}^{\infty}\,\frac{\sof{6\,a_{f}^{n_{t}}\of{\beta}\,\bar{L}}^{2 k-l_1}}{k!\,\of{k-l_1}!}\,\sum\limits_{m=0}^{2 k-l_1}\,\binom{2 k-l_1}{m}\,I_{\of{k-l_2-m}}\sof{12\,a_{f}^{n_{t}}\of{\beta}\,\bar{L}}\label{eq:exactmfpartfq}
\end{multline}
respectively, where the coefficients $C_{l_1,l_2}\sof{\mu_{u},\mu_{d},\kappa_{u},\kappa_{d},n_{t}}$ are given by
\begin{multline}
H\of{\theta_{1},\theta_{2}}\Det{D\of{\theta_{1},\theta_{2};\mu_{u},\kappa_{u},n_{t}}}\,\Det{D\of{\theta_{1},\theta_{2};\mu_{d},\kappa_{d},n_{t}}}\,=\\
\sum\limits_{l_1,l_2}\,C_{l_1,l_2}\sof{\mu_{u},\mu_{d},\kappa_{u},\kappa_{d},n_{t}}\,\e^{\ii\,\of{l_1\,\theta_1\,+\,l_2\,\theta_2}}\ .
\end{multline}
Unfortunately, the evaluation of \eqref{eq:exactmfpartf} and \eqref{eq:exactmfpartfq} to the required accuracy is numerically rather expensive and it turned out to be more efficient to use direct numerical integration to solve for $\bar{L}$ in \eqref{eq:mfselfconseq} and for computing observables. Expressions \eqref{eq:exactmfpartf} and \eqref{eq:exactmfpartfq} were therefore merely used for cross-check purposes.\\

An expression for \eqref{eq:exactmfpartf} in the more general case of $\bar{L}^{*}\neq \bar{L}$ could be found along the lines of \cite{Splittorff} where the mean-field action (including correction terms) is derived for the case of a pure $\SU{N}$ Polyakov line model with a chemical potential.

\subsection{Mean-Field Heavy-Dense Phase Diagram and Finite $\beta$ Effects}\label{ssec:meanfieldhdphasediagram}
\begin{figure}[h]
\centering
\begin{minipage}[t]{0.485\linewidth}
\centering
\includegraphics[width=\linewidth]{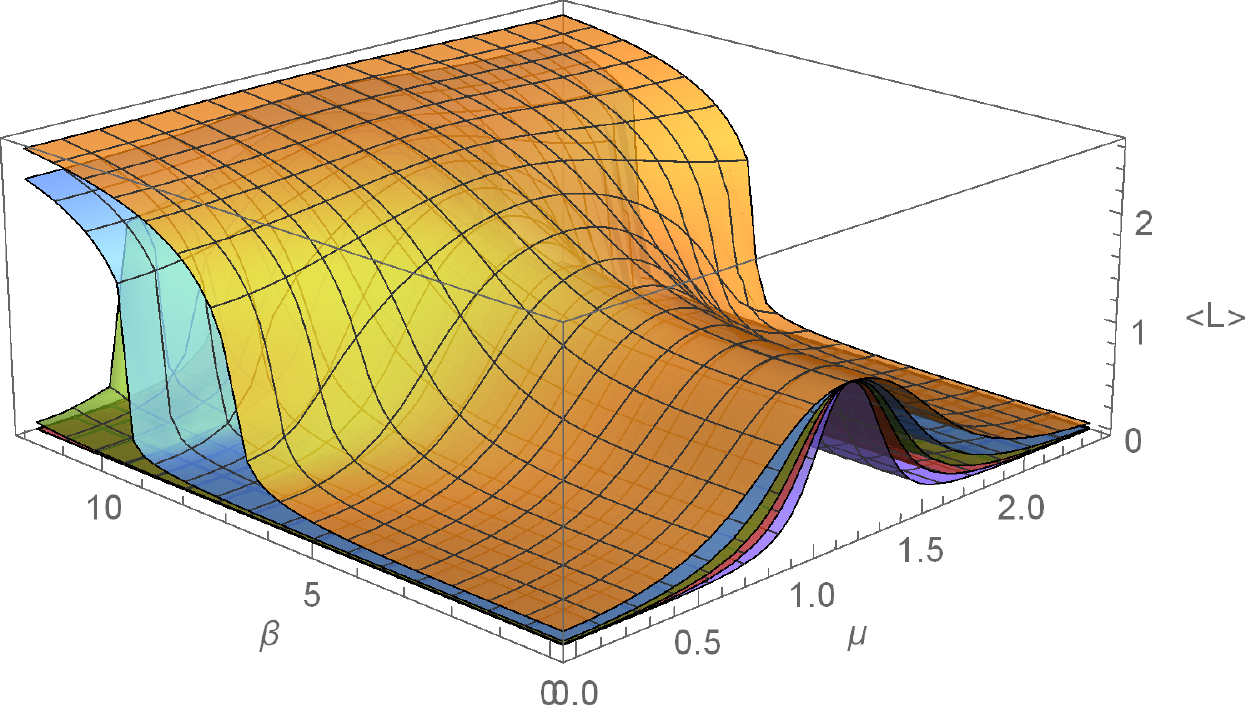}
\end{minipage}\hfill
\begin{minipage}[t]{0.485\linewidth}
\centering
\includegraphics[width=\linewidth]{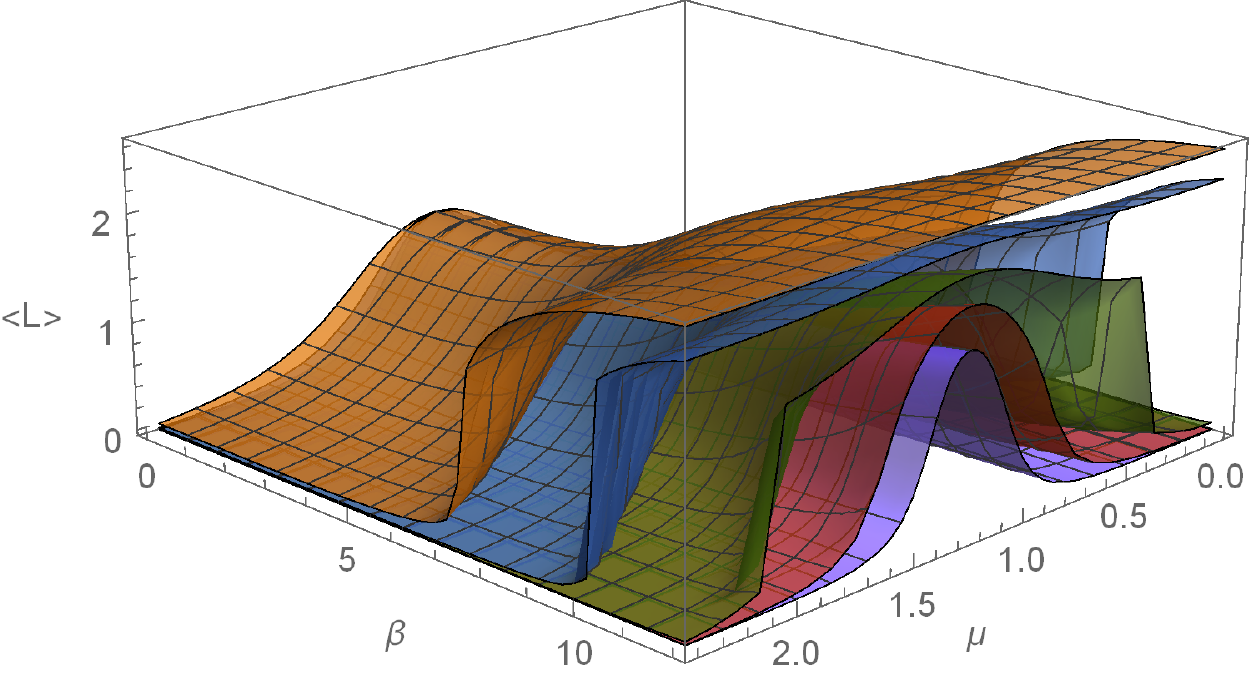}
\end{minipage}
\caption{Average Polyakov loop as a function of isospin chemical potential $\mu=\mu_u=-\mu_d$ and inverse coupling $\beta$ for five different temporal system sizes: $n_{t}=3$ (orange), $n_{t}=4$ (blue), $n_{t}=5$ (green), $n_{t}=6$ (red) and $n_{t}=8$ (purple), computed according to \eqref{eq:mfselfconseq} with $\kappa_{u}=\kappa_{d}=0.15$. The two figures show different viewpoints on the same data.}
  \label{fig:hdphasediagmubeta}
\end{figure}

In figure \ref{fig:hdphasediagmubeta} we show the average Polyakov loop for the mass degenerate two-flavor case as a function of isospin chemical potential $\mu=\mu_{u}=-\mu_{d}$ and inverse coupling $\beta$ for five different temperatures, $n_{t}^{-1}$, as obtained with our mean-field method described in the previous section. As can be seen, for sufficiently large $\beta$ a deconfinement transition occurs and the value of $\beta$ at which this transition happens becomes smaller with increasing temperature (decreasing $n_{t}$).\\

\begin{figure}[h]
\centering
\begin{minipage}[t]{0.485\linewidth}
\centering
\includegraphics[width=\linewidth]{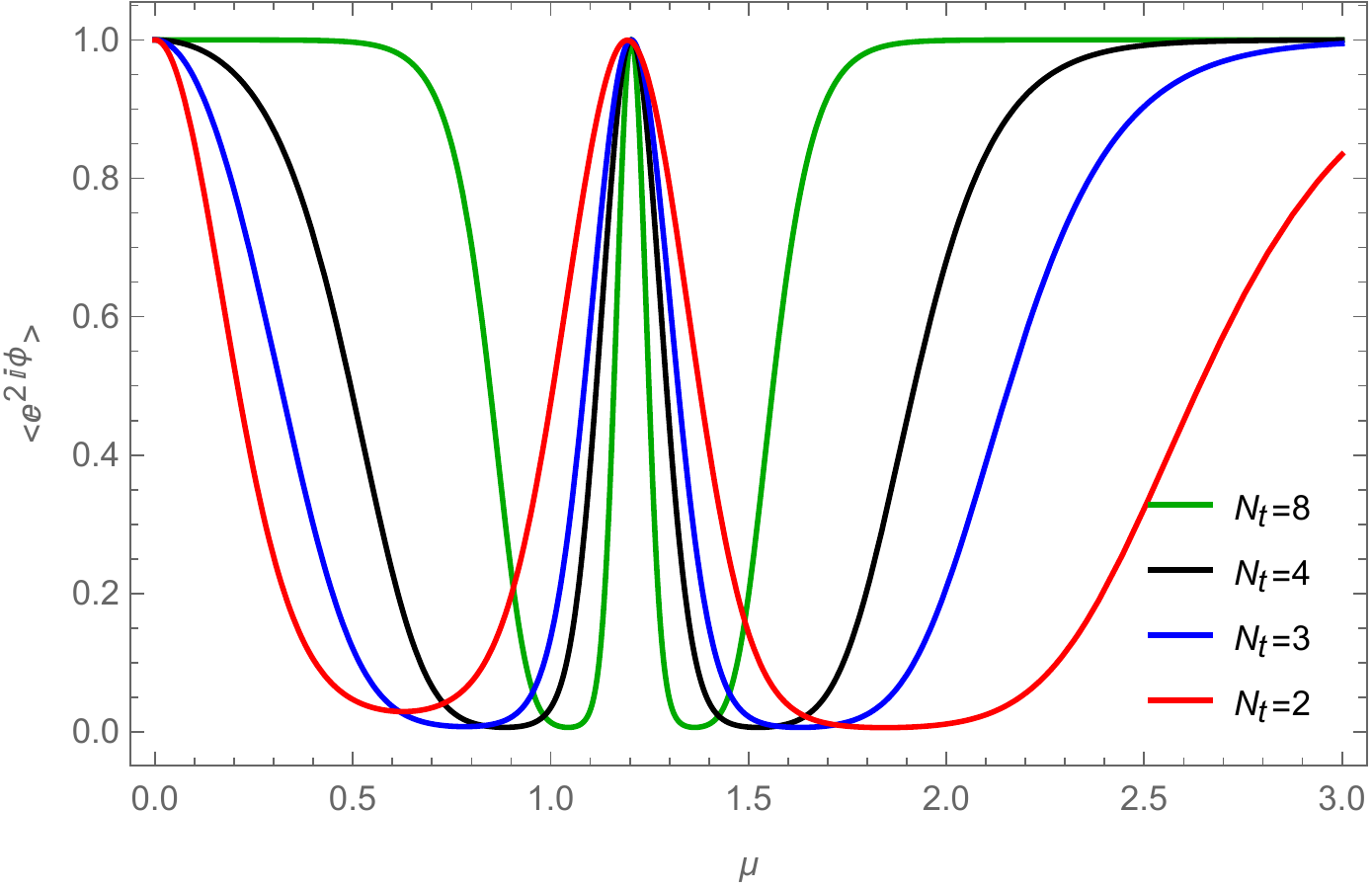}
\end{minipage}\hfill
\begin{minipage}[t]{0.485\linewidth}
\centering
\includegraphics[width=\linewidth]{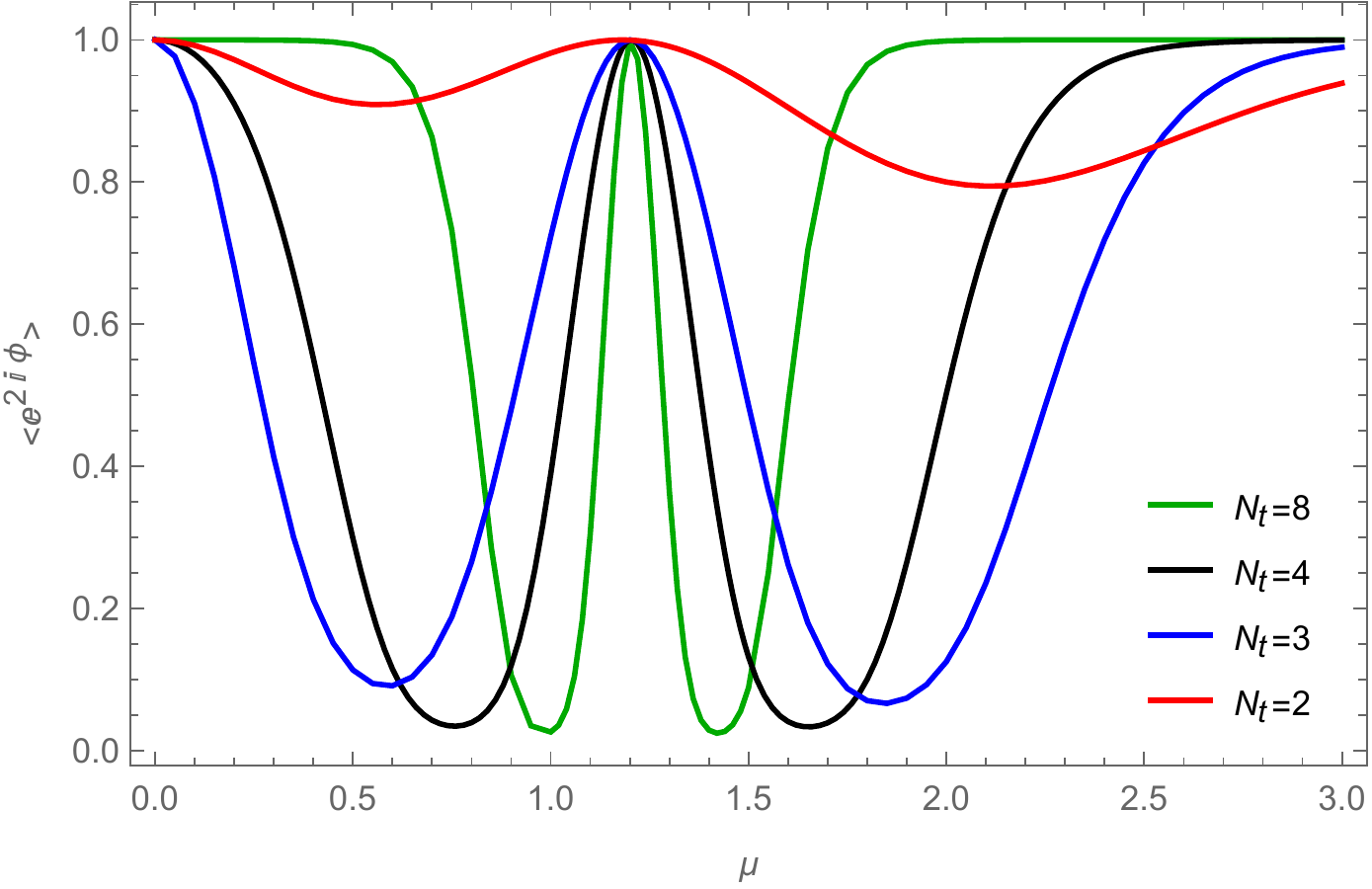}
\end{minipage}
\caption{Mean-field results for the average sign as a function of $\mu=\mu_u=\mu_d$ for a mass degenerate ($\kappa=0.15$) heavy-dense two-flavor system of spatial volume $V=L^3=4^3$ and temporal extents $N_{t}=2,\,3,\,4,\,8$. The left-hand part shows the strong-coupling limit, $\beta=0$, while the right-hand part shows the $\beta=5.0$ case.}
\label{fig:hdavsignvsmueffsc}
\end{figure}

A finite value of the inverse gauge coupling $\beta$ has also an effect on the average sign, which is illustrated in figure \ref{fig:hdavsignvsmueffsc}. There we show mean-field results for the average sign as a function of $\mu=\mu_u=\mu_d$ for a mass degenerate ($\kappa=0.15$) heavy-dense two-flavor system of spatial size $V=L^3=4^3$ and temporal extent $N_{t}=2,\,3,\,4,\,8$. The left-hand part shows the strong-coupling limit, $\beta=0$, while the right-hand part shows the $\beta=5.0$ case. For static quarks, the location of the half-filling point is not affected by the finite $\beta$ value, but the overall temperature dependency of the average sign clearly changes: at finite $\beta$ the sign-problem becomes weaker with increasing temperature and almost disappears after the deconfinement transition.\\
Note that the computation of the average sign with our mean-field method comes with a slight complication: the active site couples to its six nearest neighbors and there are therefore $6$ interaction terms entering the mean-field action. But the ratio of links to sites should be $3:1$ on a periodic 3-dimensional lattice, which means that by using simply a formula of the form of \eqref{eq:avsign} to compute the average sign with our mean-field partition functions \eqref{eq:qcd2fpartfmf} and \eqref{eq:qcd2fpartfmfq}, i.e.
\[
\avof{\e^{2\ii\,\phi}}\of{\mu,\kappa,\beta,n_{t},n_{x}\,n_{y}\,n_{z}}\,=\,\of{\frac{Z_{s}\sof{\mu,\kappa,n_{t},\bar{L}\of{\mu,\kappa,n_{t}}}}{Z_{s,q}\sof{\mu,\kappa,n_{t},\bar{L}\of{\mu,\kappa,n_{t}}}}}^{n_{x}\,n_{y}\,n_{z}}\ ,\label{eq:avsigneffwrong}
\]
we would get a wrong result. To get the correct answer, we have to include another reweighting factor in \eqref{eq:avsigneffwrong}, coming from one of the passive sites (a pair of one active and one passive site, together with 6 links connecting an active and a passive site, describes the smallest representative subset of a larger system, in which for example even sites are active and odd sites are passive or vice versa, and the computed sign will correspond to such a system). This factor is obtained by taking the ratio of the average fermion determinants in the non-phase-quenched and phase-quenched case. The average sign then becomes:
\begin{multline}
\avof{\e^{2\ii\,\phi}}\of{\mu,\kappa,\beta,n_{t},n_{x}\,n_{y}\,n_{z}}\,=\\
\of{\frac{Z_{s}\sof{\mu,\kappa,n_{t},\bar{L}\of{\mu,\kappa,n_{t}}}}{Z_{s,q}\sof{\mu,\kappa,n_{t},\bar{L}\of{\mu,\kappa,n_{t}}}}\frac{\avof{\Det{D}\Det{D}}\sof{\mu,\kappa,n_{t},\bar{L}\of{\mu,\kappa,n_{t}}}}{\avof{\abs{\Det{D}\Det{D}}}_{q}\sof{\mu,\kappa,n_{t},\bar{L}\of{\mu,\kappa,n_{t}}}}}^{\sfrac{n_{x}\,n_{y}\,n_{z}}{2}}\ ,\label{eq:avsigneff}
\end{multline}
where $\Det{D}$ is the static quark determinant \eqref{eq:fermiondetprod} and the factor of $1/2$ in the exponent in \eqref{eq:avsigneff} is there because the term inside the bracket is now the reweighting factor for two sites.
 
\begin{figure}[h]
\centering
\begin{minipage}[t]{0.485\linewidth}
\centering
\includegraphics[width=\linewidth]{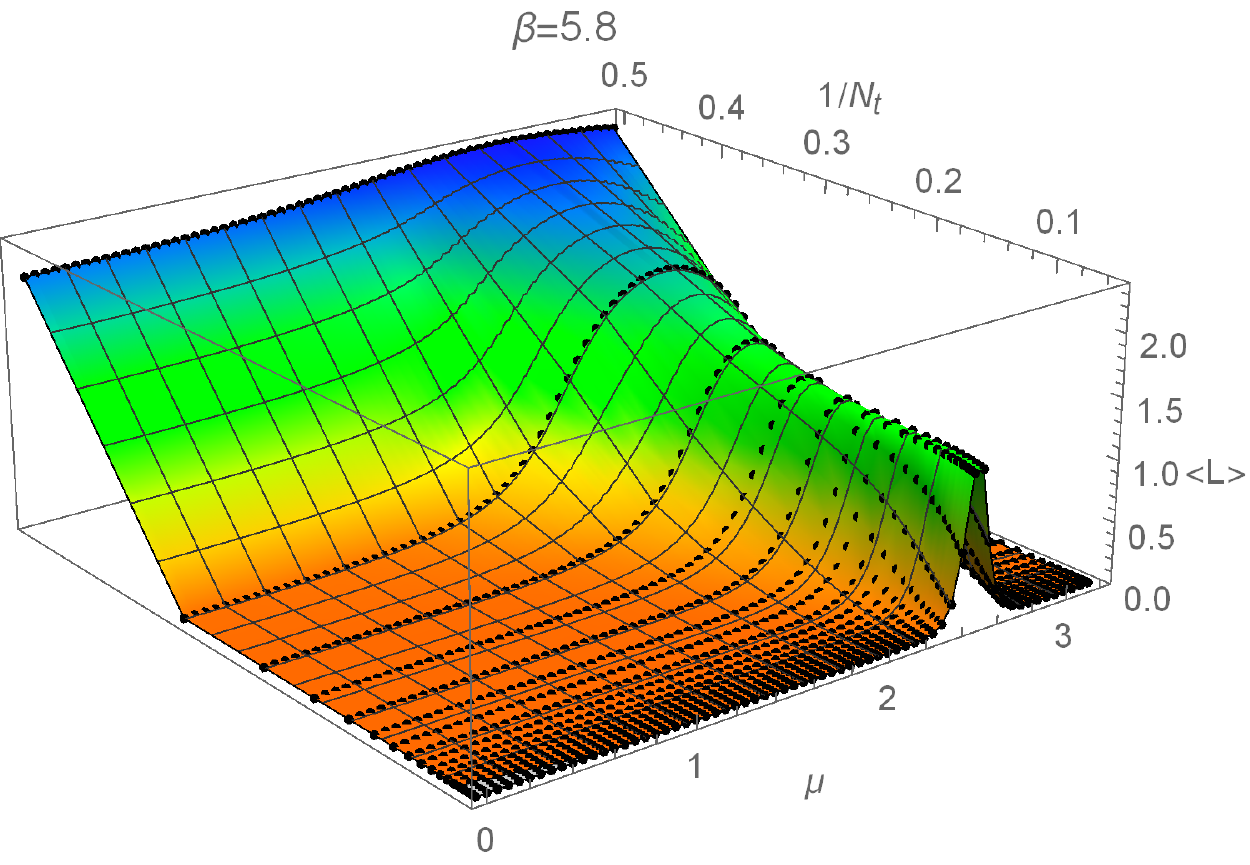}
\end{minipage}\hfill
\begin{minipage}[t]{0.485\linewidth}
\centering
\includegraphics[width=\linewidth]{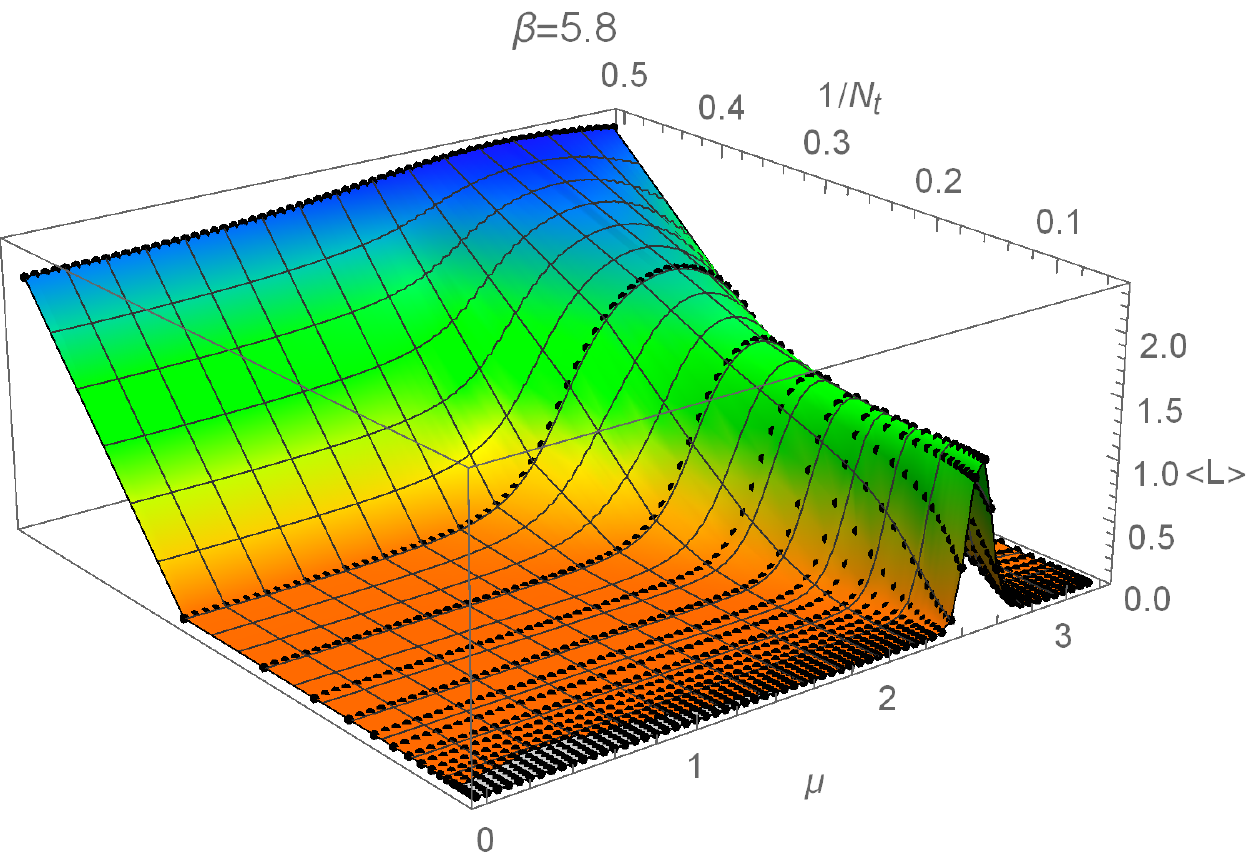}
\end{minipage}
\caption{Average Polyakov loop as a function of temperature $1/N_{t}$ and isospin (left) and quark (right) chemical potential respectively at $\beta=5.8$, computed according to \eqref{eq:mfselfconseq} and \eqref{eq:mfpolyakovl} respectively with $\kappa_{u}=\kappa_{d}=0.04$.}
  \label{fig:hdphasediagmutempb7}
\end{figure}

\begin{figure}[h]
\centering
\includegraphics[width=0.6\linewidth]{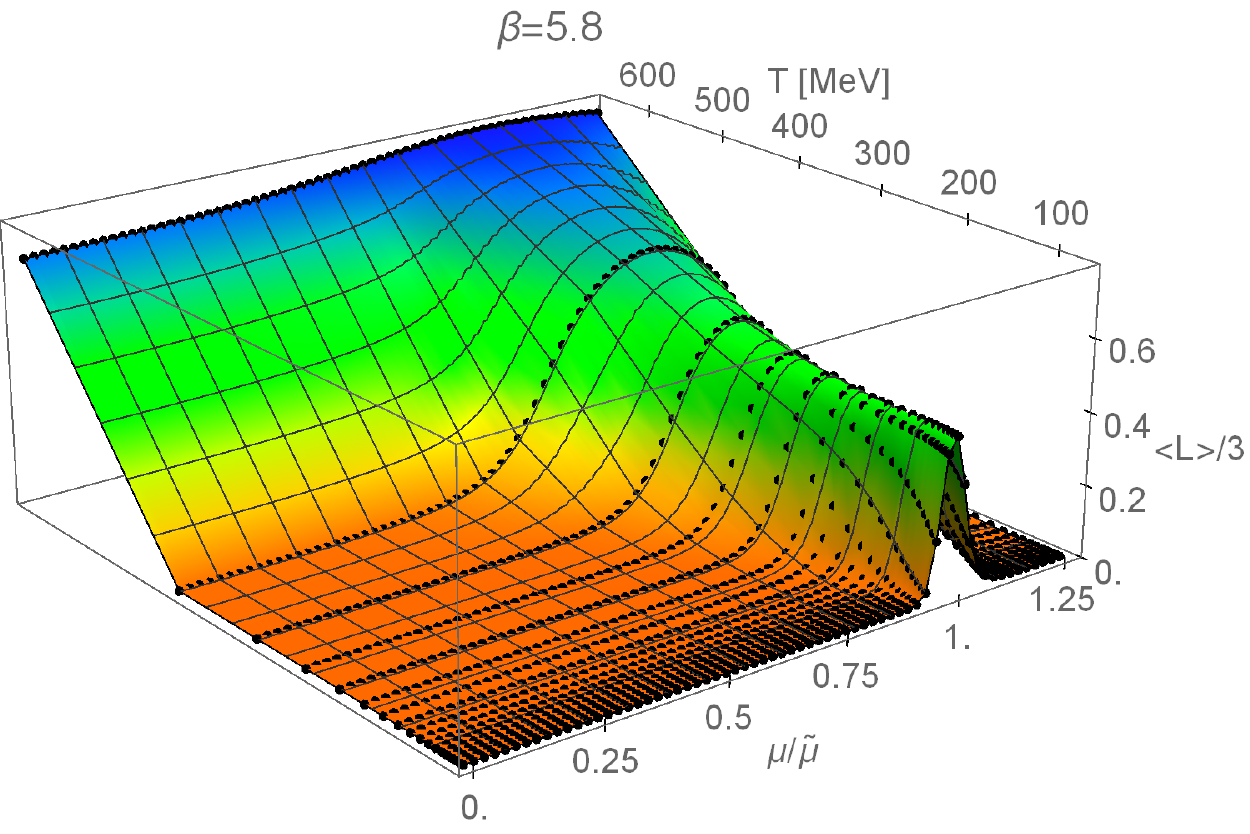}
\caption{Average Polyakov loop as a function of temperature $T$ and normalized quark chemical potential $\mu/\tilde{\mu}$, where $\tilde{\mu}=-\log{2\kappa}$ is the half-filling value of $\mu$ in the heavy-dense limit, with $\kappa=0.04$. It is the same data as in the right-hand part of Fig. \ref{fig:hdphasediagmutempb7}, expressed in the units used in \cite[Fig. 2]{Aarts}, in order to allow for a better comparison.}
  \label{fig:hdphasediagmutempb7phys}
\end{figure}

Figure \ref{fig:hdphasediagmutempb7} finally shows the average Polyakov loop as a function of temperature and isospin (left) or quark (right) chemical potential i.e. it essentially shows the $\of{T,\mu}$ phase diagram of our simplified model in the phase-quenched and un-quenched case: one can read off how the pseudo-critical temperature for the deconfinement transition changes as a function of the chemical potential. The right-hand part of Fig. \ref{fig:hdphasediagmutempb7}, corresponding to the un-quenched case, can be compared, e.g. to the phase diagram in \cite[Fig. 2]{Aarts}, obtained by complex Langevin. In order to simplify the comparison, Fig. \ref{fig:hdphasediagmutempb7phys} shows the data from Fig. \ref{fig:hdphasediagmutempb7} using the same scales used in \cite[Fig. 2]{Aarts}, assuming that the lattice spacing of $a=0.15$ fm, determined in \cite{Aarts}, should apply equally well to our system, as we used the same simulation parameters $\kappa=0.04$, $\beta=5.8$.  In our effective model, we take the gauge field only in the fundamental representation into account. But at high temperature, effects coming from higher representations are much less suppressed than at low temperature (the coefficients $a_{r}\of{\beta}$, if included in the effective Polyakov loop action, would appear only with small exponents, leading to a weaker suppression of the higher order terms), which explains why in our figure the deconfinement transition is generally shifted towards larger temperatures compared to \cite[Fig. 2]{Aarts}. 

\subsection{From the Heavy-Dense to the Full Fermion Determinant}\label{ssec:fromhdtofulldet}
\begin{figure}[H]
\centering
\begin{minipage}[t]{\linewidth}
\centering
\begin{tikzpicture}[scale=0.7,nodes={inner sep=0}]
  \pgfpointtransformed{\pgfpointxy{1}{1}};
  \pgfgetlastxy{\vx}{\vy}
  \begin{scope}[node distance=\vx and \vy]
  \node at (0,0) (x0) {};
  \node at (2,0) (x1) {};
  \node at (3,1) (x2) {};
  \node at (1,1) (x3) {};
  \node at (0,7) (x0t) {};
  \node at (2,7) (x1t) {};
  \node at (3,8) (x2t) {};
  \node at (1,8) (x3t) {};
  \node at (0,2) (x0s1) {};
  \node at (2,2) (x1e1) {};
  \node at (2,6) (x1s1) {};
  \node at (3,7) (x2e1) {};
  \node at (3,5) (x2s1) {};
  \node at (1,5) (x3e1) {};
  \node at (1,4) (x3s1) {};
  \node at (0,3) (x0e1) {};
  \node at (0,5.5) (x0e1f) {};
  
  \node at (2,2.5) (x1e1t) {};
  \node at (3,7.5) (x2e1t) {};
  \node at (1,5.5) (x3e1t) {};
  \node at (0,3.5) (x0e1t) {};
  
  \node at (2,1.5) (x1e1b) {};
  \node at (3,6.5) (x2e1b) {};
  \node at (1,4.5) (x3e1b) {};
  \node at (0,2.5) (x0e1b) {};
  
  \node[left=0.1 of x0] (s0) {$\bar{x}_{0}$};
  \node[right=0.1 of x1] (s1) {$\bar{x}_{1}$};
  \node[right=0.1 of x2] (s2) {$\bar{x}_{2}$};
  \node[left=0.1 of x3] (s3) {$\bar{x}_{3}$};
  \draw[->,color=blue] (x0) edge (x1) (x1) edge (x2) (x2) edge (x3) (x3) edge (x0);
  \draw[-,color=black] (x0) edge (x0t);
  \draw[-,color=black] (x1) edge (x1t);
  \draw[-,color=black] (x2) edge (x2t);
  \draw[-,color=black] (x3) edge (x3t); 
  \draw[->,color=red] (x0s1) edge (x1e1) (x1s1) edge (x2e1) (x2s1) edge (x3e1) (x3s1) edge (x0e1);
  \draw[<->,color=red,dotted] (x0e1b) edge (x0e1t);
  \draw[<->,color=red,dotted] (x1e1b) edge (x1e1t);
  \draw[<->,color=red,dotted] (x2e1b) edge (x2e1t);
  \draw[<->,color=red,dotted] (x3e1b) edge (x3e1t); 
  \node[left=0.1 of x0s1,color=red] (s0s1) {$x_{4,s}$};
  \node[left=0.1 of x0e1f,color=red] (s0e1) {$x_{4,e}$};
  \node[left=0.1 of x1s1,color=red] (s1i) {$x_{4,i_{1}}$};
  \node[right=0.1 of x2s1,color=red] (s2i) {$x_{4,i_{2}}$};
  \node[left=0.1 of x3s1,color=red] (s3i) {$x_{4,i_{3}}$};
  \node[draw,circle,inner sep=0.75pt,fill,color=red] at (x0s1) {};
  \node[draw,circle,inner sep=0.75pt,fill,color=red] at (x0e1f) {};
  \node[draw,circle,inner sep=0.75pt,fill,color=red] at (x1s1) {};
  \node[draw,circle,inner sep=0.75pt,fill,color=red] at (x2s1) {};
  \node[draw,circle,inner sep=0.75pt,fill,color=red] at (x3s1) {};
  \node at (-5,3) (l1) {$\sof{\tilde{M}_{C_{4}}}_{a,I,x_{4,s};b,J,x_{4,e}}$};
  \end{scope}
\end{tikzpicture}
\end{minipage}
\caption{The figure illustrates the meaning of the \emph{spatial loop matrix}, introduced in eq. \eqref{eq:spatloopmatrix}, for the case of a contour $C_{s_0}$ with $s_0=4$, which is spanned by the four spatial sites $\rbar{x}_{0}$, $\rbar{x}_{1}$, $\rbar{x}_{2}$ and $\rbar{x}_{3}$. In the example, the loop starts and ends at the spatial site $\rbar{x}_{0}$, but as the matrices $\tilde{M}_{C_{s_0}}$ always appears as $\det_{c,d,t}\ssof{\id-\kappa_{s}^{s_0}\tilde{M}_{C_{s_0}}}$ (or inside a trace $\trace_{c,d,t}\ssof{\tilde{M}_{C_{s_0}}}$), it is clear that the base point of the spatial loop is arbitrary.\\
If we look at a particular matrix element of $\tilde{M}_{C_4}$ with time indices $x_{4,s}$ and $x_{4,e}$, it corresponds to the sum over all possible ways to draw a quark world-line (example depicted in red), connecting $\of{\rbar{x}_{0},x_{4,s}}$ and $\of{\rbar{x}_{0},x_{4,e}}$ by winding exactly once around the spatial (oriented) loop $C_{4}$ (depicted in blue): it starts with a spatial hop from $\of{\rbar{x}_{0},x_{4,s}}$ to $\of{\rbar{x}_{1},x_{4,s}}$. The next step involves the static quark time-propagator, which is the sum over all possible ways to go from $\of{\rbar{x}_{1},x_{4,s}}$ to let us say $\of{\rbar{x}_{1},x_{4,i_1}}$ (including all possible positive and negative windings around the time direction). It continues with another spatial hop from $\of{\rbar{x}_{1},x_{4,i_1}}$ to $\of{\rbar{x}_{2},x_{4,i_1}}$, followed by another static time-propagation from $\of{\rbar{x}_{2},x_{4,i_1}}$ to $\of{\rbar{x}_{2},x_{4,i_2}}$, and so on. The final step consists of a spatial hop from $\of{\rbar{x}_{3},x_{4,i_3}}$ to $\of{\rbar{x}_{0},x_{4,i_3}}$, followed by a final static time-propagation from $\of{\rbar{x}_{0},x_{4,i_3}}$ to $\of{\rbar{x}_{0},x_{4,e}}$. The full matrix element $\ssof{\tilde{M}_{C_4}}_{x_{4,s},x_{4,e}}$ (color and Dirac indices suppressed) is obtained by summing over all intermediate time indices $x_{4,i_1}$, $x_{4,i_2}$ and $x_{4,i_3}$.\\
Similarly, a matrix element of the $n$-th power of $\tilde{M}_{C_4}$, e.g. $\ssof{\tilde{M}_{C_4}^{n}}_{x_{4,s},x_{4,e}}$, corresponds to the sum over all possible ways to draw a quark world-line, connecting $\of{\rbar{x}_{0},x_{4,s}}$ and $\of{\rbar{x}_{0},x_{4,e}}$ by winding exactly $n$ times around the spatial loop $C_4$.}
  \label{fig:spatialloopexpandex}
\end{figure}
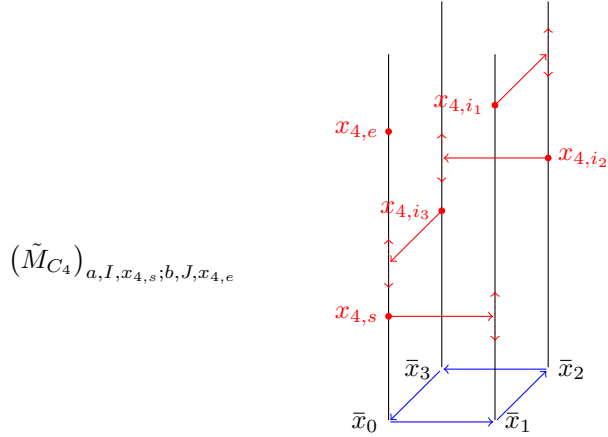

The heavy-dense fermion determinant is the leading term in the so-called \emph{spatial hopping expansion} of the full fermion determinant which, for one flavor, can be obtained as follows:
\begin{align}
\Det{D\of{\mu}}\,=&\,\Det{\id-\kappa\,T-\kappa_{s}\,S}\,=\,\Det{\id-\kappa_{s}\,S\of{\id-\kappa\,T}^{-1}}\,\Det{\id-\kappa\,T}\nonumber\\
=&\,\exp\bcof{-\sum\limits_{k=1}^{\infty}\,\frac{\kappa_{s}^{k}}{k}\Trace{\sof{S\,\of{\id-\kappa\,T}^{-1}}^{k}}}\cdot\Det{\id-\kappa\,T}\ ,\label{eq:spathoppingexpand1}
\end{align}
where $S$ and $T$ are the spatial and temporal hopping terms as defined in \eqref{eq:wilsondiracop} and we have added a subscript $s$ to the spatial hopping parameter, just for book-keeping purposes. The matrix $\of{\id-\kappa\,T}^{-1}$ is the heavy dense quark propagator given by \eqref{eq:hdquarkprop}. We can now proceed in a similar way as in \eqref{eq:fullfermiondet} to write \eqref{eq:spathoppingexpand1} in terms of a product of smaller determinants of "loop matrices", but this time, the loops are purely spatial and the matrices carry also a time-index, i.e.
\begin{align}
\Det{D\of{\mu}}\,=&\,\exp\bcof{-\sum\limits_{s_{0}}\sum\limits_{\cof{C_{s_{0}}}}\sum\limits_{n=1}^{\infty}\,\frac{\of{\kappa_{s}^{s_{0}}}^{n}}{n}\trace_{c,d,t}\of{\sof{\tilde{M}_{C_{s_{0}}}}^{n}}}\cdot\Det{\id-\kappa\,T}\nonumber\\
=&\,\exp\bcof{\sum\limits_{s_{0}}\sum\limits_{\cof{C_{s_{0}}}}\,\trace_{c,d,t}\of{\log\sof{\id-\kappa_{s}^{s_{0}}\tilde{M}_{C_{s_{0}}}}}}\cdot\Det{\id-\kappa\,T}\nonumber\\
=&\,\bof{\prod\limits_{s_{0}}\prod\limits_{\cof{C_{s_{0}}}}\,\det_{c,d,t}\sof{\id-\kappa_{s}^{s_{0}}\tilde{M}_{C_{s_{0}}}}}\cdot\Det{\id-\kappa\,T}\ ,\label{eq:spathoppingexpand}
\end{align}
where the subscripts in $\trace_{c,d,t}$, $\det_{c,d,t}$ indicate that these operators act on color, Dirac and time indices. $C_{s_{0}}$ describes a closed spatial path of length $s_{0}$ (where, in contrast to \eqref{eq:fullfermiondet}, backtracking is now allowed) and the matrix $\tilde{M}_{C_{s_{0}}}$ is given by the (ordered) product of the matrices $S\of{\id-\kappa\,T}^{-1}$ along that spatial path (see Fig. \ref{fig:spatialloopexpandex}),
\[
\tilde{M}_{C_{s_{0}}}\,=\,\prod\limits_{\overset{\scriptstyle i=0}{\bar{x}_{i}\in C_{s_{0}}}}^{s_{0}-1}\,S_{\bar{x}_{i},\bar{x}_{\ssof{\ssof{i+1}\bmod s_0}}}\,\of{\id-\kappa\,T}^{-1}_{\bar{x}_{\ssof{\ssof{i+1}\bmod s_0}}}\ ,\label{eq:spatloopmatrix}
\]
where each of the $S_{\bar{x}_{i},\bar{x}_{i+1}}\of{\id-\kappa\,T}^{-1}_{\bar{x}_{i+1}}$ is understood to be a matrix with color, Dirac and time indices.\\

The final form of \eqref{eq:spathoppingexpand}, which could be called a \emph{spatial loop expansion}, is particularly useful if one is aiming towards including spatial fermion hoppings into the mean-field treatment. In this case one can just restrict the product over $s_{0}$ in \eqref{eq:spathoppingexpand} to the factors with $s_{0}\,=\,0,\,2$ but still get the full interaction between nearest-neighboring sites. In contrast, in the original spatial hopping expansion \eqref{eq:spathoppingexpand1}, a nearest-neighbor truncation would limit the accuracy to $\order{\kappa_{s}^{2}}$.\\

\begin{figure}[H]
\centering
\begin{minipage}[t]{0.485\linewidth}
\centering
\includegraphics[width=\linewidth]{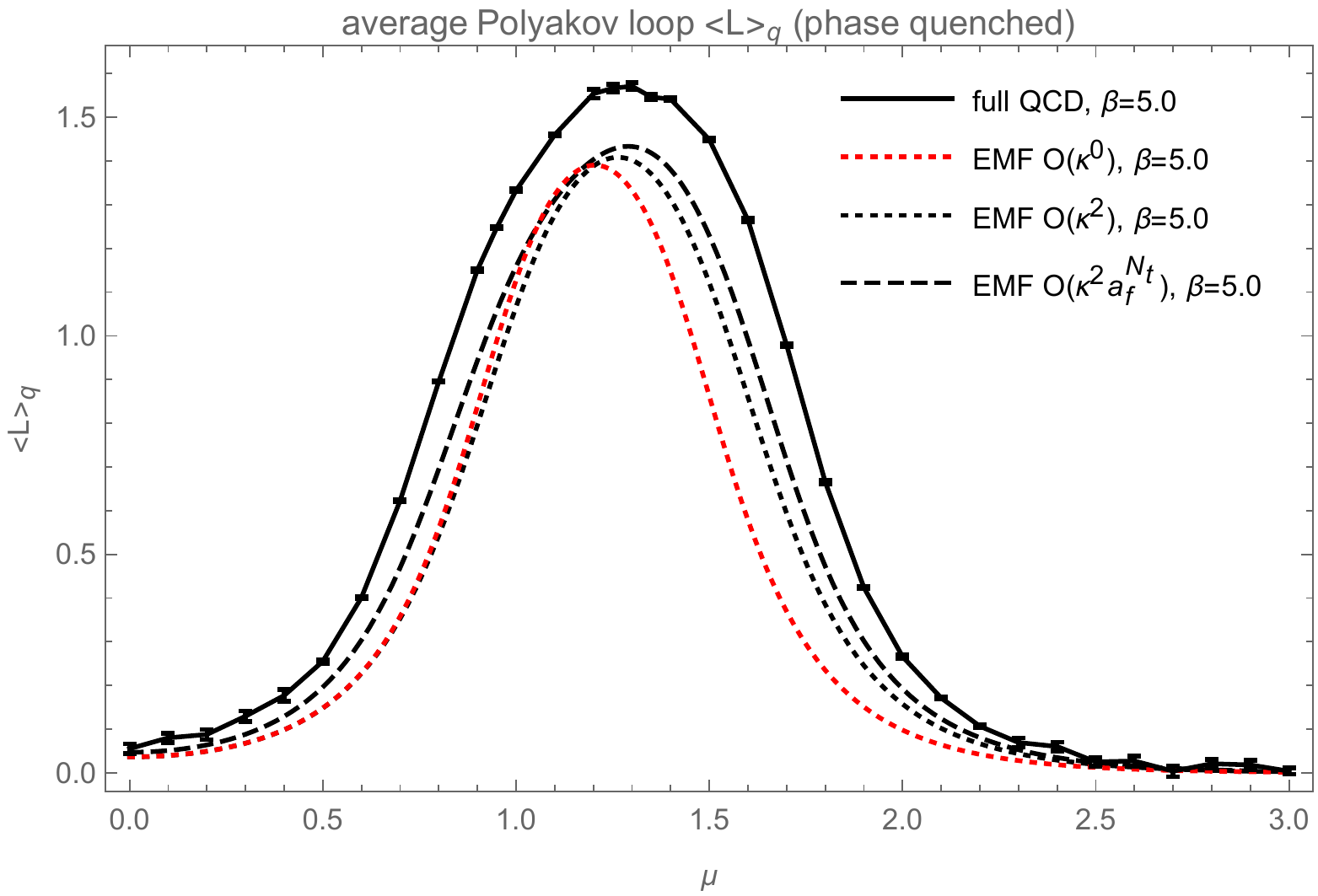}
\end{minipage}\hfill
\begin{minipage}[t]{0.485\linewidth}
\centering
\includegraphics[width=\linewidth]{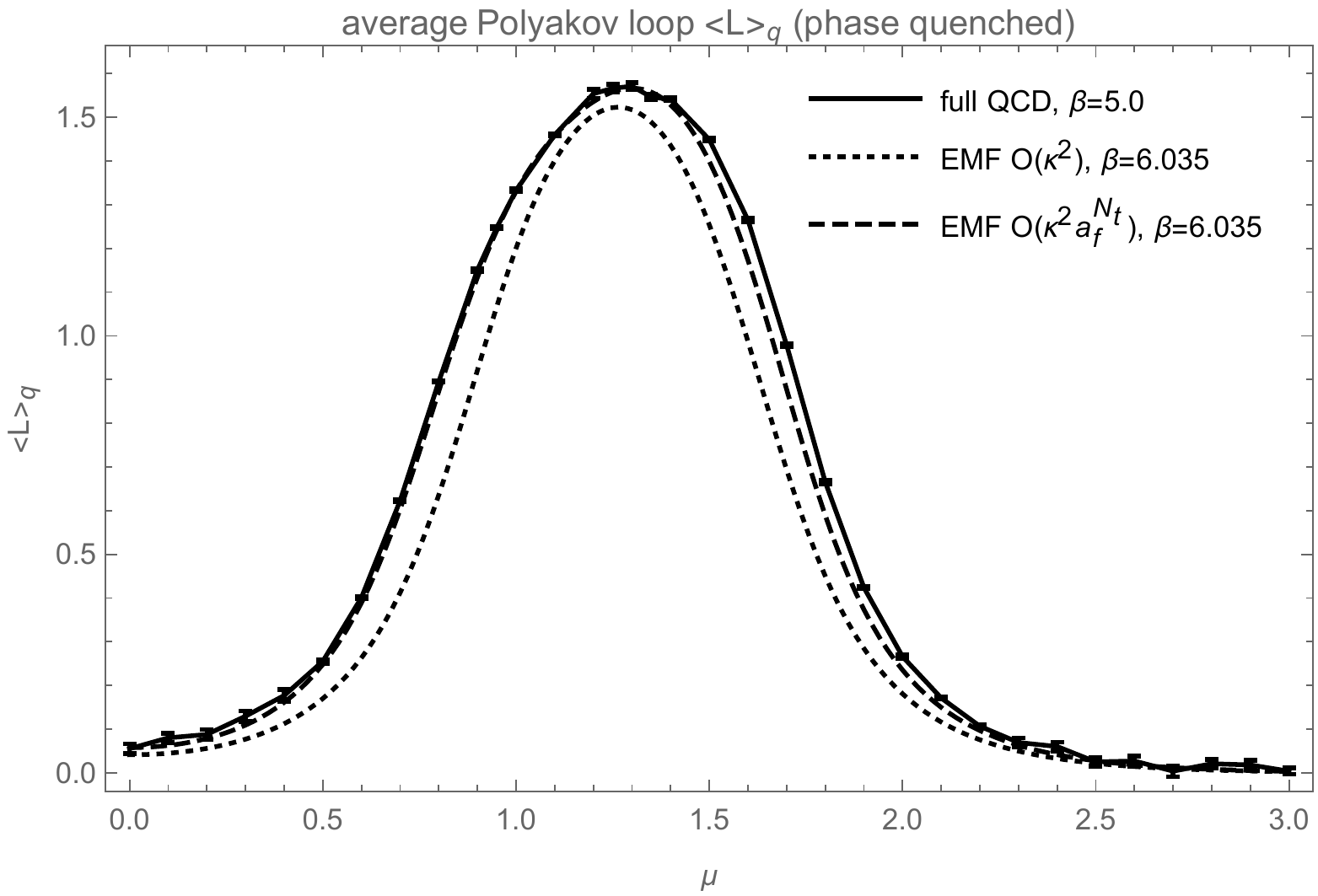}
\end{minipage}
\caption{The left-hand figure shows the average Polyakov loop for a phase quenched, mass-degenerate two-flavor system with $n_{x}n_{y}n_{z}n_{t}=4^4$, $\kappa=0.15$ and $\beta=5.0$. The solid curve guides the eye through the data points coming from a full Monte Carlo simulation, while the dashed line corresponds to the result of a corresponding mean-field calculation with our effective nearest-neighbor Polyakov loop action (see appendix \ref{sec:kappasqerms}), including terms up to order $\order{\kappa_{s}^{2}a_{f}^{n_{t}}\of{\beta}}$. For comparison, the dotted curves show the corresponding results when using only $\order{\kappa_{s}^{2}}$ (black dotted curve) or $\order{\kappa_{s}^{0}}$ (red dotted curve) terms in the fermionic part of the effective action, which results in larger deviations from the Monte Carlo data. The right-hand figure shows the same Monte Carlo data (solid curve), but this time it is compared to mean-field results obtained at a slightly larger value of the inverse gauge coupling $\beta=6.035$ (trying to compensate for the neglected piece of the full gauge interaction). The dashed line, corresponding to the $\order{\kappa_{s}^{2}a_{f}^{n_{t}}\of{\beta}}$ improved effective nearest-neighbor Polyakov loop action, then almost coincides with the Monte Carlo data.}
  \label{fig:polyakovloopfulleffcmp}
\end{figure}

\begin{figure}[H]
\centering
\begin{minipage}[t]{0.485\linewidth}
\centering
\includegraphics[width=\linewidth]{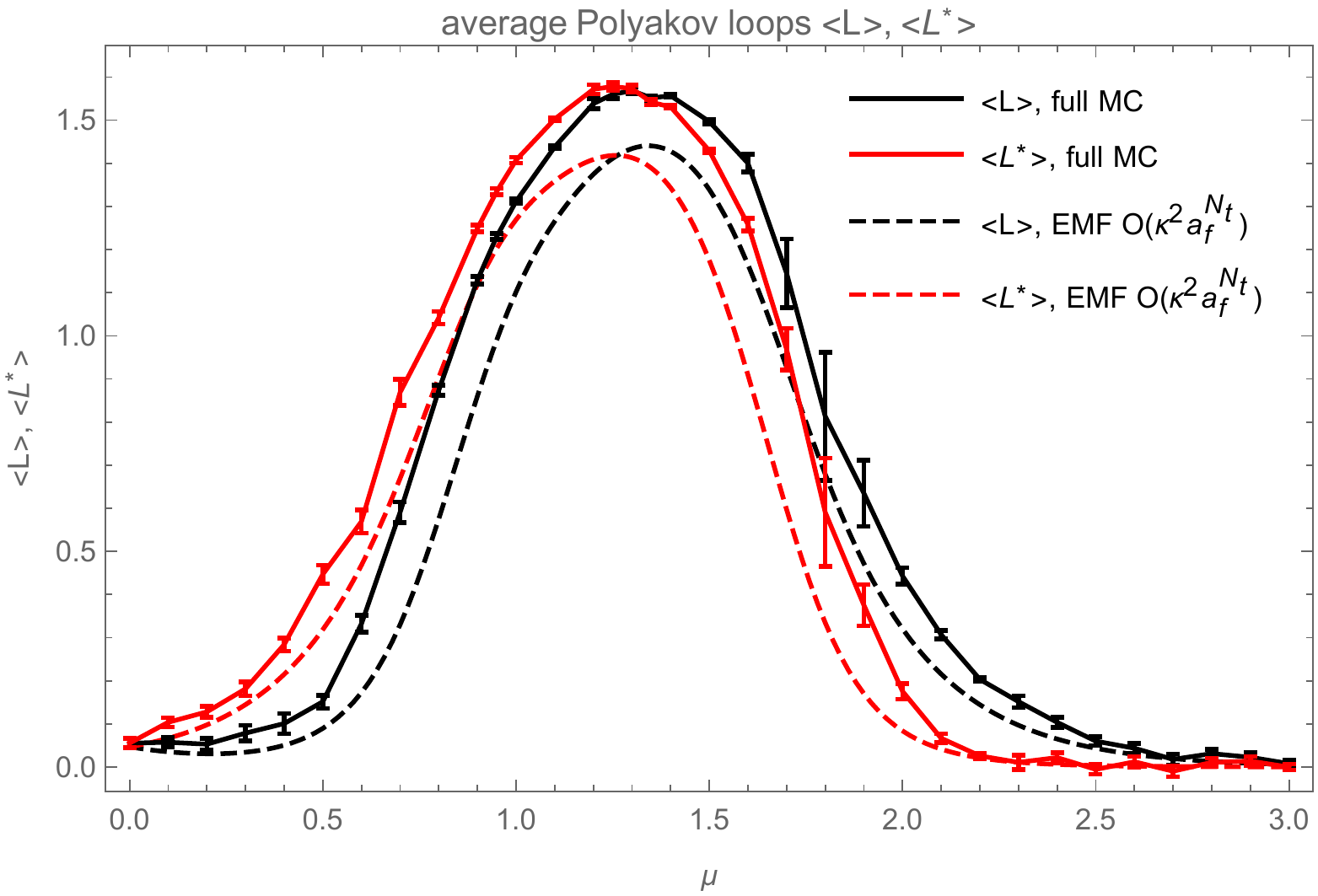}
\end{minipage}\hfill
\begin{minipage}[t]{0.485\linewidth}
\centering
\includegraphics[width=\linewidth]{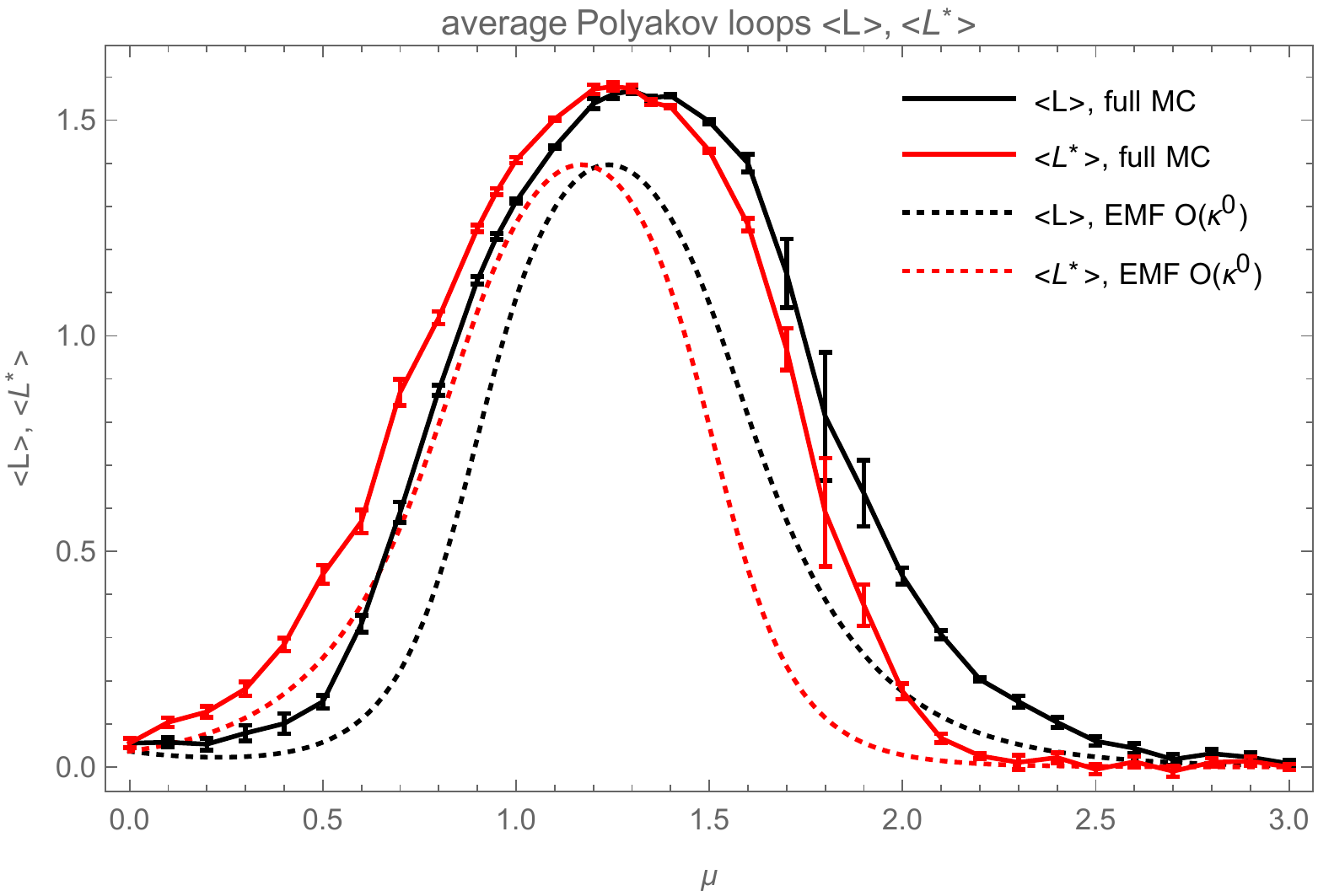}
\end{minipage}
\caption{Average Polyakov loop (black) and complex conjugate Polyakov loop (red) in a non-phase quenched two-flavor system with $n_{x}n_{y}n_{z}n_{t}=4^4$, $\kappa=0.15$ and $\beta=5.0$. The solid curves are to guide the eye through the data points coming from a Monte Carlo simulation of full LQCD, while the dashed lines in the left-hand figure show the result of a corresponding mean-field calculation analogous to the one that lead to Fig. \ref{fig:hdpolyakovlvspolyakovlc}, but using an effective action that takes terms up to order $\order{\kappa_{s}^{2}a_{f}^{n_{t}}\of{\beta}}$ into account (see appendix \ref{sec:kappasqerms}). For comparison, the right-hand figure shows again the data from Fig. \ref{fig:hdpolyakovlvspolyakovlc} (dotted lines) together with the full LQCD Monte Carlo data.}
  \label{fig:polyakovloopfulleffcmpnq}
\end{figure}

In order to see how the presence of spatial hoppings changes the properties of the static-quark system, it is sufficient to take just the lowest order terms in $\kappa_{s}$ of \eqref{eq:spathoppingexpand} into account. These terms lead, after integrating out the spatial links, to corrections up to order $\order{\kappa_{s}^{2}\,a_{f}^{n_{t}}\of{\beta}}$ (see appendix \ref{sec:kappasqerms}). Using this improved effective nearest-neighbor Polyakov loop action within the mean-field framework described above in Sec. \ref{ssec:mfmethod}, we can for example compute the average Polyakov loop for a mass degenerate two-flavor system. The result is shown in the left-hand parts of the figures \ref{fig:polyakovloopfulleffcmp} (phase-quenched case) and \ref{fig:polyakovloopfulleffcmpnq} (non-phase-quenched case), together with the corresponding data obtained from Monte Carlo simulations of full LQCD. More precisely, the left-hand part of Fig. \ref{fig:polyakovloopfulleffcmp} shows the full LQCD result together with mean-field results obtained with three different effective nearest-neighbor Polyakov loop actions: 
\begin{enumerate}
\item the one just mentioned, which takes fermion hopping into account up to order $\order{\kappa_{s}^{2}\,a_{f}^{n_{t}}\of{\beta}}$ (dashed black curve),
\item one that takes fermion hopping into account only at order $\order{\kappa_{s}^2}$ (dotted black curve),
\item and the one from Sec. \ref{ssec:mfmethod}, which neglects fermion hopping completely (dotted red curve).
\end{enumerate}
As can be seen, allowing for spatial fermion hopping in the effective theory shifts the value of $\mu$ where the average Polyakov loop is maximal closer to the corresponding value found in full LQCD. With increasing approximation order also the magnitude of the average Polyakov loop gets closer to the LQCD result. In the right-hand part of Fig. \ref{fig:polyakovloopfulleffcmp} we tried to compensate for the truncated spatial nearest-neighbor interaction in the effective theory by increasing the value of the inverse gauge coupling $\beta$. Setting $\beta=6.035$ in the effective theory, the average Polyakov loop obtained by mean-field with the $\order{\kappa_{s}^{2}\,a_{f}^{n_{t}}\of{\beta}}$ effective action from appendix \ref{sec:kappasqerms} then almost coincides with the corresponding LQCD result for $\beta=5.0$.


\begin{figure}[h]
\centering
\begin{minipage}[t]{0.485\linewidth}
\centering
\includegraphics[width=\linewidth]{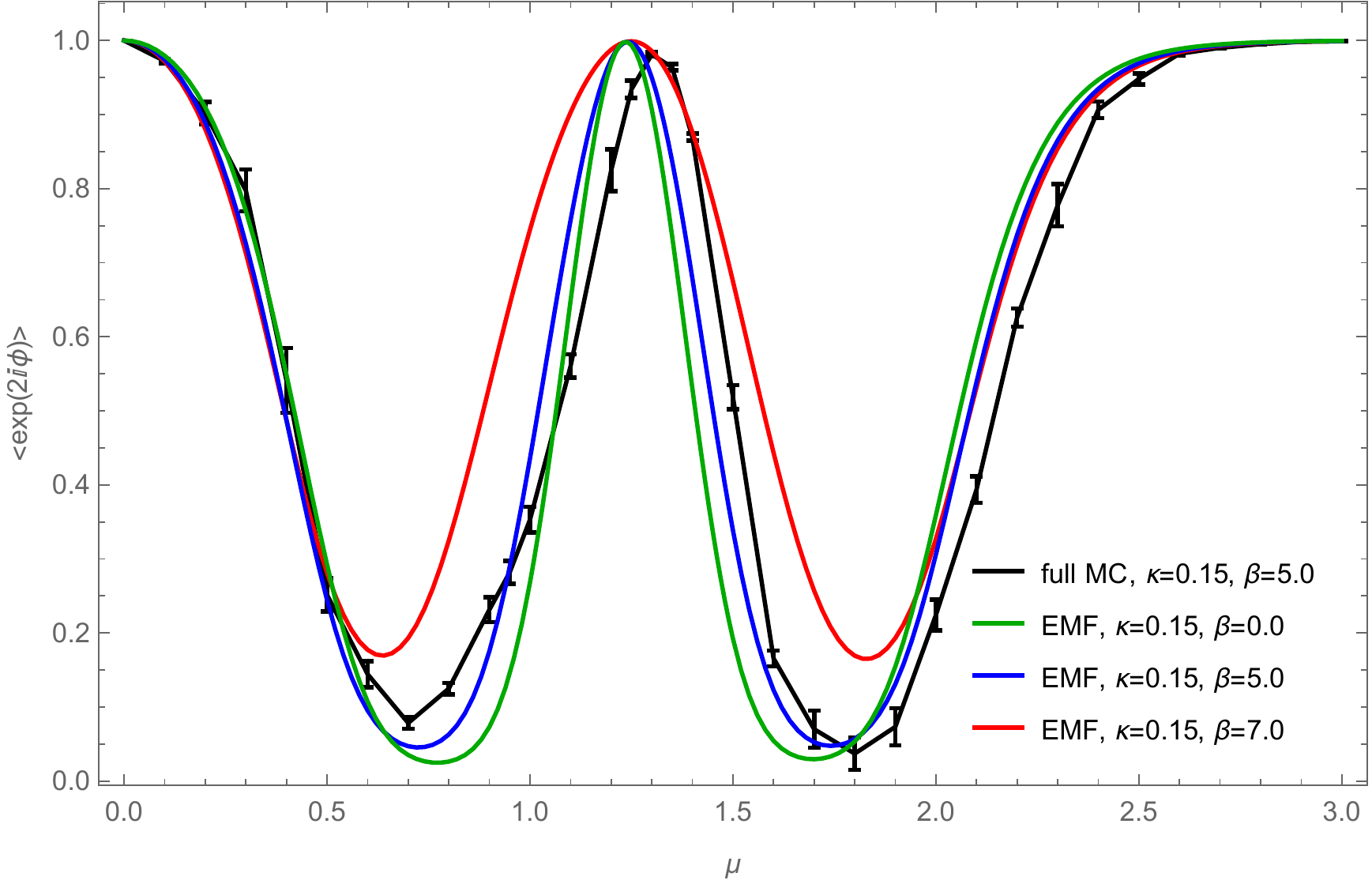}
\end{minipage}\hfill
\begin{minipage}[t]{0.485\linewidth}
\centering
\includegraphics[width=\linewidth]{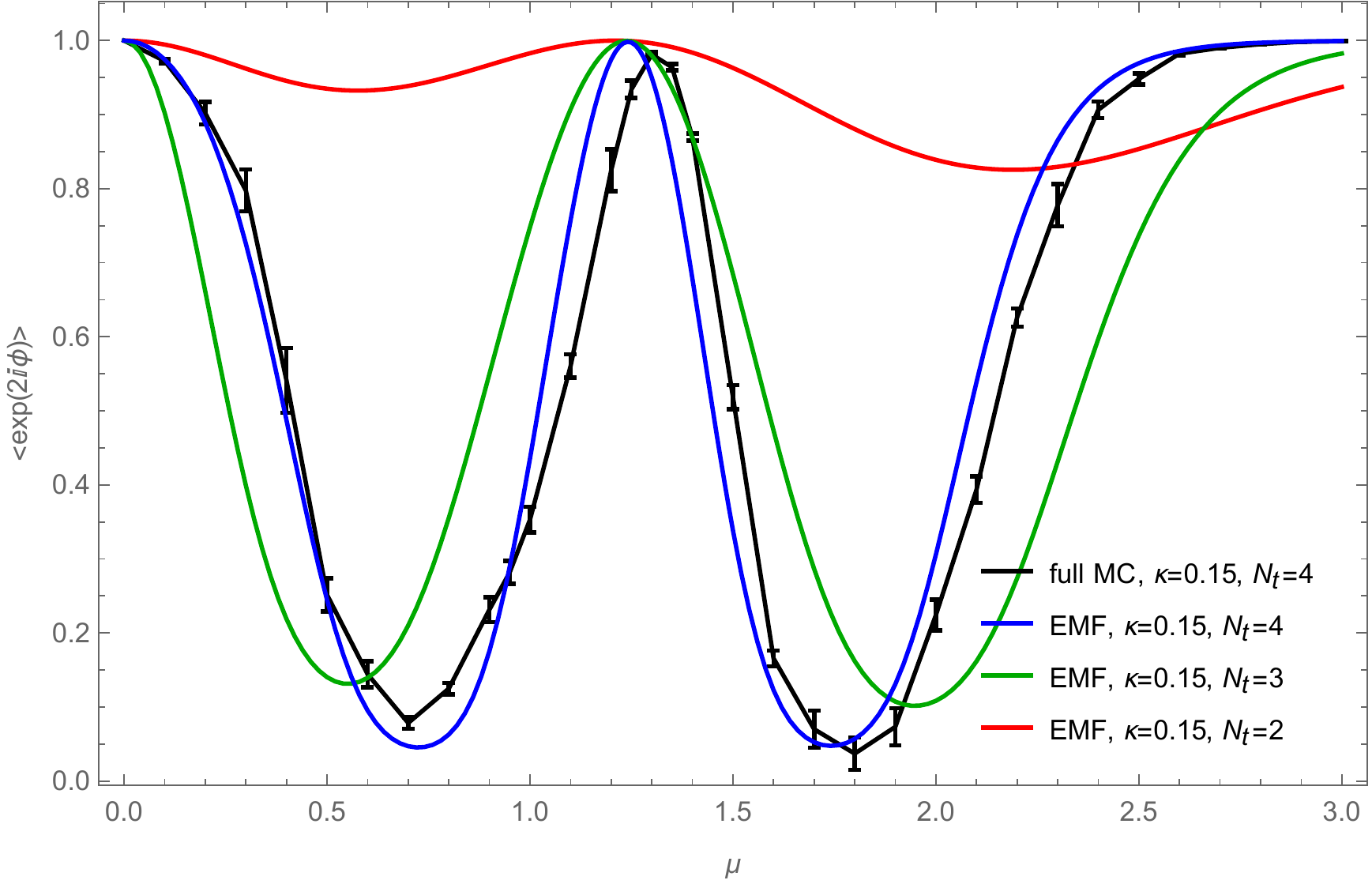}
\end{minipage}
\caption{Average sign as a function of $\mu$ for a two-flavor system of spatial volume $V=L^3=4^3$ and with $\kappa=0.15$. The solid line in both figures guides the eye through the data points obtained by a Monte Carlo simulation of full two-flavor LQCD at inverse gauge coupling $\beta=5.0$ and with $N_{t}=4$. The other curves were obtained by a mean-field calculation using our improved effective action and \eqref{eq:avsigneff}. The left-hand figure shows results for the effective model for three different values of $\beta$, with fixed $N_{t}=4$, while in the right-hand figure, the different curves correspond to different values of $N_{t}$ with $\beta=5.0$ fixed. The corrections up to order $\order{\kappa_{s}^{2}a_{f}^{n_{t}}\of{\beta}}$ in the effective action shift the half-filling point from the static quark value, $\mu\approx 1.2$, a bit closer towards the value $\mu\approx 1.3$, found in full LQCD.}
  \label{fig:effavsignvsmubeta}
\end{figure}

Finally, figure \ref{fig:effavsignvsmubeta} shows the average sign \eqref{eq:avsign} as a function of $\mu$, again for a mass-degenerate ($\kappa=5.0$) two-flavor system of spatial size $V=L^3=4^3$, computed by mean-field, using our improved effective action and \eqref{eq:avsigneff}. The left-hand part shows results for three different values of the inverse gauge coupling $\beta=0,\,0.5,\,0.7$ for fixed $N_{t}=4$, while in the right-hand figure, results for different values of $N_{t}=2,3,4$ are shown for $\beta=5.0$ fixed. For comparison, the figures also include a curve showing the situation in full LQCD with $\kappa=0.15$, $N_{t} L^3=4^4$ and $\beta=5.0$. The corrections up to order $\order{\kappa_{s}^{2}a_{f}^{n_{t}}\of{\beta}}$ in the effective action shift the half-filling point in the effective model from the static quark value, $\mu\approx 1.2$, a bit closer towards $\mu\approx 1.3$, the value found in full LQCD.\\

However, the value of the hopping parameter $\kappa=0.15$ used in Figs. \ref{fig:polyakovloopfulleffcmp} and \ref{fig:polyakovloopfulleffcmpnq} is already almost too large to be used in a calculation based on a $\order{\kappa_{s}^{2}a_{f}^{n_{t}}\of{\beta}}$ truncated nearest-neighbor interaction. For larger values of $\kappa$ or lower temperatures (i.e. larger $N_{t}$), the deviation of the mean-field result from the Monte Carlo data becomes larger and it is no longer possible to match the data by adjusting the inverse gauge coupling in the effective model as was done in the right-hand part of Fig. \ref{fig:polyakovloopfulleffcmp}. This is not surprising as, concerning the physics, the $\order{\kappa_{s}^{2}a_{f}^{n_{t}}\of{\beta}}$ truncation of the nearest-neighbor interaction term, $\det_{c,d,t}\sof{\id-\kappa_{s}^{2}\tilde{M}_{C_{2}}}$, implies that mesons and baryons (the low energy degrees of freedom of the theory) remain essentially static and cannot undergo spatial hoppings. But when lowering the temperature and/or the quark mass, it is precisely the hopping of these low energy degrees of freedom that becomes more important. From a more technical point of view, looking at the expansion of the fermionic nearest-neighbor interaction term,
\begin{align}
\det_{c,d,t}\sof{\id-\kappa_{s}^{2}\tilde{M}_{C_{2}}}\,=\,1\,&-\,\kappa_{s}^{2}\,\trace_{c,d,t}\sof{\tilde{M}_{C_{2}}}\nonumber\\
&+\,\frac{\kappa_{s}^{4}}{2}\,\of{\trace^{2}_{c,d,t}\sof{\tilde{M}_{C_{2}}}-\trace_{c,d,t}\sof{\tilde{M}^{2}_{C_{2}}}}\nonumber\\
&-\,\frac{\kappa_{s}^{6}}{6}\,\of{\trace^{3}_{c,d,t}\sof{\tilde{M}_{C_{2}}}-3\trace_{c,d,t}\sof{\tilde{M}_{C_{2}}}\trace_{c,d,t}\sof{\tilde{M}^{2}_{C_{2}}}+2\trace_{c,d,t}\sof{\tilde{M}^{3}_{C_{2}}}}\nonumber\\
&+\,\ldots\nonumber\\
&=\,\sum\limits_{k=0}^{12\,n_{t}}\kappa_{s}^{2\,k}\,c_{12\,n_{t}-k}\sof{\tilde{M}_{C_{2}}}\ ,\label{eq:nearestneighbordetexpansion}
\end{align}
where the $c_{k}\of{A}$ for a rank-$n$ matrix $A$ are defined by the characteristic polynomial of $A$,
\[
\chi_{A}\of{\lambda}\,=\,\det\of{\id\lambda-A}\,=\,\sum\limits_{k=0}^{n}\,\lambda^{k}\,c_{k}\sof{A}\ ,
\]
and can be obtained with the already mentioned Leverrier-Faddeev method, i.e. by starting with a matrix $B_{0}\of{A}=0$ and the constant $c_{n}\of{A}=1$, and then using the recursion
\[
B_{k}\of{A}\,=\,A\,B_{k-1}\of{A}\,+\,\id\,c_{n-k+1}\of{A}\quad,\quad c_{n-k}\of{A}\,=\,-\frac{1}{k}\,\trace\of{A\,B_{k}\of{A}}\quad,\quad k\,=\,1,\ldots,n\ ,
\]
it becomes clear that the $\kappa_{s}^{2}$ term does not necessarily give the dominant contribution to \eqref{eq:nearestneighbordetexpansion} as the $c_{k}\sof{\tilde{M}_{C_{2}}}$ can become rather big (especially if $n_{t}$ is large) and to guarantee that the $\kappa_{s}^{2}$-term dominates the expansion, one would have to choose $\kappa_{s}$ much smaller than we did.\\
To incorporate spatial meson and baryon hopping into the mean-field calculation, one should include at least the $\kappa_{s}^6$-terms in the expansion of \eqref{eq:nearestneighbordetexpansion}. Although this is probably feasible, already at order $\kappa_{s}^4$ the computation by hand of the corresponding terms in the effective action becomes a delicate issue and it would be desirable to automate the combinatorics required to get the correct dependency on $a_{f}\of{\beta}$.\\

\begin{figure}[h]
\centering
\begin{minipage}[t]{0.465\linewidth}
\centering
\includegraphics[width=\linewidth]{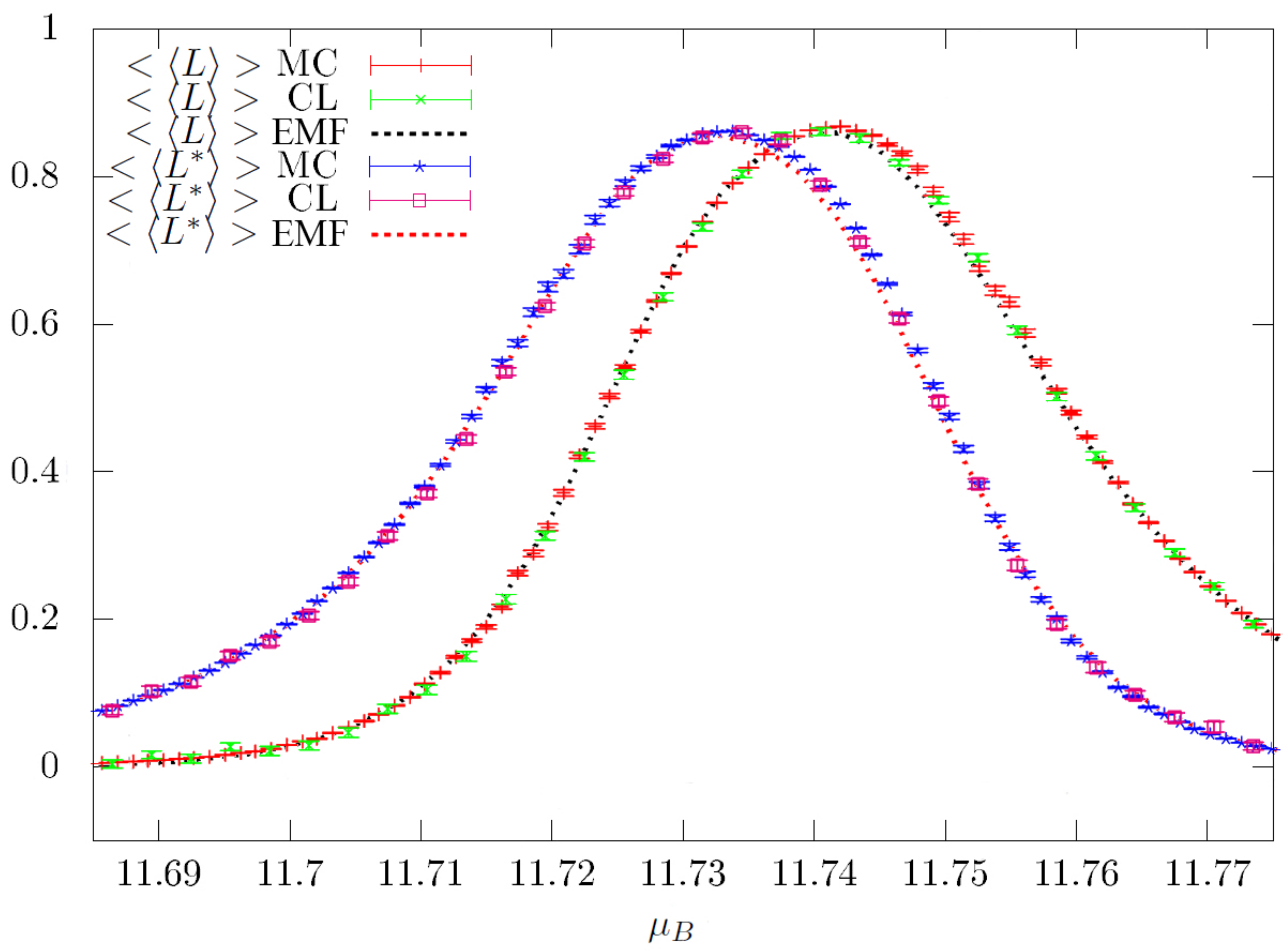}
\end{minipage}\hfill
\begin{minipage}[t]{0.52\linewidth}
\centering
\includegraphics[width=\linewidth]{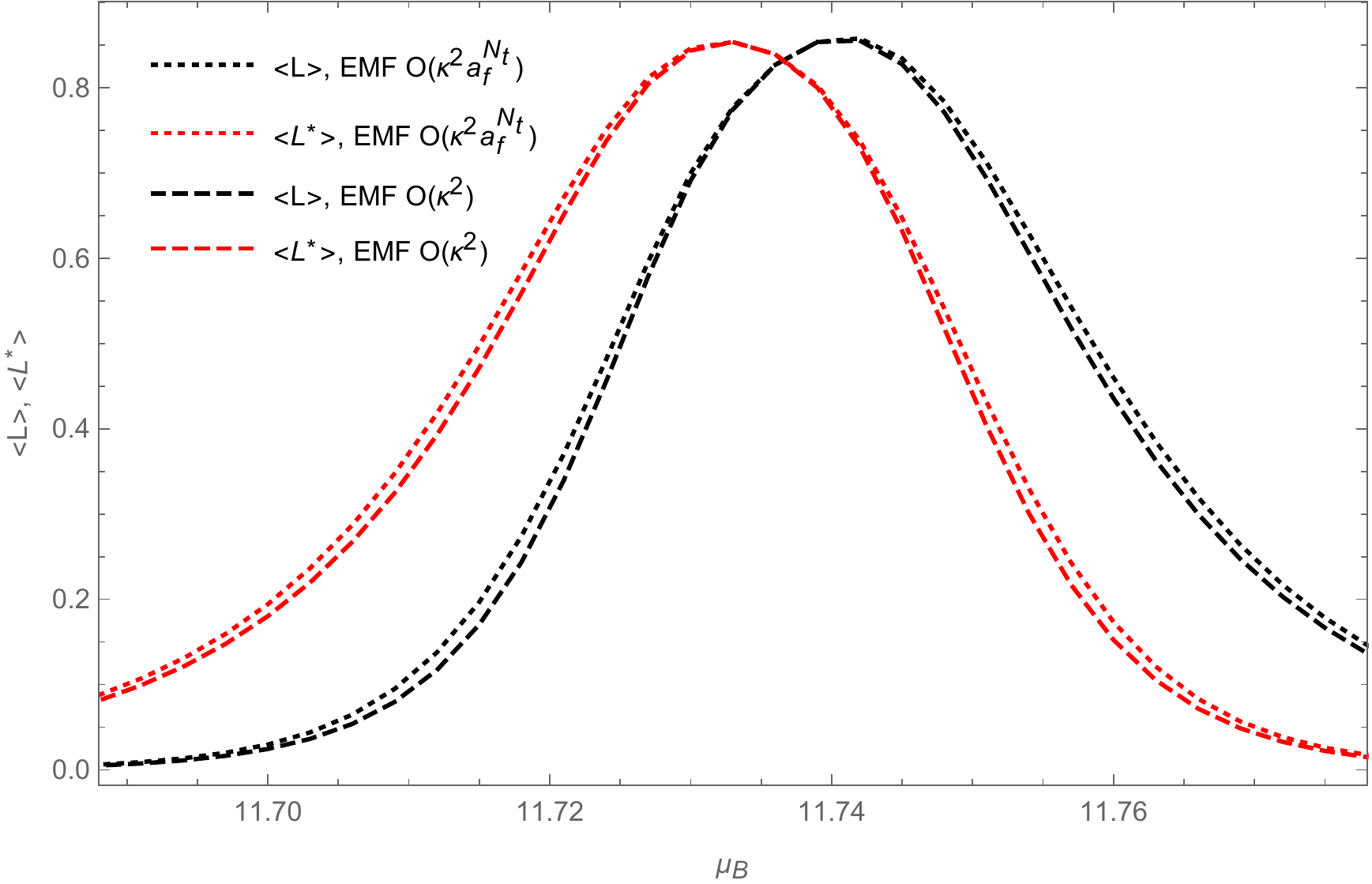}
\end{minipage}
\caption{The left-hand figure shows a comparison of \cite[Fig. 6]{Langelage}, showing the Polyakov loop $\avof{L}$ and inverse Polyakov loop $\avof{L^{*}}$ as a function of a baryon chemical potential $\mu_{B}=3\,\mu$ ($\mu$: quark chemical potential), obtained once by a Monte Carlo (MC) and once by a complex Langevin (CL) simulation of the effective model derived in \cite{Langelage} with $\kappa=0.01$, $N_{t}=200$ and $\beta=5.7$, and the corresponding result obtained with our mean-field method (dotted lines). As can be seen, our mean-field result coincides remarkably well with the Monte Carlo and complex Langevin data. The figure to the right shows again the curves that were used in the left-hand figure (dotted lines), but this time, together with the mean-field result obtained using the $\order{\kappa_{s}^2}$ truncated effective fermion action (dashed lines). As can be seen, even for a small value of $\kappa=0.01$ and large $N_{t}=200$, the higher order terms in the effective fermion action lead to a clearly observable difference in the result.}
  \label{fig:mfeffvsmcefflangelage}
\end{figure}

In order to demonstrate that the deviation of the mean-field result from the Monte Carlo data in Figs. \ref{fig:polyakovloopfulleffcmp} and \ref{fig:polyakovloopfulleffcmpnq} is merely due to the truncation of the nearest-neighbor interactions and not caused by our mean-field method itself, we try to reproduce figure 6 of \cite{Langelage}, which shows $\avof{L}$ and $\avof{L^{*}}$ for a single flavor system with $\kappa=0.01$, $N_{t}=200$ and $\beta=5.7$, obtained once by a Monte Carlo and once by a Complex Langevin simulation of the effective model derived in \cite{Langelage}. Our mean-field action is obtained by applying \eqref{eq:meanfiledaction} to the partition function \eqref{eq:effnnpartf}, which is of order $\kappa_{s}^2$ in $\kappa_{s}$, while for \cite[Fig. 6]{Langelage} also terms proportional to $\kappa_{s}^4$ have been taken into account. However, as $\kappa$ is rather small, we should nevertheless get a result comparable to \cite[Fig. 6]{Langelage} if our mean-field method works correctly. In Fig. \ref{fig:mfeffvsmcefflangelage}, left, we show \cite[Fig. 6]{Langelage} superimposed with the corresponding result from our mean-field calculation, where also for this single flavor system, we determined the mean-field value $\bar{L}$ within the phase-quenched system and then used reweighting to obtain $\avof{L}$ and $\avof{L^{*}}$ in the phase-unquenched case. As the figure shows, our mean-field results (dotted lines) match remarkably well the data obtained by full Monte Carlo and complex Langevin simulations. On the right hand side of Fig. \ref{fig:mfeffvsmcefflangelage}, we also compare the mean-field result obtained with the $\order{\kappa_{s}^{2}}$ truncated effective fermion action with the corresponding result obtained with the "full" ($\order{\kappa_{s}^{2}a_{f}^{N_{t}}\of{\beta}}$ truncated) effective fermion action \eqref{eq:effplactionfull}. As can be seen, even for the small value of $\kappa=0.01$ and large $N_{t}=200$, the higher order terms in the effective fermion action lead to a clearly observable difference in the result.

\section{Summary}
After a short introduction to the sign problem, we have shown that the half-filling point, i.e. the value of the chemical potential where the fermion density assumes half its maximal value, has similar properties in LQCD as in e.g. the fermionic Hubbard model used in solid states physics: the sign problem almost disappears at half-filling and the system possesses an apparent particle-hole symmetry. We traced the origin of these properties by analyzing the heavy-dense fermion determinant and the corresponding strong coupling partition function. However, in contrast to the Hubbard model, where the lattice and the half-filling state are physical, in LQCD the lattice is just a regulator that has to be removed in order to extract continuum physics and the state corresponding to half-filling of that lattice is therefore of no physical relevance.\\
As the half-filling state in LQCD and the corresponding maximum in the average sign are just lattice artifacts, it seems sensible that the first appearance of a minimum in the average sign (as a function of increasing chemical potential) indicates the point where lattice artifacts start to become dominant and measurements of observables taken at larger values of the chemical potential should therefore be considered with caution.\\
Finally we used a mean-field calculation to obtain in an analytic way the $\of{T,\mu}$ phase diagram for two-flavor LQCD (for both, isospin and quark chemical potential) in the static quark limit but at finite inverse gauge coupling $\beta$, where the $\beta$ dependency was implemented by an effective Polyakov loop action based on the leading terms of the character expanded gauge-field Boltzmann factor. We also showed how this calculation can be improved by taking into account spatial fermion hoppings.

\section{Acknowledgment}
We would like to thank J. Langelage for introducing us to the effective theory approach to finite density LQCD \cite{Langelage} and bringing the spatial hopping expansion to our attention. We would also like to thank C. Nonaka for many discussions at an early stage of this work, and A. Nakamura for enabling us to access the RCNP supercomputer, which we gratefully acknowledge for computing resources. Ph. de F. thanks KITP for hospitality while this paper was completed. This research was supported in part by the National Science Foundation under Grant No. NSF PHY11-25915.

\newpage
\setcounter{arabicPagenumber}{\value{page}} 
\pagenumbering{Roman}
\setcounter{page}{\theromanPagenumber}
\appendix
\section{Single Site Partition Function for Mass Degenerate QCD in the Strong Coupling Limit}\label{sec:isospinssitepartf}
For degenerate quark masses, i.e. $\kappa_{u}=\kappa_{d}=\kappa$, the strong coupling single site partition function \eqref{eq:qcd2fpartf} with an isospin chemical potential $\mu=\mu_{u}=-\mu_{d}$, reads:
\begin{multline}
Z_{s,q}\of{\mu,\kappa,n_{t}}\,=\\
\,\bof{\int\limits_{0}^{2\pi}\int\limits_{0}^{2\pi}\dd{\theta_{1}}\dd{\theta_{2}}\,H\of{\theta_{1},\theta_{2}}\Det{D\of{\theta_{1},\theta_{2};\mu,\kappa,n_{t}}}\Det{D\of{\theta_{1},\theta_{2};-\mu,\kappa,n_{t}}}}\\
=\,\bof{\int\limits_{0}^{2\pi}\int\limits_{0}^{2\pi}\dd{\theta_{1}}\dd{\theta_{2}}\,H\of{\theta_{1},\theta_{2}}\abs{\Det{D\of{\theta_{1},\theta_{2};\mu_{I},\kappa,n_{t}}}}^{2}}\\
\,=\,\bfof{\bigg.\of{h^{12}+1} \sof{\big.\bar{h}^{12}+4 \bar{h}^{10}+8 \bar{h}^9+10 \bar{h}^8+12 \bar{h}^7\\
+22 \bar{h}^6+12 \bar{h}^5+10 \bar{h}^4+8 \bar{h}^3+4 \bar{h}^2+1\big.}\\
+4\,\sof{h^{11}+h} \sof{\big.2 \bar{h}^{11}+3 \bar{h}^{10}+8 \bar{h}^9+15 \bar{h}^8+22 \bar{h}^7\\
+19 \bar{h}^6+22 \bar{h}^5+15 \bar{h}^4+8 \bar{h}^3+3 \bar{h}^2+2 \bar{h}\big.}\\
+2\,\sof{h^{10}+h^2} \sof{\big.2 \bar{h}^{12}+6 \bar{h}^{11}+26 \bar{h}^{10}+54 \bar{h}^9+102 \bar{h}^8+134\bar{h}^7\\
+163 \bar{h}^6+134 \bar{h}^5+102 \bar{h}^4+54 \bar{h}^3+26 \bar{h}^2+6 \bar{h}+2\big.} \\
+4\,\sof{h^9+h^3} \sof{\big.2 \bar{h}^{12}+8 \bar{h}^{11}+27 \bar{h}^{10}+72 \bar{h}^9+115 \bar{h}^8+172 \bar{h}^7\\
+193 \bar{h}^6+172 \bar{h}^5+115 \bar{h}^4+72 \bar{h}^3+27 \bar{h}^2+8 \bar{h}+2\big.} \\
 +\sof{h^8+h^4} \sof{\big.10 \bar{h}^{12}+60 \bar{h}^{11}+204 \bar{h}^{10}+460 \bar{h}^9+907 \bar{h}^8+1232\bar{h}^7\\
 +1398 \bar{h}^6+1232 \bar{h}^5+907 \bar{h}^4+460 \bar{h}^3+204 \bar{h}^2+60 \bar{h}+10\big.} \\
 +4 \sof{h^7+h^5} \sof{\big.3 \bar{h}^{12}+22 \bar{h}^{11}+67 \bar{h}^{10}+172 \bar{h}^9+308 \bar{h}^8+450 \bar{h}^7\\
 +487 \bar{h}^6+450 \bar{h}^5+308 \bar{h}^4+172 \bar{h}^3+67 \bar{h}^2+22 \bar{h}+3\big.} \\
 +2\,h^6\,\sof{\big.11 \bar{h}^{12}+38 \bar{h}^{11}+163 \bar{h}^{10}+386 \bar{h}^9+699 \bar{h}^8+974 \bar{h}^7\\
 +1161 \bar{h}^6+974 \bar{h}^5+699 \bar{h}^4+386 \bar{h}^3+163 \bar{h}^2+38 \bar{h}+11\big.}\bigg.}_{{}^{\scriptstyle h=\ssof{2\kappa\e^{\mu}}^{n_{t}}}_{\scriptstyle\bar{h}=\ssof{2\kappa\e^{-\mu}}^{n_{t}}}}\ .
\end{multline}
The corresponding partition function with a quark chemical potential reads:
\begin{multline}
Z_{s}\of{\mu,\kappa,n_{t}}\,=\\
\,\bof{\int\limits_{0}^{2\pi}\int\limits_{0}^{2\pi}\dd{\theta_{1}}\dd{\theta_{2}}\,H\of{\theta_{1},\theta_{2}}\Det{D\of{\theta_{1},\theta_{2};\mu,\kappa,n_{t}}}\Det{D\of{\theta_{1},\theta_{2};\mu,\kappa,n_{t}}}}\\
=\,\bfof{\bigg.h^{12} \sof{\bar{h}^{12}+20 \bar{h}^9+50 \bar{h}^6+20 \bar{h}^3+1} \\
 +h^{11} \sof{16 \bar{h}^{11}+180 \bar{h}^8+240 \bar{h}^5+40 \bar{h}^2} \\
 +h^{10} \sof{136 \bar{h}^{10}+816 \bar{h}^7+570 \bar{h}^4+40 \bar{h}} \\
 +h^9 \sof{20 \bar{h}^{12}+816 \bar{h}^9+2320 \bar{h}^6+800 \bar{h}^3+20} \\
 +h^8 \sof{180 \bar{h}^{11}+2651 \bar{h}^8+3720 \bar{h}^5+570 \bar{h}^2} \\
 +h^7 \sof{816 \bar{h}^{10}+5312 \bar{h}^7+3720 \bar{h}^4+240 \bar{h}} \\
 +h^6 \sof{50 \bar{h}^{12}+2320 \bar{h}^9+6832 \bar{h}^6+2320 \bar{h}^3+50} \\
 +h^5 \sof{240 \bar{h}^{11}+3720 \bar{h}^8+5312 \bar{h}^5+816 \bar{h}^2} \\
 +h^4 \sof{570 \bar{h}^{10}+3720 \bar{h}^7+2651 \bar{h}^4+180 \bar{h}} \\
 +h^3 \sof{20 \bar{h}^{12}+800 \bar{h}^9+2320 \bar{h}^6+816 \bar{h}^3+20} \\
 +h^2 \sof{40 \bar{h}^{11}+570 \bar{h}^8+816 \bar{h}^5+136 \bar{h}^2} \\
 +h \sof{40 \bar{h}^{10}+240 \bar{h}^7+180 \bar{h}^4+16 \bar{h}} \\
 +\sof{\bar{h}^{12}+20 \bar{h}^9+50 \bar{h}^6+20 \bar{h}^3+1}\bigg.}_{{}^{\scriptstyle h=\ssof{2\kappa\e^{\mu}}^{n_{t}}}_{\scriptstyle\bar{h}=\ssof{2\kappa\e^{-\mu}}^{n_{t}}}}
\end{multline}

\section{Higher Order Corrections to the Effective Polyakov Loop Action}\label{sec:kappasqerms}
In this section we derive an effective nearest-neighbor Polyakov loop action that includes not just the nearest-neighbor effects coming from the gauge field action, but also those coming from the fermion determinant.\\

We start with the single flavor partition function
\[
Z\,=\,\int\DD{U}\,\Det{D}\,\e^{-S_{g}}\ ,\label{eq:sfpartf}
\]
which, using the spatial loop expansion \eqref{eq:spathoppingexpand} for the fermion determinant and the character expansion \eqref{eq:strongcouplingexp} for the gauge field Boltzmann factor, can be written as:
\[
Z\,=\,\int\DD{U}\,\Det{\id-\kappa\,T}\,\prod\limits_{s_{0}}\prod\limits_{\cof{C_{s_{0}}}}\,\det_{c,d,t}\sof{\id-\kappa_{s}^{s_{0}}\tilde{M}_{C_{s_{0}}}}\,\prod\limits_{p}\bcof{1\,+\,\sum\limits_{r\neq 0}\,d_{r}\,a_{r}\of{\beta}\,\chi_{r}\ssof{U_{p}}}\ .\label{eq:sfpartfexp}
\]
We are only interested in the nearest-neighbor interactions and restrict the loop lengths $s_{0}$ in \eqref{eq:sfpartfexp} to $s_{0}\leq 2$. The integrand can then be written as
\[
\Det{\id-\kappa T}\,\prod\limits_{\avof{\bar{x},\bar{y}}}\det_{c,d,t}\sof{\id-\kappa_{s}^{2}\tilde{M}_{\bar{x},\bar{y}}}\,\prod\limits_{p}\bcof{1\,+\,\sum\limits_{r\neq 0}\,d_{r}\,a_{r}\of{\beta}\,\chi_{r}\ssof{U_{p}}}\ ,\label{eq:sfpartfexpnn}
\]
where $\bar{x}$, $\bar{y}$ are nearest-neighboring sites spanning a path $C_{2}$. Equation \eqref{eq:sfpartfexpnn} would capture the full spectrum of nearest-neighbor interactions but is unfortunately still rather complicated. We simplify the expression further by expanding the determinants $\det_{c,d,t}\sof{\id-\kappa_{s}^{2}\tilde{M}_{\bar{x},\bar{y}}}$ to order $\kappa_{s}^{2}$ and by considering the gauge field only in the fundamental representation and only along temporal plaquettes $p_t$. With these simplifications, the two products in \eqref{eq:sfpartfexpnn} turn into the following expression:
\[
\prod\limits_{\avof{\bar{x},\bar{y}}}\bcof{1\,-\,\kappa_{s}^{2}\,\trace_{c,d,t}\sof{\tilde{M}_{\bar{x},\bar{y}}}}\,\prod\limits_{p_t}\bcof{1\,+\,d_{f}\,a_{f}\of{\beta}\,\trace_{c}\ssof{U_{p_t}}}\ .\label{eq:nneffpartf}
\]
The Boltzmann factor for the desired effective action should then coincide with \eqref{eq:nneffpartf} up to order $\order{\kappa_{s}^{2}\,a^{n_{t}}_{f}\of{\beta}}$ (the single site/non-interaction terms coming from the factor $\Det{\id-\kappa T}$ in \eqref{eq:sfpartfexp} are not considered as part of the effective action).\\

\begin{figure}[h]
\centering
\begin{minipage}[t]{0.25\linewidth}
\centering
\begin{tikzpicture}[scale=0.75]
  \begin{scope}[nodes={inner sep=0cm}]
  \node at (-0.5cm,-2cm) (s1) {};
  \node[below=1mm of s1] (s1l) {$\bar{x}$};
  \node at (-0.5cm,-0.05cm) (i1a) {};
  \node at (-0.5cm,0.05cm) (i1b) {}; 
  \node[below left=0mm and 1mm of i1b] (i1al) {$x_{4}$};
  \node at (-0.5cm,2cm) (e1) {};
  \node at (0.5cm,-2cm) (s2) {};
  \node[below=1mm of s2] (s2l) {$\bar{y}$};
  \node at (0.5cm,-0.05cm) (i2a) {};
  \node at (0.5cm,0.05cm) (i2b) {};
  \node at (0.5cm,2cm) (e2) {};
  \draw[->,color=blue] (s1) edge (i1a) (i1a) edge (i2b) (i2b) edge (e2);
  \draw[->,color=blue] (s2) edge (i2a) (i2a) edge (i1b) (i1b) edge (e1);
  \end{scope}
  
  \begin{scope}[nodes={inner sep=0cm}]
  \node at (1.5cm,-2cm) (s1) {};
  \node[below=1mm of s1] (s1l) {$\bar{x}$};
  \node at (1.5cm,2cm) (e1) {};
  \node at (2.5cm,-2cm) (s2) {};
  \node[below=1mm of s2] (s2l) {$\bar{y}$};
  \node at (2.5cm,2cm) (e2) {};
  \draw[->,color=blue] (s1) edge (e1);
  \draw[->,color=red] (s2) edge (e2);
  \node at (0.75cm,0cm) (c1) {};
  \node at (1.25cm,0cm) (c2) {};
  \draw[->] (c1) edge (c2);
  \end{scope}
\end{tikzpicture}
\end{minipage}\hfill
\begin{minipage}[t]{0.25\linewidth}
\centering
\begin{tikzpicture}[scale=0.75]
  \begin{scope}[nodes={inner sep=0cm}]
  \node at (-0.5cm,-2cm) (s1) {};
  \node[below=1mm of s1] (s1l) {$\bar{x}$};
  \node at (-0.5cm,-0.05cm) (i1a) {};
  \node at (-0.5cm,0.05cm) (i1b) {}; 
  \node[below left=0mm and 1mm of i1b] (i1al) {$x_{4}$};
  \node at (-0.5cm,2cm) (e1) {};
  \node at (0.5cm,-2cm) (s2) {};
  \node[below=1mm of s2] (s2l) {$\bar{y}$};
  \node at (0.5cm,-0.05cm) (i2a) {};
  \node at (0.5cm,0.05cm) (i2b) {};
  \node at (0.5cm,2cm) (e2) {};
  \draw[<-,color=blue] (s1) edge (i1a) (i1a) edge (i2a) (i2a) edge (s2);
  \draw[<-,color=blue] (e2) edge (i2b) (i2b) edge (i1b) (i1b) edge (e1);
  \end{scope}
  
  \begin{scope}[nodes={inner sep=0cm}]
  \node at (1.5cm,-2cm) (s1) {};
  \node[below=1mm of s1] (s1l) {$\bar{x}$};
  \node at (1.5cm,2cm) (e1) {};
  \node at (2.5cm,-2cm) (s2) {};
  \node[below=1mm of s2] (s2l) {$\bar{y}$};
  \node at (2.5cm,2cm) (e2) {};
  \draw[<-,color=blue] (s1) edge (e1);
  \draw[->,color=red] (s2) edge (e2);
  \node at (0.75cm,0cm) (c1) {};
  \node at (1.25cm,0cm) (c2) {};
  \draw[->] (c1) edge (c2);
  \end{scope}
\end{tikzpicture}
\end{minipage}\hfill
\begin{minipage}[t]{0.25\linewidth}
\centering
\begin{tikzpicture}[scale=0.75]
  \begin{scope}[nodes={inner sep=0cm}]
  \node at (-0.5cm,-2cm) (s1) {};
  \node[below=1mm of s1] (s1l) {$\bar{x}$};
  \node at (-0.5cm,-0.05cm) (i1a) {};
  \node at (-0.5cm,0.05cm) (i1b) {}; 
  \node[below left=0mm and 1mm of i1b] (i1al) {$x_{4}$};
  \node at (-0.5cm,2cm) (e1) {};
  \node at (0.5cm,-2cm) (s2) {};
  \node[below=1mm of s2] (s2l) {$\bar{y}$};
  \node at (0.5cm,-0.05cm) (i2a) {};
  \node at (0.5cm,0.05cm) (i2b) {};
  \node at (0.5cm,2cm) (e2) {};
  \draw[->,color=blue] (s1) edge (i1a) (i1a) edge (i2a) (i2a) edge (s2);
  \draw[->,color=blue] (e2) edge (i2b) (i2b) edge (i1b) (i1b) edge (e1);
  \end{scope}
  
  \begin{scope}[nodes={inner sep=0cm}]
  \node at (1.5cm,-2cm) (s1) {};
  \node[below=1mm of s1] (s1l) {$\bar{x}$};
  \node at (1.5cm,2cm) (e1) {};
  \node at (2.5cm,-2cm) (s2) {};
  \node[below=1mm of s2] (s2l) {$\bar{y}$};
  \node at (2.5cm,2cm) (e2) {};
  \draw[->,color=blue] (s1) edge (e1);
  \draw[<-,color=red] (s2) edge (e2);
  \node at (0.75cm,0cm) (c1) {};
  \node at (1.25cm,0cm) (c2) {};
  \draw[->] (c1) edge (c2);
  \end{scope}
\end{tikzpicture}
\end{minipage}\hfill
\begin{minipage}[t]{0.25\linewidth}
\centering
\begin{tikzpicture}[scale=0.75]
  \begin{scope}[nodes={inner sep=0cm}]
  \node at (-0.5cm,-2cm) (s1) {};
  \node[below=1mm of s1] (s1l) {$\bar{x}$};
  \node at (-0.5cm,-0.05cm) (i1a) {};
  \node at (-0.5cm,0.05cm) (i1b) {}; 
  \node[below left=0mm and 1mm of i1b] (i1al) {$x_{4}$};
  \node at (-0.5cm,2cm) (e1) {};
  \node at (0.5cm,-2cm) (s2) {};
  \node[below=1mm of s2] (s2l) {$\bar{y}$};
  \node at (0.5cm,-0.05cm) (i2a) {};
  \node at (0.5cm,0.05cm) (i2b) {};
  \node at (0.5cm,2cm) (e2) {};
  \draw[<-,color=blue] (s1) edge (i1a) (i1a) edge (i2b) (i2b) edge (e2);
  \draw[<-,color=blue] (s2) edge (i2a) (i2a) edge (i1b) (i1b) edge (e1);
  \end{scope}

  \begin{scope}[nodes={inner sep=0cm}]
  \node at (1.5cm,-2cm) (s1) {};
  \node[below=1mm of s1] (s1l) {$\bar{x}$};
  \node at (1.5cm,2cm) (e1) {};
  \node at (2.5cm,-2cm) (s2) {};
  \node[below=1mm of s2] (s2l) {$\bar{y}$};
  \node at (2.5cm,2cm) (e2) {};
  \draw[<-,color=blue] (s1) edge (e1);
  \draw[<-,color=red] (s2) edge (e2);
  \node at (0.75cm,0cm) (c1) {};
  \node at (1.25cm,0cm) (c2) {};
  \draw[->] (c1) edge (c2);
  \end{scope}
\end{tikzpicture}
\end{minipage}
\caption{The figure shows the four types of diagrams corresponding to the four terms in \eqref{eq:sqtermsum} that give non-vanishing contributions after integrating out the spatial links in the fundamental representation (left-hand sides of the little black arrows), together with the corresponding configuration after the spatial links have been integrated out (righ-hand sides of the little black arrows).}
\label{fig:ksqb0}
\end{figure}
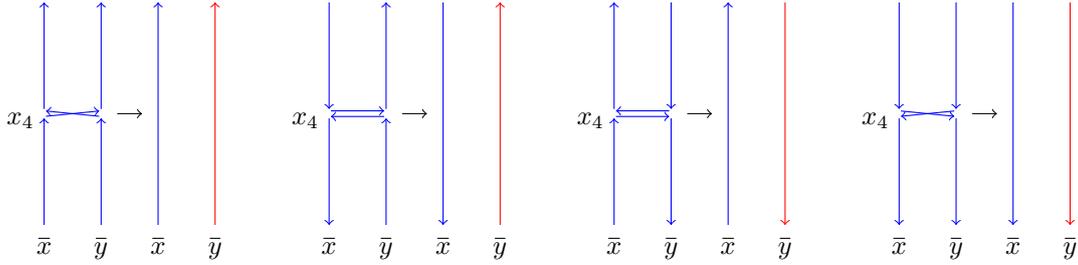

To find such an action, we start by writing out $\trace_{c,d,t}\sof{\tilde{M}_{\bar{x},\bar{y}}}$ with $\bar{y}=\bar{x}+\hat{i}$ and $i\in\cof{1,2,3}$:
\begin{multline}
\trace_{c,d,t}\sof{\tilde{M}_{\bar{x},\bar{y}}}\,=\,\trace_{c,d,t}\sof{S_{\bar{x},\bar{y}}\,\of{\id-\kappa\,T}^{-1}_{\bar{y}}\,S_{\bar{y},\bar{x}}\,\of{\id-\kappa\,T}^{-1}_{\bar{x}}}\\
=\,\sum\limits_{x_{4},y_{4}=0}^{n_{t}-1}\trace_{c,d}\bcof{\bigg.\of{\id-\gamma_{i}}\otimes U_{i}\of{\bar{x},x_{4}}\\
\frac{1}{2}\bof{\bigg.\of{2\,\kappa\e^{\mu}}^{\umod\of{y_{4}-x_{4},n_{t}}}\of{-1}^{\theta\of{x_{4}>y_{4}}}\,\of{\id-\gamma_{4}}\otimes\sof{P_{\bar{y}}\of{x_{4},y_{4}}\,\sof{\id+\of{2\,\kappa\e^{\mu}}^{n_{t}}P_{\bar{y}}\of{y_{4}}}^{-1}}\\
+\,\bigg.\of{2\,\kappa\e^{-\mu}}^{\umod\of{x_{4}-y_{4},n_{t}}}\of{-1}^{\theta\of{y_{4}>x_{4}}}\,\of{\id+\gamma_{4}}\otimes\sof{P^{\dagger}_{\bar{y}}\of{y_{4},x_{4}}\,\sof{\id+\of{2\,\kappa\e^{-\mu}}^{n_{t}}P^{\dagger}_{\bar{y}}\of{y_{4}}}^{-1}}}\\
\cdot\of{\id+\gamma_{i}}\otimes U^{\dagger}_{i}\of{\bar{x},y_{4}}\\
\frac{1}{2}\bof{\bigg.\of{2\,\kappa\e^{\mu}}^{\umod\of{x_{4}-y_{4},n_{t}}}\of{-1}^{\theta\of{y_{4}>x_{4}}}\,\of{\id-\gamma_{4}}\otimes\sof{P_{\bar{x}}\of{y_{4},x_{4}}\,\sof{\id+\of{2\,\kappa\e^{\mu}}^{n_{t}}P_{\bar{x}}\of{x_{4}}}^{-1}}\\
+\,\bigg.\bigg.\of{2\,\kappa\e^{-\mu}}^{\umod\of{y_{4}-x_{4},n_{t}}}\of{-1}^{\theta\of{x_{4}>y_{4}}}\,\of{\id+\gamma_{4}}\otimes\sof{P^{\dagger}_{\bar{x}}\of{x_{4},y_{4}}\,\sof{\id+\of{2\,\kappa\e^{-\mu}}^{n_{t}}P^{\dagger}_{\bar{x}}\of{x_{4}}}^{-1}}}}\ .
\end{multline}
After carrying out the trace in Dirac space this yields:
\begin{multline}
\trace_{c,d,t}\sof{\tilde{M}_{\bar{x},\bar{y}}}\,=\,2\,\sum\limits_{x_{4},y_{4}=0}^{n_{t}-1}\,\bof{\of{2\,\kappa\e^{\mu}}^{\umod\of{y_{4}-x_{4},n_{t}}}\of{2\,\kappa\e^{\mu}}^{\umod\of{x_{4}-y_{4},n_{t}}}\of{-1}^{\theta\of{x_{4}>y_{4}}+\theta\of{y_{4}>x_{4}}}\bigg.\\
\cdot\trace_{c}\of{\right.U_{i}\of{\bar{x},x_{4}}\,P_{\bar{y}}\of{x_{4},y_{4}}\,\sof{\id+\of{2\,\kappa\e^{\mu}}^{n_{t}}P_{\bar{y}}\of{y_{4}}}^{-1}\\
\quad\quad\cdot U_{i}^{\dagger}\of{\bar{x},y_{4}}\,P_{\bar{x}}\of{y_{4},x_{4}}\,\sof{\id+\of{2\,\kappa\e^{\mu}}^{n_{t}}P_{\bar{x}}\of{x_{4}}}^{-1}\left.}\\
-\of{2\,\kappa\e^{\mu}}^{\umod\of{y_{4}-x_{4},n_{t}}}\of{2\,\kappa\e^{-\mu}}^{\umod\of{y_{4}-x_{4},n_{t}}}\of{-1}^{\theta\of{x_{4}>y_{4}}+\theta\of{y_{4}<x_{4}}}\bigg.\\
\quad\cdot\trace_{c}\of{\right.U_{i}\of{\bar{x},x_{4}}\,P_{\bar{y}}\of{x_{4},y_{4}}\,\sof{\id+\of{2\,\kappa\e^{\mu}}^{n_{t}}P_{\bar{y}}\of{y_{4}}}^{-1}\\
\quad\quad\cdot U_{i}^{\dagger}\of{\bar{x},y_{4}}\,P^{\dagger}_{\bar{x}}\of{x_{4},y_{4}}\,\sof{\id+\of{2\,\kappa\e^{-\mu}}^{n_{t}}P^{\dagger}_{\bar{x}}\of{x_{4}}}^{-1}\left.}\\
-\of{2\,\kappa\e^{-\mu}}^{\umod\of{x_{4}-y_{4},n_{t}}}\of{2\,\kappa\e^{\mu}}^{\umod\of{x_{4}-y_{4},n_{t}}}\of{-1}^{\theta\of{y_{4}>x_{4}}+\theta\of{x_{4}<y_{4}}}\bigg.\\
\quad\cdot\trace_{c}\of{\right.U_{i}\of{\bar{x},x_{4}}\,P^{\dagger}_{\bar{y}}\of{y_{4},x_{4}}\,\sof{\id+\of{2\,\kappa\e^{-\mu}}^{n_{t}}P^{\dagger}_{\bar{y}}\of{y_{4}}}^{-1}\\
\quad\quad\cdot U_{i}^{\dagger}\of{\bar{x},y_{4}}\,P_{\bar{x}}\of{y_{4},x_{4}}\,\sof{\id+\of{2\,\kappa\e^{\mu}}^{n_{t}}P_{\bar{x}}\of{x_{4}}}^{-1}\left.}\\
+\of{2\,\kappa\e^{-\mu}}^{\umod\of{x_{4}-y_{4},n_{t}}}\of{2\,\kappa\e^{-\mu}}^{\umod\of{y_{4}-x_{4},n_{t}}}\of{-1}^{\theta\of{y_{4}>x_{4}}+\theta\of{x_{4}>y_{4}}}\bigg.\\
\quad\cdot\trace_{c}\of{\right.U_{i}\of{\bar{x},x_{4}}\,P^{\dagger}_{\bar{y}}\of{y_{4},x_{4}}\,\sof{\id+\of{2\,\kappa\e^{-\mu}}^{n_{t}}P^{\dagger}_{\bar{y}}\of{y_{4}}}^{-1}\\
\quad\quad\cdot U_{i}^{\dagger}\of{\bar{x},y_{4}}\,P^{\dagger}_{\bar{x}}\of{x_{4},y_{4}}\,\sof{\id+\of{2\,\kappa\e^{-\mu}}^{n_{t}}P^{\dagger}_{\bar{x}}\of{x_{4}}}^{-1}\left.}\bigg.}\ .\label{eq:sqtermsum}
\end{multline}
Integrating over the spatial links in \eqref{eq:sqtermsum} at order $\order{a_{f}^{0}\of{\beta}}$ leads to the constraint $y_{4}=x_{4}$ (see Fig. \ref{fig:ksqb0}) as for $\SU{3}$ integrals, we have:
\[
\int\idd{U}{}\, U_{a,b} = 0 \qquad , \qquad \int\idd{U}{}\, U_{a,b}\,U_{c,d}^{\dagger} = \frac{1}{3}\,\delta_{a,d}\,\delta_{b,c}\ .\label{eq:su3int}
\]
According to the definition of $P_{\bar{x}}\of{x_{4},y_{y}}$ given below eq. \eqref{eq:hdquarkproppos}, we then have $P_{\bar{x}}\of{x_{4},x_{4}}=\id$ and find therefore:
\begin{multline}
\trace_{c,d,t}\sof{\tilde{M}_{\bar{x},\bar{y}}}\,=\,\frac{2\,n_{t}}{3}\of{\trace_{c}\sof{\sof{\id+\of{2\,\kappa\e^{\mu}}^{n_{t}}P_{\bar{x}}}^{-1}}-\trace_{c}\sof{\sof{\id+\of{2\,\kappa\e^{-\mu}}^{n_{t}}P^{\dagger}_{\bar{x}}}^{-1}}}\\
\of{\trace_{c}\sof{\sof{\id+\of{2\,\kappa\e^{\mu}}^{n_{t}}P_{\bar{y}}}^{-1}}-\trace_{c}\sof{\sof{\id+\of{2\,\kappa\e^{-\mu}}^{n_{t}}P^{\dagger}_{\bar{y}}}^{-1}}}\,+\,\order{a_{f}\of{\beta}}\ ,\label{eq:effpksqb0}
\end{multline}
which is the same as eq. (2.28) of \cite{Langelage}, as can be seen by noting that:
\[
\trace_{c}\sof{\of{2\,\kappa\e^{\mu}}^{n_{t}}P_{\bar{x}}\sof{\id+\of{2\,\kappa\e^{\mu}}^{n_{t}}P_{\bar{x}}}^{-1}}=3-\trace_{c}\sof{\sof{\id+\of{2\,\kappa\e^{\mu}}^{n_{t}}P_{\bar{x}}}^{-1}}\ ,\label{eq:effpl1}
\]
and
\[
\trace_{c}\sof{\of{2\,\kappa\e^{-\mu}}^{n_{t}}P^{\dagger}_{\bar{x}}\sof{\id+\of{2\,\kappa\e^{-\mu}}^{n_{t}}P^{\dagger}_{\bar{x}}}^{-1}}=3-\trace_{c}\sof{\sof{\id+\of{2\,\kappa\e^{-\mu}}^{n_{t}}P^{\dagger}_{\bar{x}}}^{-1}}\ .\label{eq:effpl2}
\]

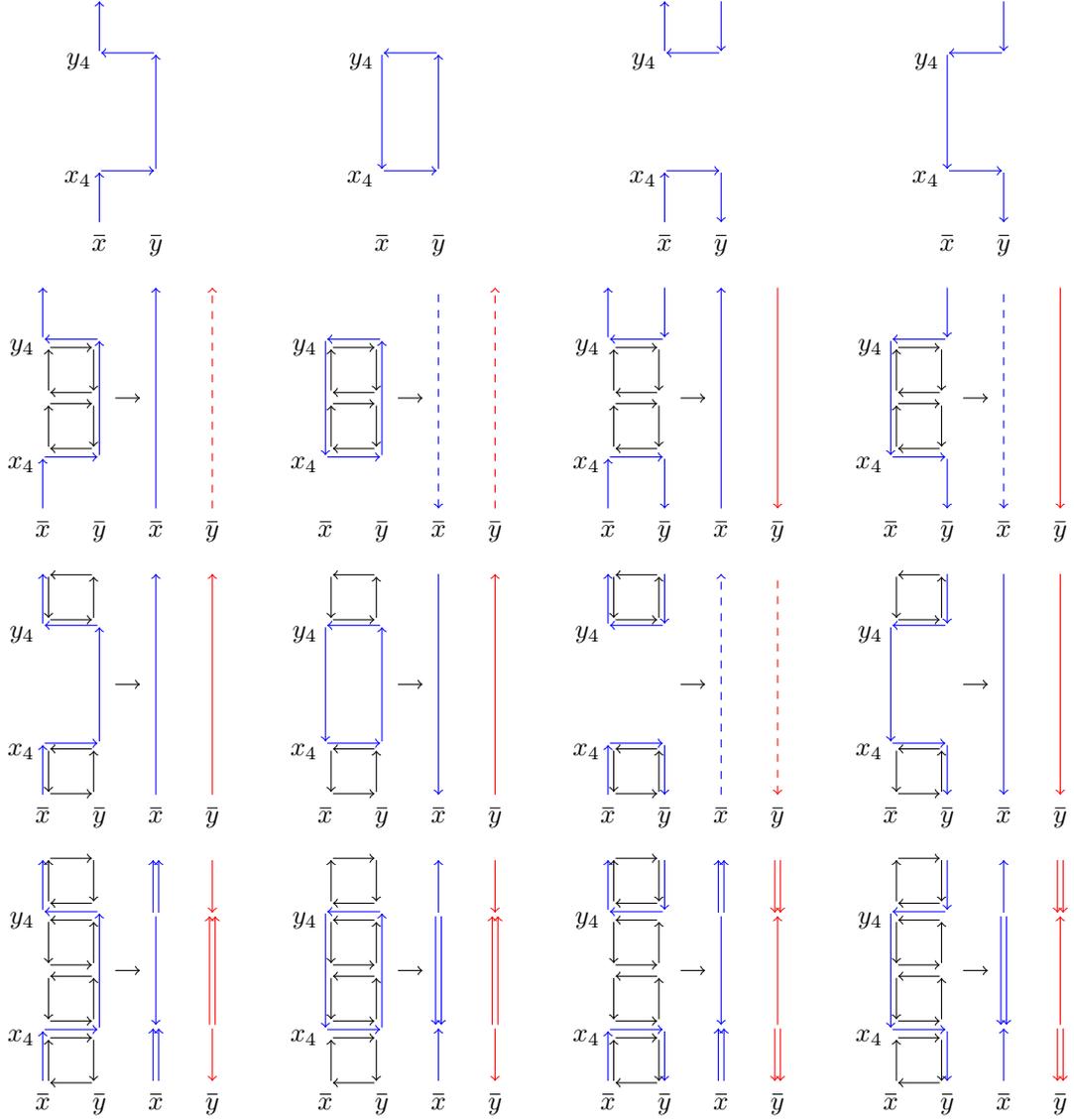
\begin{figure}[h!]
\centering
\begin{minipage}[t]{0.25\linewidth}
\centering
\begin{tikzpicture}[scale=0.75,nodes={inner sep=0}]
  \pgfpointtransformed{\pgfpointxy{1}{1}};
  \pgfgetlastxy{\vx}{\vy}
  \begin{scope}[node distance=\vx and \vy]
  \node at (-0.5,-2) (s1) {};
  \node[below=0.1 of s1] (s1l) {$\bar{x}$};
  \node at (-0.5,-1.05) (i1a) {};
  \node at (-0.5,1.05) (i1b) {}; 
  \node[below left=0 and 0.1 of i1a.center] (i1al) {$x_{4}$};
  \node[below left=0 and 0.1 of i1b.center] (i1bl) {$y_{4}$};
  \node at (-0.5,2) (e1) {};
  \node at (0.5,-2) (s2) {};
  \node[below=0.1 of s2] (s2l) {$\bar{y}$};
  \node at (0.5,-1.05) (i2a) {};
  \node at (0.5,1.05) (i2b) {};
  \node at (0.5,2) (e2) {};
  \draw[->,color=blue] (s1) edge (i1a) (i1a) edge (i2a) (i2a) edge (i2b) (i2b) edge (i1b) (i1b) edge (e1);
  \end{scope}
\end{tikzpicture}
\end{minipage}\hfill
\begin{minipage}[t]{0.25\linewidth}
\centering
\begin{tikzpicture}[scale=0.75,nodes={inner sep=0}]
  \pgfpointtransformed{\pgfpointxy{1}{1}};
  \pgfgetlastxy{\vx}{\vy}
  \begin{scope}[node distance=\vx and \vy]
  \node at (-0.5,-2) (s1) {};
  \node[below=0.1 of s1] (s1l) {$\bar{x}$};
  \node at (-0.5,-1.05) (i1a) {};
  \node at (-0.5,1.05) (i1b) {}; 
  \node[below left=0 and 0.1 of i1a.center] (i1al) {$x_{4}$};
  \node[below left=0 and 0.1 of i1b.center] (i1bl) {$y_{4}$};
  \node at (-0.5,2) (e1) {};
  \node at (0.5,-2) (s2) {};
  \node[below=0.1 of s2] (s2l) {$\bar{y}$};
  \node at (0.5,-1.05) (i2a) {};
  \node at (0.5,1.05) (i2b) {};
  \node at (0.5,2) (e2) {};
  \draw[->,color=blue] (i1a) edge (i2a) (i2a) edge (i2b) (i2b) edge (i1b) (i1b) edge (i1a);
  \end{scope}
\end{tikzpicture}
\end{minipage}\hfill
\begin{minipage}[t]{0.25\linewidth}
\centering
\begin{tikzpicture}[scale=0.75,nodes={inner sep=0}]
  \pgfpointtransformed{\pgfpointxy{1}{1}};
  \pgfgetlastxy{\vx}{\vy}
  \begin{scope}[node distance=\vx and \vy]
  \node at (-0.5,-2) (s1) {};
  \node[below=0.1 of s1] (s1l) {$\bar{x}$};
  \node at (-0.5,-1.05) (i1a) {};
  \node at (-0.5,1.05) (i1b) {}; 
  \node[below left=0 and 0.1 of i1a.center] (i1al) {$x_{4}$};
  \node[below left=0 and 0.1 of i1b.center] (i1bl) {$y_{4}$};
  \node at (-0.5,2) (e1) {};
  \node at (0.5,-2) (s2) {};
  \node[below=0.1 of s2] (s2l) {$\bar{y}$};
  \node at (0.5,-1.05) (i2a) {};
  \node at (0.5,1.05) (i2b) {};
  \node at (0.5,2) (e2) {};
  \draw[->,color=blue] (s1) edge (i1a) (i1a) edge (i2a) (i2a) edge (s2);
  \draw[->,color=blue] (e2) edge (i2b) (i2b) edge (i1b) (i1b) edge (e1);
  \end{scope}
\end{tikzpicture}
\end{minipage}\hfill
\begin{minipage}[t]{0.25\linewidth}
\centering
\begin{tikzpicture}[scale=0.75,nodes={inner sep=0}]
  \pgfpointtransformed{\pgfpointxy{1}{1}};
  \pgfgetlastxy{\vx}{\vy}
  \begin{scope}[node distance=\vx and \vy]
  \node at (-0.5,-2) (s1) {};
  \node[below=0.1 of s1] (s1l) {$\bar{x}$};
  \node at (-0.5,-1.05) (i1a) {};
  \node at (-0.5,1.05) (i1b) {}; 
  \node[below left=0 and 0.1 of i1a.center] (i1al) {$x_{4}$};
  \node[below left=0 and 0.1 of i1b.center] (i1bl) {$y_{4}$};
  \node at (-0.5,2) (e1) {};
  \node at (0.5,-2) (s2) {};
  \node[below=0.1 of s2] (s2l) {$\bar{y}$};
  \node at (0.5,-1.05) (i2a) {};
  \node at (0.5,1.05) (i2b) {};
  \node at (0.5,2) (e2) {};
  \draw[<-,color=blue] (s2) edge (i2a) (i2a) edge (i1a) (i1a) edge (i1b) (i1b) edge (i2b) (i2b) edge (e2);
  \end{scope}
\end{tikzpicture}
\end{minipage}\\
\vspace{10pt}
\begin{minipage}[t]{0.25\linewidth}
\centering
\begin{tikzpicture}[scale=0.75,nodes={inner sep=0}]
  \pgfpointtransformed{\pgfpointxy{1}{1}};
  \pgfgetlastxy{\vx}{\vy}
  \begin{scope}[node distance=\vx and \vy]
  \node at (-0.5,-2) (s1) {};
  \node[below=0.1 of s1] (s1l) {$\bar{x}$};
  \node at (-0.5,-1.05) (i1a) {};
  \node at (-0.5,1.05) (i1b) {}; 
  \node[below left=0 and 0.1 of i1a.center] (i1al) {$x_{4}$};
  \node[below left=0 and 0.1 of i1b.center] (i1bl) {$y_{4}$};
  \node at (-0.5,2) (e1) {};
  \node at (0.5,-2) (s2) {};
  \node[below=0.1 of s2] (s2l) {$\bar{y}$};
  \node at (0.5,-1.05) (i2a) {};
  \node at (0.5,1.05) (i2b) {};
  \node at (0.5,2) (e2) {};
  \draw[->,color=blue] (s1) edge (i1a) (i1a) edge (i2a) (i2a) edge (i2b) (i2b) edge (i1b) (i1b) edge (e1);
  \begin{scope}
  \node at ($(i1a)+(0.1,1.15)$) (p1) {};
  \node at ($(p1)+(0.8,0)$) (p2) {};
  \node at ($(p2)+(0,0.8)$) (p3) {};
  \node at ($(p3)-(0.8,0)$) (p4) {};
  \draw[<-,color=black] (p1) edge (p2) (p2) edge (p3) (p3) edge (p4) (p4) edge (p1);
  \end{scope}
  \begin{scope}
  \node at ($(i1a)+(0.1,0.15)$) (p1) {};
  \node at ($(p1)+(0.8,0)$) (p2) {};
  \node at ($(p2)+(0,0.8)$) (p3) {};
  \node at ($(p3)-(0.8,0)$) (p4) {};
  \draw[<-,color=black] (p1) edge (p2) (p2) edge (p3) (p3) edge (p4) (p4) edge (p1);
  \end{scope}
  \end{scope}
  
  \begin{scope}[node distance=\vx and \vy]
  \node at (1.5,-2) (s1) {};
  \node[below=0.1 of s1] (s1l) {$\bar{x}$};
  \node at (1.5,2) (e1) {};
  \node at (2.5,-2) (s2) {};
  \node[below=0.1 of s2] (s2l) {$\bar{y}$};
  \node at (2.5,2) (e2) {};
  \draw[->,color=blue] (s1) edge (e1);
  \draw[->,color=red,dashed] (s2) edge (e2);
  \node at (0.75,0) (c1) {};
  \node at (1.25,0) (c2) {};
  \draw[->] (c1) edge (c2);
  \end{scope}
\end{tikzpicture}
\end{minipage}\hfill
\begin{minipage}[t]{0.25\linewidth}
\centering
\begin{tikzpicture}[scale=0.75,nodes={inner sep=0}]
  \pgfpointtransformed{\pgfpointxy{1}{1}};
  \pgfgetlastxy{\vx}{\vy}
  \begin{scope}[node distance=\vx and \vy]
  \node at (-0.5,-2) (s1) {};
  \node[below=0.1 of s1] (s1l) {$\bar{x}$};
  \node at (-0.5,-1.05) (i1a) {};
  \node at (-0.5,1.05) (i1b) {}; 
  \node[below left=0 and 0.1 of i1a.center] (i1al) {$x_{4}$};
  \node[below left=0 and 0.1 of i1b.center] (i1bl) {$y_{4}$};
  \node at (-0.5,2) (e1) {};
  \node at (0.5,-2) (s2) {};
  \node[below=0.1 of s2] (s2l) {$\bar{y}$};
  \node at (0.5,-1.05) (i2a) {};
  \node at (0.5,1.05) (i2b) {};
  \node at (0.5,2) (e2) {};
  \draw[->,color=blue] (i1a) edge (i2a) (i2a) edge (i2b) (i2b) edge (i1b) (i1b) edge (i1a);
  
  \begin{scope}
  \node at ($(i1a)+(0.1,1.15)$) (p1) {};
  \node at ($(p1)+(0.8,0)$) (p2) {};
  \node at ($(p2)+(0,0.8)$) (p3) {};
  \node at ($(p3)-(0.8,0)$) (p4) {};
  \draw[<-,color=black] (p1) edge (p2) (p2) edge (p3) (p3) edge (p4) (p4) edge (p1);
  \end{scope}
  \begin{scope}
  \node at ($(i1a)+(0.1,0.15)$) (p1) {};
  \node at ($(p1)+(0.8,0)$) (p2) {};
  \node at ($(p2)+(0,0.8)$) (p3) {};
  \node at ($(p3)-(0.8,0)$) (p4) {};
  \draw[<-,color=black] (p1) edge (p2) (p2) edge (p3) (p3) edge (p4) (p4) edge (p1);
  \end{scope}
  \end{scope}
  
  \begin{scope}[node distance=\vx and \vy]
  \node at (1.5,-2) (s1) {};
  \node[below=0.1 of s1] (s1l) {$\bar{x}$};
  \node at (1.5,2) (e1) {};
  \node at (2.5,-2) (s2) {};
  \node[below=0.1 of s2] (s2l) {$\bar{y}$};
  \node at (2.5,2) (e2) {};
  \draw[<-,color=blue,dashed] (s1) edge (e1);
  \draw[->,color=red,dashed] (s2) edge (e2);
  \node at (0.75,0) (c1) {};
  \node at (1.25,0) (c2) {};
  \draw[->] (c1) edge (c2);
  \end{scope}
\end{tikzpicture}
\end{minipage}\hfill
\begin{minipage}[t]{0.25\linewidth}
\centering
\begin{tikzpicture}[scale=0.75,nodes={inner sep=0}]
  \pgfpointtransformed{\pgfpointxy{1}{1}};
  \pgfgetlastxy{\vx}{\vy}
  \begin{scope}[node distance=\vx and \vy]
  \node at (-0.5,-2) (s1) {};
  \node[below=0.1 of s1] (s1l) {$\bar{x}$};
  \node at (-0.5,-1.05) (i1a) {};
  \node at (-0.5,1.05) (i1b) {}; 
  \node[below left=0 and 0.1 of i1a.center] (i1al) {$x_{4}$};
  \node[below left=0 and 0.1 of i1b.center] (i1bl) {$y_{4}$};
  \node at (-0.5,2) (e1) {};
  \node at (0.5,-2) (s2) {};
  \node[below=0.1 of s2] (s2l) {$\bar{y}$};
  \node at (0.5,-1.05) (i2a) {};
  \node at (0.5,1.05) (i2b) {};
  \node at (0.5,2) (e2) {};
  \draw[->,color=blue] (s1) edge (i1a) (i1a) edge (i2a) (i2a) edge (s2);
  \draw[->,color=blue] (e2) edge (i2b) (i2b) edge (i1b) (i1b) edge (e1);
  
  \begin{scope}
  \node at ($(i1a)+(0.1,1.15)$) (p1) {};
  \node at ($(p1)+(0.8,0)$) (p2) {};
  \node at ($(p2)+(0,0.8)$) (p3) {};
  \node at ($(p3)-(0.8,0)$) (p4) {};
  \draw[<-,color=black] (p1) edge (p2) (p2) edge (p3) (p3) edge (p4) (p4) edge (p1);
  \end{scope}
  \begin{scope}
  \node at ($(i1a)+(0.1,0.15)$) (p1) {};
  \node at ($(p1)+(0.8,0)$) (p2) {};
  \node at ($(p2)+(0,0.8)$) (p3) {};
  \node at ($(p3)-(0.8,0)$) (p4) {};
  \draw[<-,color=black] (p1) edge (p2) (p2) edge (p3) (p3) edge (p4) (p4) edge (p1);
  \end{scope}
  \end{scope}
  
  \begin{scope}[node distance=\vx and \vy]
  \node at (1.5,-2) (s1) {};
  \node[below=0.1 of s1] (s1l) {$\bar{x}$};
  \node at (1.5,2) (e1) {};
  \node at (2.5,-2) (s2) {};
  \node[below=0.1 of s2] (s2l) {$\bar{y}$};
  \node at (2.5,2) (e2) {};
  \draw[->,color=blue] (s1) edge (e1);
  \draw[<-,color=red] (s2) edge (e2);
  \node at (0.75,0) (c1) {};
  \node at (1.25,0) (c2) {};
  \draw[->] (c1) edge (c2);
  \end{scope}
\end{tikzpicture}
\end{minipage}\hfill
\begin{minipage}[t]{0.25\linewidth}
\centering
\begin{tikzpicture}[scale=0.75,nodes={inner sep=0}]
  \pgfpointtransformed{\pgfpointxy{1}{1}};
  \pgfgetlastxy{\vx}{\vy}
  \begin{scope}[node distance=\vx and \vy]
  \node at (-0.5,-2) (s1) {};
  \node[below=0.1 of s1] (s1l) {$\bar{x}$};
  \node at (-0.5,-1.05) (i1a) {};
  \node at (-0.5,1.05) (i1b) {}; 
  \node[below left=0 and 0.1 of i1a.center] (i1al) {$x_{4}$};
  \node[below left=0 and 0.1 of i1b.center] (i1bl) {$y_{4}$};
  \node at (-0.5,2) (e1) {};
  \node at (0.5,-2) (s2) {};
  \node[below=0.1 of s2] (s2l) {$\bar{y}$};
  \node at (0.5,-1.05) (i2a) {};
  \node at (0.5,1.05) (i2b) {};
  \node at (0.5,2) (e2) {};
  \draw[<-,color=blue] (s2) edge (i2a) (i2a) edge (i1a) (i1a) edge (i1b) (i1b) edge (i2b) (i2b) edge (e2);
  
  \begin{scope}
  \node at ($(i1a)+(0.1,1.15)$) (p1) {};
  \node at ($(p1)+(0.8,0)$) (p2) {};
  \node at ($(p2)+(0,0.8)$) (p3) {};
  \node at ($(p3)-(0.8,0)$) (p4) {};
  \draw[<-,color=black] (p1) edge (p2) (p2) edge (p3) (p3) edge (p4) (p4) edge (p1);
  \end{scope}
  \begin{scope}
  \node at ($(i1a)+(0.1,0.15)$) (p1) {};
  \node at ($(p1)+(0.8,0)$) (p2) {};
  \node at ($(p2)+(0,0.8)$) (p3) {};
  \node at ($(p3)-(0.8,0)$) (p4) {};
  \draw[<-,color=black] (p1) edge (p2) (p2) edge (p3) (p3) edge (p4) (p4) edge (p1);
  \end{scope}
  \end{scope}
  
  \begin{scope}[node distance=\vx and \vy]
  \node at (1.5,-2) (s1) {};
  \node[below=0.1 of s1] (s1l) {$\bar{x}$};
  \node at (1.5,2) (e1) {};
  \node at (2.5,-2) (s2) {};
  \node[below=0.1 of s2] (s2l) {$\bar{y}$};
  \node at (2.5,2) (e2) {};
  \draw[<-,color=blue,dashed] (s1) edge (e1);
  \draw[<-,color=red] (s2) edge (e2);
  \node at (0.75,0) (c1) {};
  \node at (1.25,0) (c2) {};
  \draw[->] (c1) edge (c2);
  \end{scope}
\end{tikzpicture}
\end{minipage}\\
\vspace{10pt}
\begin{minipage}[t]{0.25\linewidth}
\centering
\begin{tikzpicture}[scale=0.75,nodes={inner sep=0}]
  \pgfpointtransformed{\pgfpointxy{1}{1}};
  \pgfgetlastxy{\vx}{\vy}
  \begin{scope}[node distance=\vx and \vy]
  \node at (-0.5,-2) (s1) {};
  \node[below=0.1 of s1] (s1l) {$\bar{x}$};
  \node at (-0.5,-1.05) (i1a) {};
  \node at (-0.5,1.05) (i1b) {}; 
  \node[below left=0 and 0.1 of i1a.center] (i1al) {$x_{4}$};
  \node[below left=0 and 0.1 of i1b.center] (i1bl) {$y_{4}$};
  \node at (-0.5,2) (e1) {};
  \node at (0.5,-2) (s2) {};
  \node[below=0.1 of s2] (s2l) {$\bar{y}$};
  \node at (0.5,-1.05) (i2a) {};
  \node at (0.5,1.05) (i2b) {};
  \node at (0.5,2) (e2) {};
  \draw[->,color=blue] (s1) edge (i1a) (i1a) edge (i2a) (i2a) edge (i2b) (i2b) edge (i1b) (i1b) edge (e1);
  \begin{scope}
  \node at ($(i1b)+(0.1,0.1)$) (p1) {};
  \node at ($(p1)+(0.8,0)$) (p2) {};
  \node at ($(p2)+(0,0.8)$) (p3) {};
  \node at ($(p3)-(0.8,0)$) (p4) {};
  \draw[->,color=black] (p1) edge (p2) (p2) edge (p3) (p3) edge (p4) (p4) edge (p1);
  \end{scope}
  \begin{scope}
  \node at ($(i1a)+(0.1,-0.9)$) (p1) {};
  \node at ($(p1)+(0.8,0)$) (p2) {};
  \node at ($(p2)+(0,0.8)$) (p3) {};
  \node at ($(p3)-(0.8,0)$) (p4) {};
  \draw[->,color=black] (p1) edge (p2) (p2) edge (p3) (p3) edge (p4) (p4) edge (p1);
  \end{scope}
  \end{scope}
  
  \begin{scope}[node distance=\vx and \vy]
  \node at (1.5,-2) (s1) {};
  \node[below=0.1 of s1] (s1l) {$\bar{x}$};
  \node at (1.5,2) (e1) {};
  \node at (2.5,-2) (s2) {};
  \node[below=0.1 of s2] (s2l) {$\bar{y}$};
  \node at (2.5,2) (e2) {};
  \draw[->,color=blue] (s1) edge (e1);
  \draw[->,color=red] (s2) edge (e2);
  \node at (0.75,0) (c1) {};
  \node at (1.25,0) (c2) {};
  \draw[->] (c1) edge (c2);
  \end{scope}
\end{tikzpicture}
\end{minipage}\hfill
\begin{minipage}[t]{0.25\linewidth}
\centering
\begin{tikzpicture}[scale=0.75,nodes={inner sep=0}]
  \pgfpointtransformed{\pgfpointxy{1}{1}};
  \pgfgetlastxy{\vx}{\vy}
  \begin{scope}[node distance=\vx and \vy]
  \node at (-0.5,-2) (s1) {};
  \node[below=0.1 of s1] (s1l) {$\bar{x}$};
  \node at (-0.5,-1.05) (i1a) {};
  \node at (-0.5,1.05) (i1b) {}; 
  \node[below left=0 and 0.1 of i1a.center] (i1al) {$x_{4}$};
  \node[below left=0 and 0.1 of i1b.center] (i1bl) {$y_{4}$};
  \node at (-0.5,2) (e1) {};
  \node at (0.5,-2) (s2) {};
  \node[below=0.1 of s2] (s2l) {$\bar{y}$};
  \node at (0.5,-1.05) (i2a) {};
  \node at (0.5,1.05) (i2b) {};
  \node at (0.5,2) (e2) {};
  \draw[->,color=blue] (i1a) edge (i2a) (i2a) edge (i2b) (i2b) edge (i1b) (i1b) edge (i1a);
  
  \begin{scope}
  \node at ($(i1b)+(0.1,0.1)$) (p1) {};
  \node at ($(p1)+(0.8,0)$) (p2) {};
  \node at ($(p2)+(0,0.8)$) (p3) {};
  \node at ($(p3)-(0.8,0)$) (p4) {};
  \draw[->,color=black] (p1) edge (p2) (p2) edge (p3) (p3) edge (p4) (p4) edge (p1);
  \end{scope}
  \begin{scope}
  \node at ($(i1a)+(0.1,-0.9)$) (p1) {};
  \node at ($(p1)+(0.8,0)$) (p2) {};
  \node at ($(p2)+(0,0.8)$) (p3) {};
  \node at ($(p3)-(0.8,0)$) (p4) {};
  \draw[->,color=black] (p1) edge (p2) (p2) edge (p3) (p3) edge (p4) (p4) edge (p1);
  \end{scope}
  \end{scope}
  
  \begin{scope}[node distance=\vx and \vy]
  \node at (1.5,-2) (s1) {};
  \node[below=0.1 of s1] (s1l) {$\bar{x}$};
  \node at (1.5,2) (e1) {};
  \node at (2.5,-2) (s2) {};
  \node[below=0.1 of s2] (s2l) {$\bar{y}$};
  \node at (2.5,2) (e2) {};
  \draw[<-,color=blue] (s1) edge (e1);
  \draw[->,color=red] (s2) edge (e2);
  \node at (0.75,0) (c1) {};
  \node at (1.25,0) (c2) {};
  \draw[->] (c1) edge (c2);
  \end{scope}
\end{tikzpicture}
\end{minipage}\hfill
\begin{minipage}[t]{0.25\linewidth}
\centering
\begin{tikzpicture}[scale=0.75,nodes={inner sep=0}]
  \pgfpointtransformed{\pgfpointxy{1}{1}};
  \pgfgetlastxy{\vx}{\vy}
  \begin{scope}[node distance=\vx and \vy]
  \node at (-0.5,-2) (s1) {};
  \node[below=0.1 of s1] (s1l) {$\bar{x}$};
  \node at (-0.5,-1.05) (i1a) {};
  \node at (-0.5,1.05) (i1b) {}; 
  \node[below left=0 and 0.1 of i1a.center] (i1al) {$x_{4}$};
  \node[below left=0 and 0.1 of i1b.center] (i1bl) {$y_{4}$};
  \node at (-0.5,2) (e1) {};
  \node at (0.5,-2) (s2) {};
  \node[below=0.1 of s2] (s2l) {$\bar{y}$};
  \node at (0.5,-1.05) (i2a) {};
  \node at (0.5,1.05) (i2b) {};
  \node at (0.5,2) (e2) {};
  \draw[->,color=blue] (s1) edge (i1a) (i1a) edge (i2a) (i2a) edge (s2);
  \draw[->,color=blue] (e2) edge (i2b) (i2b) edge (i1b) (i1b) edge (e1);
  
  \begin{scope}
  \node at ($(i1b)+(0.1,0.1)$) (p1) {};
  \node at ($(p1)+(0.8,0)$) (p2) {};
  \node at ($(p2)+(0,0.8)$) (p3) {};
  \node at ($(p3)-(0.8,0)$) (p4) {};
  \draw[->,color=black] (p1) edge (p2) (p2) edge (p3) (p3) edge (p4) (p4) edge (p1);
  \end{scope}
  \begin{scope}
  \node at ($(i1a)+(0.1,-0.9)$) (p1) {};
  \node at ($(p1)+(0.8,0)$) (p2) {};
  \node at ($(p2)+(0,0.8)$) (p3) {};
  \node at ($(p3)-(0.8,0)$) (p4) {};
  \draw[->,color=black] (p1) edge (p2) (p2) edge (p3) (p3) edge (p4) (p4) edge (p1);
  \end{scope}
  \end{scope}
  
  \begin{scope}[node distance=\vx and \vy]
  \node at (1.5,-2) (s1) {};
  \node[below=0.1 of s1] (s1l) {$\bar{x}$};
  \node at (1.5,2) (e1) {};
  \node at (2.5,-2) (s2) {};
  \node[below=0.1 of s2] (s2l) {$\bar{y}$};
  \node at (2.5,2) (e2) {};
  \draw[->,color=blue,dashed] (s1) edge (e1);
  \draw[<-,color=red,dashed] (s2) edge (e2);
  \node at (0.75,0) (c1) {};
  \node at (1.25,0) (c2) {};
  \draw[->] (c1) edge (c2);
  \end{scope}
\end{tikzpicture}
\end{minipage}\hfill
\begin{minipage}[t]{0.25\linewidth}
\centering
\begin{tikzpicture}[scale=0.75,nodes={inner sep=0}]
  \pgfpointtransformed{\pgfpointxy{1}{1}};
  \pgfgetlastxy{\vx}{\vy}
  \begin{scope}[node distance=\vx and \vy]
  \node at (-0.5,-2) (s1) {};
  \node[below=0.1 of s1] (s1l) {$\bar{x}$};
  \node at (-0.5,-1.05) (i1a) {};
  \node at (-0.5,1.05) (i1b) {}; 
  \node[below left=0 and 0.1 of i1a.center] (i1al) {$x_{4}$};
  \node[below left=0 and 0.1 of i1b.center] (i1bl) {$y_{4}$};
  \node at (-0.5,2) (e1) {};
  \node at (0.5,-2) (s2) {};
  \node[below=0.1 of s2] (s2l) {$\bar{y}$};
  \node at (0.5,-1.05) (i2a) {};
  \node at (0.5,1.05) (i2b) {};
  \node at (0.5,2) (e2) {};
  \draw[<-,color=blue] (s2) edge (i2a) (i2a) edge (i1a) (i1a) edge (i1b) (i1b) edge (i2b) (i2b) edge (e2);
  
  \begin{scope}
  \node at ($(i1b)+(0.1,0.1)$) (p1) {};
  \node at ($(p1)+(0.8,0)$) (p2) {};
  \node at ($(p2)+(0,0.8)$) (p3) {};
  \node at ($(p3)-(0.8,0)$) (p4) {};
  \draw[->,color=black] (p1) edge (p2) (p2) edge (p3) (p3) edge (p4) (p4) edge (p1);
  \end{scope}
  \begin{scope}
  \node at ($(i1a)+(0.1,-0.9)$) (p1) {};
  \node at ($(p1)+(0.8,0)$) (p2) {};
  \node at ($(p2)+(0,0.8)$) (p3) {};
  \node at ($(p3)-(0.8,0)$) (p4) {};
  \draw[->,color=black] (p1) edge (p2) (p2) edge (p3) (p3) edge (p4) (p4) edge (p1);
  \end{scope}
  \end{scope}
  
  \begin{scope}[node distance=\vx and \vy]
  \node at (1.5,-2) (s1) {};
  \node[below=0.1 of s1] (s1l) {$\bar{x}$};
  \node at (1.5,2) (e1) {};
  \node at (2.5,-2) (s2) {};
  \node[below=0.1 of s2] (s2l) {$\bar{y}$};
  \node at (2.5,2) (e2) {};
  \draw[<-,color=blue] (s1) edge (e1);
  \draw[<-,color=red] (s2) edge (e2);
  \node at (0.75,0) (c1) {};
  \node at (1.25,0) (c2) {};
  \draw[->] (c1) edge (c2);
  \end{scope}
\end{tikzpicture}
\end{minipage}\\
\vspace{10pt}
\begin{minipage}[t]{0.25\linewidth}
\centering
\begin{tikzpicture}[scale=0.75,nodes={inner sep=0}]
  \pgfpointtransformed{\pgfpointxy{1}{1}};
  \pgfgetlastxy{\vx}{\vy}
  \begin{scope}[node distance=\vx and \vy]
  \node at (-0.5,-2) (s1) {};
  \node[below=0.1 of s1] (s1l) {$\bar{x}$};
  \node at (-0.5,-1.05) (i1a) {};
  \node at (-0.5,1.05) (i1b) {}; 
  \node[below left=0 and 0.1 of i1a.center] (i1al) {$x_{4}$};
  \node[below left=0 and 0.1 of i1b.center] (i1bl) {$y_{4}$};
  \node at (-0.5,2) (e1) {};
  \node at (0.5,-2) (s2) {};
  \node[below=0.1 of s2] (s2l) {$\bar{y}$};
  \node at (0.5,-1.05) (i2a) {};
  \node at (0.5,1.05) (i2b) {};
  \node at (0.5,2) (e2) {};
  \draw[->,color=blue] (s1) edge (i1a) (i1a) edge (i2a) (i2a) edge (i2b) (i2b) edge (i1b) (i1b) edge (e1);
  \begin{scope}
  \node at ($(i1a)+(0.1,2.25)$) (p1) {};
  \node at ($(p1)+(0.8,0)$) (p2) {};
  \node at ($(p2)+(0,0.8)$) (p3) {};
  \node at ($(p3)-(0.8,0)$) (p4) {};
  \draw[<-,color=black] (p1) edge (p2) (p2) edge (p3) (p3) edge (p4) (p4) edge (p1);
  \end{scope}
  \begin{scope}
  \node at ($(i1a)+(0.1,1.15)$) (p1) {};
  \node at ($(p1)+(0.8,0)$) (p2) {};
  \node at ($(p2)+(0,0.8)$) (p3) {};
  \node at ($(p3)-(0.8,0)$) (p4) {};
  \draw[->,color=black] (p1) edge (p2) (p2) edge (p3) (p3) edge (p4) (p4) edge (p1);
  \end{scope}
  \begin{scope}
  \node at ($(i1a)+(0.1,0.15)$) (p1) {};
  \node at ($(p1)+(0.8,0)$) (p2) {};
  \node at ($(p2)+(0,0.8)$) (p3) {};
  \node at ($(p3)-(0.8,0)$) (p4) {};
  \draw[->,color=black] (p1) edge (p2) (p2) edge (p3) (p3) edge (p4) (p4) edge (p1);
  \end{scope}
  \begin{scope}
  \node at ($(i1a)+(0.1,-0.95)$) (p1) {};
  \node at ($(p1)+(0.8,0)$) (p2) {};
  \node at ($(p2)+(0,0.8)$) (p3) {};
  \node at ($(p3)-(0.8,0)$) (p4) {};
  \draw[<-,color=black] (p1) edge (p2) (p2) edge (p3) (p3) edge (p4) (p4) edge (p1);
  \end{scope}
  \end{scope}
  
  \begin{scope}[node distance=\vx and \vy]
  \node at (1.5,-2) (s1) {};
  \node at (1.45,-2) (s1l) {};
  \node at (1.55,-2) (s1r) {};
  \node[below=0.1 of s1] (s1lb) {$\bar{x}$};
  \node at (1.5,2) (e1) {};
  \node at (1.45,2) (e1l) {};
  \node at (1.55,2) (e1r) {};
  \node at (1.5,-1) (i11) {};
  \node at (1.5,1) (i12) {};
  \node at (1.45,-1) (i11l) {};
  \node at (1.55,-1) (i11r) {};
  \node at (1.45,1) (i12l) {};
  \node at (1.55,1) (i12r) {};
  \node at (2.5,-2) (s2) {};
  \node at (2.45,-2) (s2l) {};
  \node at (2.55,-2) (s2r) {};
  \node[below=0.1 of s2] (s2lb) {$\bar{y}$};
  \node at (2.5,2) (e2) {};
  \node at (2.45,2) (e2l) {};
  \node at (2.55,2) (e2r) {};
  \node at (2.5,-1) (i21) {};
  \node at (2.5,1) (i22) {};
  \node at (2.45,-1) (i21l) {};
  \node at (2.55,-1) (i21r) {};
  \node at (2.45,1) (i22l) {};
  \node at (2.55,1) (i22r) {};
  \draw[->,color=blue] (s1l) edge (i11l);
  \draw[->,color=blue] (s1r) edge (i11r);
  \draw[<-,color=blue] (i11) edge (i12);
  \draw[->,color=blue] (i12l) edge (e1l);
  \draw[->,color=blue] (i12r) edge (e1r);
  \draw[<-,color=red] (s2) edge (i21);
  \draw[->,color=red] (i21l) edge (i22l);
  \draw[->,color=red] (i21r) edge (i22r);
  \draw[<-,color=red] (i22) edge (e2);
  \node at (0.75,0) (c1) {};
  \node at (1.25,0) (c2) {};
  \draw[->] (c1) edge (c2);
  \end{scope}
\end{tikzpicture}
\end{minipage}\hfill
\begin{minipage}[t]{0.25\linewidth}
\centering
\begin{tikzpicture}[scale=0.75,nodes={inner sep=0}]
  \pgfpointtransformed{\pgfpointxy{1}{1}};
  \pgfgetlastxy{\vx}{\vy}
  \begin{scope}[node distance=\vx and \vy]
  \node at (-0.5,-2) (s1) {};
  \node[below=0.1 of s1] (s1l) {$\bar{x}$};
  \node at (-0.5,-1.05) (i1a) {};
  \node at (-0.5,1.05) (i1b) {}; 
  \node[below left=0 and 0.1 of i1a.center] (i1al) {$x_{4}$};
  \node[below left=0 and 0.1 of i1b.center] (i1bl) {$y_{4}$};
  \node at (-0.5,2) (e1) {};
  \node at (0.5,-2) (s2) {};
  \node[below=0.1 of s2] (s2l) {$\bar{y}$};
  \node at (0.5,-1.05) (i2a) {};
  \node at (0.5,1.05) (i2b) {};
  \node at (0.5,2) (e2) {};
  \draw[->,color=blue] (i1a) edge (i2a) (i2a) edge (i2b) (i2b) edge (i1b) (i1b) edge (i1a);
  
  \begin{scope}
  \node at ($(i1a)+(0.1,2.25)$) (p1) {};
  \node at ($(p1)+(0.8,0)$) (p2) {};
  \node at ($(p2)+(0,0.8)$) (p3) {};
  \node at ($(p3)-(0.8,0)$) (p4) {};
  \draw[<-,color=black] (p1) edge (p2) (p2) edge (p3) (p3) edge (p4) (p4) edge (p1);
  \end{scope}
  \begin{scope}
  \node at ($(i1a)+(0.1,1.15)$) (p1) {};
  \node at ($(p1)+(0.8,0)$) (p2) {};
  \node at ($(p2)+(0,0.8)$) (p3) {};
  \node at ($(p3)-(0.8,0)$) (p4) {};
  \draw[->,color=black] (p1) edge (p2) (p2) edge (p3) (p3) edge (p4) (p4) edge (p1);
  \end{scope}
  \begin{scope}
  \node at ($(i1a)+(0.1,0.15)$) (p1) {};
  \node at ($(p1)+(0.8,0)$) (p2) {};
  \node at ($(p2)+(0,0.8)$) (p3) {};
  \node at ($(p3)-(0.8,0)$) (p4) {};
  \draw[->,color=black] (p1) edge (p2) (p2) edge (p3) (p3) edge (p4) (p4) edge (p1);
  \end{scope}
  \begin{scope}
  \node at ($(i1a)+(0.1,-0.95)$) (p1) {};
  \node at ($(p1)+(0.8,0)$) (p2) {};
  \node at ($(p2)+(0,0.8)$) (p3) {};
  \node at ($(p3)-(0.8,0)$) (p4) {};
  \draw[<-,color=black] (p1) edge (p2) (p2) edge (p3) (p3) edge (p4) (p4) edge (p1);
  \end{scope}
  \end{scope}
  
  \begin{scope}[node distance=\vx and \vy]
  \node at (1.5,-2) (s1) {};
  \node at (1.45,-2) (s1l) {};
  \node at (1.55,-2) (s1r) {};
  \node[below=0.1 of s1] (s1lb) {$\bar{x}$};
  \node at (1.5,2) (e1) {};
  \node at (1.45,2) (e1l) {};
  \node at (1.55,2) (e1r) {};
  \node at (1.5,-1) (i11) {};
  \node at (1.5,1) (i12) {};
  \node at (1.45,-1) (i11l) {};
  \node at (1.55,-1) (i11r) {};
  \node at (1.45,1) (i12l) {};
  \node at (1.55,1) (i12r) {};
  \node at (2.5,-2) (s2) {};
  \node at (2.45,-2) (s2l) {};
  \node at (2.55,-2) (s2r) {};
  \node[below=0.1 of s2] (s2lb) {$\bar{y}$};
  \node at (2.5,2) (e2) {};
  \node at (2.45,2) (e2l) {};
  \node at (2.55,2) (e2r) {};
  \node at (2.5,-1) (i21) {};
  \node at (2.5,1) (i22) {};
  \node at (2.45,-1) (i21l) {};
  \node at (2.55,-1) (i21r) {};
  \node at (2.45,1) (i22l) {};
  \node at (2.55,1) (i22r) {};
  \draw[->,color=blue] (s1) edge (i11);
  \draw[<-,color=blue] (i11l) edge (i12l);
  \draw[<-,color=blue] (i11r) edge (i12r);
  \draw[->,color=blue] (i12) edge (e1);
  \draw[<-,color=red] (s2) edge (i21);
  \draw[->,color=red] (i21l) edge (i22l);
  \draw[->,color=red] (i21r) edge (i22r);
  \draw[<-,color=red] (i22) edge (e2);
  \node at (0.75,0) (c1) {};
  \node at (1.25,0) (c2) {};
  \draw[->] (c1) edge (c2);
  \end{scope}
\end{tikzpicture}
\end{minipage}\hfill
\begin{minipage}[t]{0.25\linewidth}
\centering
\begin{tikzpicture}[scale=0.75,nodes={inner sep=0}]
  \pgfpointtransformed{\pgfpointxy{1}{1}};
  \pgfgetlastxy{\vx}{\vy}
  \begin{scope}[node distance=\vx and \vy]
  \node at (-0.5,-2) (s1) {};
  \node[below=0.1 of s1] (s1l) {$\bar{x}$};
  \node at (-0.5,-1.05) (i1a) {};
  \node at (-0.5,1.05) (i1b) {}; 
  \node[below left=0 and 0.1 of i1a.center] (i1al) {$x_{4}$};
  \node[below left=0 and 0.1 of i1b.center] (i1bl) {$y_{4}$};
  \node at (-0.5,2) (e1) {};
  \node at (0.5,-2) (s2) {};
  \node[below=0.1 of s2] (s2l) {$\bar{y}$};
  \node at (0.5,-1.05) (i2a) {};
  \node at (0.5,1.05) (i2b) {};
  \node at (0.5,2) (e2) {};
  \draw[->,color=blue] (s1) edge (i1a) (i1a) edge (i2a) (i2a) edge (s2);
  \draw[->,color=blue] (e2) edge (i2b) (i2b) edge (i1b) (i1b) edge (e1);
  
  \begin{scope}
  \node at ($(i1a)+(0.1,2.25)$) (p1) {};
  \node at ($(p1)+(0.8,0)$) (p2) {};
  \node at ($(p2)+(0,0.8)$) (p3) {};
  \node at ($(p3)-(0.8,0)$) (p4) {};
  \draw[<-,color=black] (p1) edge (p2) (p2) edge (p3) (p3) edge (p4) (p4) edge (p1);
  \end{scope}
  \begin{scope}
  \node at ($(i1a)+(0.1,1.15)$) (p1) {};
  \node at ($(p1)+(0.8,0)$) (p2) {};
  \node at ($(p2)+(0,0.8)$) (p3) {};
  \node at ($(p3)-(0.8,0)$) (p4) {};
  \draw[->,color=black] (p1) edge (p2) (p2) edge (p3) (p3) edge (p4) (p4) edge (p1);
  \end{scope}
  \begin{scope}
  \node at ($(i1a)+(0.1,0.15)$) (p1) {};
  \node at ($(p1)+(0.8,0)$) (p2) {};
  \node at ($(p2)+(0,0.8)$) (p3) {};
  \node at ($(p3)-(0.8,0)$) (p4) {};
  \draw[->,color=black] (p1) edge (p2) (p2) edge (p3) (p3) edge (p4) (p4) edge (p1);
  \end{scope}
  \begin{scope}
  \node at ($(i1a)+(0.1,-0.95)$) (p1) {};
  \node at ($(p1)+(0.8,0)$) (p2) {};
  \node at ($(p2)+(0,0.8)$) (p3) {};
  \node at ($(p3)-(0.8,0)$) (p4) {};
  \draw[<-,color=black] (p1) edge (p2) (p2) edge (p3) (p3) edge (p4) (p4) edge (p1);
  \end{scope}
  \end{scope}
  
  \begin{scope}[node distance=\vx and \vy]
  \node at (1.5,-2) (s1) {};
  \node at (1.45,-2) (s1l) {};
  \node at (1.55,-2) (s1r) {};
  \node[below=0.1 of s1] (s1lb) {$\bar{x}$};
  \node at (1.5,2) (e1) {};
  \node at (1.45,2) (e1l) {};
  \node at (1.55,2) (e1r) {};
  \node at (1.5,-1) (i11) {};
  \node at (1.5,1) (i12) {};
  \node at (1.45,-1) (i11l) {};
  \node at (1.55,-1) (i11r) {};
  \node at (1.45,1) (i12l) {};
  \node at (1.55,1) (i12r) {};
  \node at (2.5,-2) (s2) {};
  \node at (2.45,-2) (s2l) {};
  \node at (2.55,-2) (s2r) {};
  \node[below=0.1 of s2] (s2lb) {$\bar{y}$};
  \node at (2.5,2) (e2) {};
  \node at (2.45,2) (e2l) {};
  \node at (2.55,2) (e2r) {};
  \node at (2.5,-1) (i21) {};
  \node at (2.5,1) (i22) {};
  \node at (2.45,-1) (i21l) {};
  \node at (2.55,-1) (i21r) {};
  \node at (2.45,1) (i22l) {};
  \node at (2.55,1) (i22r) {};
  \draw[->,color=blue] (s1l) edge (i11l);
  \draw[->,color=blue] (s1r) edge (i11r);
  \draw[<-,color=blue] (i11) edge (i12);
  \draw[->,color=blue] (i12l) edge (e1l);
  \draw[->,color=blue] (i12r) edge (e1r);
  \draw[<-,color=red] (s2l) edge (i21l);
  \draw[<-,color=red] (s2r) edge (i21r);
  \draw[->,color=red] (i21) edge (i22);
  \draw[<-,color=red] (i22l) edge (e2l);
  \draw[<-,color=red] (i22r) edge (e2r);
  \node at (0.75,0) (c1) {};
  \node at (1.25,0) (c2) {};
  \draw[->] (c1) edge (c2);
  \end{scope}
\end{tikzpicture}
\end{minipage}\hfill
\begin{minipage}[t]{0.25\linewidth}
\centering
\begin{tikzpicture}[scale=0.75,nodes={inner sep=0}]
  \pgfpointtransformed{\pgfpointxy{1}{1}};
  \pgfgetlastxy{\vx}{\vy}
  \begin{scope}[node distance=\vx and \vy]
  \node at (-0.5,-2) (s1) {};
  \node[below=0.1 of s1] (s1l) {$\bar{x}$};
  \node at (-0.5,-1.05) (i1a) {};
  \node at (-0.5,1.05) (i1b) {}; 
  \node[below left=0 and 0.1 of i1a.center] (i1al) {$x_{4}$};
  \node[below left=0 and 0.1 of i1b.center] (i1bl) {$y_{4}$};
  \node at (-0.5,2) (e1) {};
  \node at (0.5,-2) (s2) {};
  \node[below=0.1 of s2] (s2l) {$\bar{y}$};
  \node at (0.5,-1.05) (i2a) {};
  \node at (0.5,1.05) (i2b) {};
  \node at (0.5,2) (e2) {};
  \draw[<-,color=blue] (s2) edge (i2a) (i2a) edge (i1a) (i1a) edge (i1b) (i1b) edge (i2b) (i2b) edge (e2);
  
  \begin{scope}
  \node at ($(i1a)+(0.1,2.25)$) (p1) {};
  \node at ($(p1)+(0.8,0)$) (p2) {};
  \node at ($(p2)+(0,0.8)$) (p3) {};
  \node at ($(p3)-(0.8,0)$) (p4) {};
  \draw[<-,color=black] (p1) edge (p2) (p2) edge (p3) (p3) edge (p4) (p4) edge (p1);
  \end{scope}
  \begin{scope}
  \node at ($(i1a)+(0.1,1.15)$) (p1) {};
  \node at ($(p1)+(0.8,0)$) (p2) {};
  \node at ($(p2)+(0,0.8)$) (p3) {};
  \node at ($(p3)-(0.8,0)$) (p4) {};
  \draw[->,color=black] (p1) edge (p2) (p2) edge (p3) (p3) edge (p4) (p4) edge (p1);
  \end{scope}
  \begin{scope}
  \node at ($(i1a)+(0.1,0.15)$) (p1) {};
  \node at ($(p1)+(0.8,0)$) (p2) {};
  \node at ($(p2)+(0,0.8)$) (p3) {};
  \node at ($(p3)-(0.8,0)$) (p4) {};
  \draw[->,color=black] (p1) edge (p2) (p2) edge (p3) (p3) edge (p4) (p4) edge (p1);
  \end{scope}
  \begin{scope}
  \node at ($(i1a)+(0.1,-0.95)$) (p1) {};
  \node at ($(p1)+(0.8,0)$) (p2) {};
  \node at ($(p2)+(0,0.8)$) (p3) {};
  \node at ($(p3)-(0.8,0)$) (p4) {};
  \draw[<-,color=black] (p1) edge (p2) (p2) edge (p3) (p3) edge (p4) (p4) edge (p1);
  \end{scope}
  \end{scope}
  
  \begin{scope}[node distance=\vx and \vy]
  \node at (1.5,-2) (s1) {};
  \node at (1.45,-2) (s1l) {};
  \node at (1.55,-2) (s1r) {};
  \node[below=0.1 of s1] (s1lb) {$\bar{x}$};
  \node at (1.5,2) (e1) {};
  \node at (1.45,2) (e1l) {};
  \node at (1.55,2) (e1r) {};
  \node at (1.5,-1) (i11) {};
  \node at (1.5,1) (i12) {};
  \node at (1.45,-1) (i11l) {};
  \node at (1.55,-1) (i11r) {};
  \node at (1.45,1) (i12l) {};
  \node at (1.55,1) (i12r) {};
  \node at (2.5,-2) (s2) {};
  \node at (2.45,-2) (s2l) {};
  \node at (2.55,-2) (s2r) {};
  \node[below=0.1 of s2] (s2lb) {$\bar{y}$};
  \node at (2.5,2) (e2) {};
  \node at (2.45,2) (e2l) {};
  \node at (2.55,2) (e2r) {};
  \node at (2.5,-1) (i21) {};
  \node at (2.5,1) (i22) {};
  \node at (2.45,-1) (i21l) {};
  \node at (2.55,-1) (i21r) {};
  \node at (2.45,1) (i22l) {};
  \node at (2.55,1) (i22r) {};
  \draw[->,color=blue] (s1) edge (i11);
  \draw[<-,color=blue] (i11l) edge (i12l);
  \draw[<-,color=blue] (i11r) edge (i12r);
  \draw[->,color=blue] (i12) edge (e1);
  \draw[<-,color=red] (s2l) edge (i21l);
  \draw[<-,color=red] (s2r) edge (i21r);
  \draw[->,color=red] (i21) edge (i22);
  \draw[<-,color=red] (i22l) edge (e2l);
  \draw[<-,color=red] (i22r) edge (e2r);
  \node at (0.75,0) (c1) {};
  \node at (1.25,0) (c2) {};
  \draw[->] (c1) edge (c2);
  \end{scope}
\end{tikzpicture}
\end{minipage}
\caption{On top of the figure, we show four diagrams corresponding to the four different terms in \eqref{eq:sqtermsum} for a particular choice of $x_{4}$ and $y_{4}$ (with $y_{4}-x_{4}=2$). Below each diagram, the three possibilities of attaching fundamental plaquettes (coming from the gauge action) to the diagrams, such that they survive the process of integrating out the spatial links, are shown together with the resulting loop configuration after that integration. Solid lines correspond to terms like \eqref{eq:sl1} and \eqref{eq:sl2}, dashed lines to terms like \eqref{eq:dl1} and \eqref{eq:dl2} and the mixed solid and double solid lines to terms like \eqref{eq:ml1} and \eqref{eq:ml2}.}
\label{fig:ksqbexmpl}
\end{figure}

To obtain non-zero contributions from terms in \eqref{eq:sqtermsum} for which $y_{4}\neq x_{4}$, we have to take plaquette terms into account. To see in more detail how this works, let us as an example consider the top left diagram in Fig. \ref{fig:ksqbexmpl}. It shows a contributing diagram to the first term inside the large bracket in the sum over $x_{4},y_{4}$ in eq. \eqref{eq:sqtermsum} for the case where $y_{4}-x_{4}=2$. If we would integrate out the spatial links in the corresponding term in \eqref{eq:sqtermsum} without taking into account plaquette terms, the result would just vanish due to the left-hand identity in \eqref{eq:su3int}. But allowing for plaquette terms coming from the non-trivial gauge-field part in \eqref{eq:nneffpartf}, we can form the two left-most diagrams on the second and third row of Fig. \ref{fig:ksqbexmpl}, and get for example for the one on the second row, a term like
\begin{multline}
-\of{3\,a_{f}\of{\beta}\,\trace_{c}\of{U^{\dagger}_{p_1}}}\,\of{3\,a_{f}\of{\beta}\,\trace_{c}\of{U^{\dagger}_{p_2}}}\\
\of{2\,\kappa\e^{\mu}}^{n_{t}}\trace_{c}\of{U_{i}\of{\bar{x},x_{4}}\,P_{\bar{y}}\of{x_{4},x_{4}+2}\,\sof{\id+\of{2\,\kappa\e^{\mu}}^{n_{t}}P_{\bar{y}}\of{x_{4}+2}}^{-1}\right.\\
\,\left.U_{i}^{\dagger}\of{\bar{x},x_{4}+2}\,P_{\bar{x}}\of{x_{4}+2,x_{4}}\,\sof{\id+\of{2\,\kappa\e^{\mu}}^{n_{t}}P_{\bar{x}}\of{x_{4}}}^{-1}}\ ,
\end{multline}
where
\[
U^{\dagger}_{p_1}\,=\,U_{4}\of{\bar{x},x_{4}}\,U_{i}\of{\bar{x},x_{4}+1}\,U^{\dagger}_{4}\of{\bar{y},x_{4}}\,U^{\dagger}_{i}\of{\bar{x},x_{4}}
\]
and 
\[
U^{\dagger}_{p_2}\,=\,U_{4}\of{\bar{x},x_{4}+1}\,U_{i}\of{\bar{x},x_{4}+2}\,U^{\dagger}_{4}\of{\bar{y},x_{4}+1}\,U^{\dagger}_{i}\of{\bar{x},x_{4}+1}\ .
\]
Integrating over the three spatial links $U_{i}\of{\bar{x},x_{4}}$, $U_{i}\of{\bar{x},x_{4}+1}$ and $U_{i}\of{\bar{x},x_{4}+2}$ then yields
\begin{multline}
-2\,\of{3\,a_{f}\of{\beta}}^{2}\frac{1}{3^3}\,\of{2\,\kappa\e^{\mu}}^{n_{t}}\\
\trace_{c}\of{U_{4}^{\dagger}\of{\bar{y},x_{4}+1}\,U_{4}^{\dagger}\of{\bar{y},x_{4}}\,P_{\bar{y}}\of{x_{4},x_{4}+2}\,\sof{\id+\of{2\,\kappa\e^{\mu}}^{n_{t}}P_{\bar{y}}\of{x_{4}+2}}^{-1}}\\
\,\trace_{c}\of{U_{4}\of{\bar{x},x_{4}}\,U_{4}\of{\bar{x},x_{4}+1}\,P_{\bar{x}}\of{x_{4}+2,x_{4}}\,\sof{\id+\of{2\,\kappa\e^{\mu}}^{n_{t}}P_{\bar{x}}\of{x_{4}}}^{-1}}\ ,\label{eq:ksqbexmpl1}
\end{multline}
where, as per definition (see Sec. \ref{sssec:hdquarkprop}), 
\[
P_{\bar{y}}\of{x_{4},x_{4}+2}\,=\,U_{4}\of{\bar{y},x_{4}}\,U_{4}\of{\bar{y},x_{4}+1}
\]
and
\[
P_{\bar{x}}\of{x_{4}+2,x_{4}}\,=\,U_{4}\of{\bar{y},x_{4}+2}\,\ldots\,U_{4}\of{\bar{y},n_{t}-1}\,U_{4}\of{\bar{y},0}\,U_{4}\of{\bar{y},x_{4}-1}\ ,
\]
we can simplify
\[
U_{4}^{\dagger}\of{\bar{y},x_{4}+1}\,U_{4}^{\dagger}\of{\bar{y},x_{4}}\,P_{\bar{y}}\of{x_{4},x_{4}+2}\,=\,\id\ ,
\]
and
\[
U_{4}\of{\bar{x},x_{4}}\,U_{4}\of{\bar{x},x_{4}+1}\,P_{\bar{x}}\of{x_{4}+2,x_{4}}\,=\,P_{\bar{x}}\of{x_{4}}\ ,
\]
to get either the identity or a closed Polyakov loop. With this, we finally obtain for eq. \eqref{eq:ksqbexmpl1}:
\[
-\frac{2\,a^{2}_{f}\of{\beta}}{3}\,\trace_{c}\of{\sof{\id+\of{2\,\kappa\e^{\mu}}^{n_{t}}P_{\bar{y}}}^{-1}}\,\trace_{c}\of{\of{2\,\kappa\e^{\mu}}^{n_{t}}\,P_{\bar{x}}\,\sof{\id+\of{2\,\kappa\e^{\mu}}^{n_{t}}P_{\bar{x}}}^{-1}}\ .\label{eq:ksqbexmpl2}
\] 

In a similar way we can proceed with the other terms in \eqref{eq:sqtermsum} up to order $\order{\kappa_{s}^2\,a_{f}^{n_{t}-1}\of{\beta}}$ and find, after grouping them according to the power of their factors of $a_{f}\of{\beta}$:
\begin{multline}
-\frac{2}{3}\,\sum\limits_{x_{4}\neq y_{4}}\bcof{\bigg.\of{a_{f}\of{\beta}}^{\umod\of{y_{4}-x_{4},n_{t}}}\bof{\bigg.
\trace_{c}\sof{\sof{\id+\of{2\,\kappa\e^{\mu}}^{n_{t}}P_{\bar{y}}}^{-1}}\,\trace_{c}\sof{\of{2\,\kappa\e^{\mu}}^{n_{t}}\,P_{\bar{x}}\,\sof{\id+\of{2\,\kappa\e^{\mu}}^{n_{t}}P_{\bar{x}}}^{-1}}\\
+\trace_{c}\sof{\of{2\,\kappa\e^{-\mu}}^{n_{t}}\,P^{\dagger}_{\bar{y}}\,\sof{\id+\of{2\,\kappa\e^{-\mu}}^{n_{t}}P^{\dagger}_{\bar{y}}}^{-1}}\,\trace_{c}\sof{\sof{\id+\of{2\,\kappa\e^{-\mu}}^{n_{t}}P^{\dagger}_{\bar{x}}}^{-1}}\\
+\of{2\,\kappa}^{2\umod\of{y_{4}-x_{4},n_{t}}}\trace_{c}\sof{\sof{\id+\of{2\,\kappa\e^{\mu}}^{n_{t}}P_{\bar{y}}}^{-1}}\,\trace_{c}\sof{\sof{\id+\of{2\,\kappa\e^{-\mu}}^{n_{t}}P^{\dagger}_{\bar{x}}}^{-1}}\\
+\of{2\,\kappa}^{-2\umod\of{y_{4}-x_{4},n_{t}}}\trace_{c}\sof{\of{2\,\kappa\e^{-\mu}}^{n_{t}}\,P^{\dagger}_{\bar{y}}\,\sof{\id+\of{2\,\kappa\e^{-\mu}}^{n_{t}}P^{\dagger}_{\bar{y}}}^{-1}}\\
\trace_{c}\sof{\of{2\,\kappa\e^{\mu}}^{n_{t}}\,P_{\bar{x}}\,\sof{\id+\of{2\,\kappa\e^{\mu}}^{n_{t}}P_{\bar{x}}}^{-1}}\bigg.}\\
+\of{a_{f}\of{\beta}}^{\umod\of{x_{4}-y_{4},n_{t}}}\bof{\trace_{c}\sof{\of{2\,\kappa\e^{\mu}}^{n_{t}}\,P_{\bar{y}}\,\sof{\id+\of{2\,\kappa\e^{\mu}}^{n_{t}}P_{\bar{y}}}^{-1}}\,\trace_{c}\sof{\sof{\id+\of{2\,\kappa\e^{\mu}}^{n_{t}}P_{\bar{x}}}^{-1}}\bigg.\bigg.\\
+\trace_{c}\sof{\sof{\id+\of{2\,\kappa\e^{-\mu}}^{n_{t}}P^{\dagger}_{\bar{y}}}^{-1}}\,\trace_{c}\sof{\of{2\,\kappa\e^{-\mu}}^{n_{t}}\,P^{\dagger}_{\bar{x}}\,\sof{\id+\of{2\,\kappa\e^{-\mu}}^{n_{t}}P^{\dagger}_{\bar{x}}}^{-1}}\\
+\of{2\,\kappa}^{2\umod\of{x_{4}-y_{4},n_{t}}}\trace_{c}\sof{\sof{\id+\of{2\,\kappa\e^{-\mu}}^{n_{t}}P^{\dagger}_{\bar{y}}}^{-1}}\,\trace_{c}\sof{\sof{\id+\of{2\,\kappa\e^{\mu}}^{n_{t}}P_{\bar{x}}}^{-1}}\\
+\of{2\,\kappa}^{-2\umod\of{x_{4}-y_{4},n_{t}}}\trace_{c}\sof{\of{2\,\kappa\e^{\mu}}^{n_{t}}\,P_{\bar{y}}\,\sof{\id+\of{2\,\kappa\e^{\mu}}^{n_{t}}P_{\bar{y}}}^{-1}}\\
\trace_{c}\sof{\of{2\,\kappa\e^{-\mu}}^{n_{t}}\,P^{\dagger}_{\bar{x}}\,\sof{\id+\of{2\,\kappa\e^{-\mu}}^{n_{t}}P^{\dagger}_{\bar{x}}}^{-1}}\bigg.}\bigg.}\ ,\label{eq:ksqbsum}
\end{multline}
Note that all the traces in \eqref{eq:ksqbsum} are time-independent. The time-dependency of the individual terms in the sum is only in the different exponents of $a_{f}\of{\beta}$, $\of{2\,\kappa}^{2}$ and $\of{2\,\kappa}^{-2}$ which form essentially simple geometric sequences. The (finite) sum can therefore easily be carried out and yields:
\begin{align}
-\frac{2}{3}\,\bigg\{\frac{n_{t}\sof{a_{f}\of{\beta}-a^{n_{t}}_{f}\of{\beta}}}{1-a_{f}\of{\beta}}\bof{
&\trace_{c}\sof{\sof{\id+\of{2\,\kappa\e^{\mu}}^{n_{t}}P_{\bar{y}}}^{-1}}\bigg.\bigg.\nonumber\\
&\qquad\trace_{c}\sof{\of{2\,\kappa\e^{\mu}}^{n_{t}}\,P_{\bar{x}}\,\sof{\id+\of{2\,\kappa\e^{\mu}}^{n_{t}}P_{\bar{x}}}^{-1}}\nonumber\\
&+\trace_{c}\sof{\of{2\,\kappa\e^{\mu}}^{n_{t}}\,P_{\bar{y}}\,\sof{\id+\of{2\,\kappa\e^{\mu}}^{n_{t}}P_{\bar{y}}}^{-1}}\nonumber\\
&\qquad\trace_{c}\sof{\sof{\id+\of{2\,\kappa\e^{\mu}}^{n_{t}}P_{\bar{x}}}^{-1}}\nonumber\\
&+\trace_{c}\sof{\sof{\id+\of{2\,\kappa\e^{-\mu}}^{n_{t}}P^{\dagger}_{\bar{y}}}^{-1}}\nonumber\\
&\qquad\trace_{c}\sof{\of{2\,\kappa\e^{-\mu}}^{n_{t}}\,P^{\dagger}_{\bar{x}}\,\sof{\id+\of{2\,\kappa\e^{-\mu}}^{n_{t}}P^{\dagger}_{\bar{x}}}^{-1}}\nonumber\\
&+\trace_{c}\sof{\of{2\,\kappa\e^{-\mu}}^{n_{t}}\,P^{\dagger}_{\bar{y}}\,\sof{\id+\of{2\,\kappa\e^{-\mu}}^{n_{t}}P^{\dagger}_{\bar{y}}}^{-1}}\nonumber\\
&\qquad\trace_{c}\sof{\sof{\id+\of{2\,\kappa\e^{-\mu}}^{n_{t}}P^{\dagger}_{\bar{x}}}^{-1}}\bigg.}\nonumber\\
+\frac{n_{t}\sof{\of{2\kappa}^{2}a_{f}\of{\beta}-\of{2\kappa}^{2\,n_{t}}a^{n_{t}}_{f}\of{\beta}}}{1-\of{2\kappa}^{2}a_{f}\of{\beta}}\bof{
&\trace_{c}\sof{\sof{\id+\of{2\,\kappa\e^{\mu}}^{n_{t}}P_{\bar{y}}}^{-1}}\bigg.\bigg.\nonumber\\
&\qquad\trace_{c}\sof{\sof{\id+\of{2\,\kappa\e^{-\mu}}^{n_{t}}P^{\dagger}_{\bar{x}}}^{-1}}\nonumber\\
&+\trace_{c}\sof{\sof{\id+\of{2\,\kappa\e^{-\mu}}^{n_{t}}P^{\dagger}_{\bar{y}}}^{-1}}\nonumber\\
&\qquad\trace_{c}\sof{\sof{\id+\of{2\,\kappa\e^{\mu}}^{n_{t}}P_{\bar{x}}}^{-1}}\bigg.}\nonumber\\
+\frac{n_{t}\sof{\of{2\kappa}^{2\,n_{t}}a_{f}\of{\beta}-\of{2\kappa}^{2}a^{n_{t}}_{f}\of{\beta}}}{\of{2\kappa}^{2}-a_{f}\of{\beta}}\bof{
&\trace_{c}\sof{P_{\bar{y}}\sof{\id+\of{2\,\kappa\e^{\mu}}^{n_{t}}P_{\bar{y}}}^{-1}}\bigg.\bigg.\nonumber\\
&\qquad\trace_{c}\sof{P^{\dagger}_{\bar{x}}\sof{\id+\of{2\,\kappa\e^{-\mu}}^{n_{t}}P^{\dagger}_{\bar{x}}}^{-1}}\nonumber\\
&+\trace_{c}\sof{P^{\dagger}_{\bar{y}}\sof{\id+\of{2\,\kappa\e^{-\mu}}^{n_{t}}P^{\dagger}_{\bar{y}}}^{-1}}\nonumber\\
&\qquad\trace_{c}\sof{P_{\bar{x}}\sof{\id+\of{2\,\kappa\e^{\mu}}^{n_{t}}P_{\bar{x}}}^{-1}}\bigg.}\bigg.\bigg\}\ .
\end{align}

For $x_{4}\neq y_{4}$, there is a third class of diagrams, which are always of order $\kappa_{s}^{2}\,a_{f}^{n_{t}}\of{\beta}$, depicted in the last row of Fig. \ref{fig:ksqbexmpl}. If we take as an example again the left-most diagram on that row, we obtain by integrating out the spatial links, using the identity
\[
\int\idd{U}{}\, U_{a_1,b_1}\,U_{a_2,b_2}\,U_{a_3,b_3} = \frac{1}{3!}\,\epsilon_{a_1\,a_2\,a_3}\,\epsilon_{b_1\,b_2\,b_3}\ ,\label{eq:su3int2}
\]
a term of the following form:
\begin{multline}
-2\,\of{3\,a_{f}\of{\beta}}^{n_{t}}\,\of{2\kappa\e^{\mu}}^{n_{t}}\,\of{\frac{1}{3}}^{n_{t}-2}\of{\frac{1}{3!}}^{2}\,\epsilon_{I_1\,I_2\,I_3}\,\epsilon_{J_1\,J_2\,J_3}\,\epsilon_{M_1\,M_2\,M_3}\,\epsilon_{N_1\,N_2\,N_3}\\
P_{\bar{y}}\of{x_{4},y_{4}}_{J_1\,M_1}\,P^{\dagger}_{\bar{x}}\of{x_{4},y_{4}}_{N_1\,I_1}\,P^{\dagger}_{\bar{y}}\of{y_{4},x_{4}}_{J_2\,M_2}\,P_{\bar{x}}\of{y_{4},x_{4}}_{N_2\,I_2}\\
\sof{P_{\bar{y}}\of{x_{4},y_{4}}\sof{\id+\ssof{2\kappa\e^{\mu}}^{n_{t}}P_{\bar{y}}\of{y_{4}}}^{-1}}_{J_3\,M_3}\,\sof{P_{\bar{x}}\of{y_{4},x_{4}}\sof{\id+\ssof{2\kappa\e^{-\mu}}^{n_{t}}P_{\bar{x}}\of{x_{4}}}^{-1}}_{N_3\,I_3}\ .\label{eq:effsqbntex}
\end{multline}
By going to maximal temporal gauge, such that
\[
P_{\bar{x}}\of{x_{4},y_{4}}\,=\,\begin{cases} 
      \id &,\text{ if } x_{4}\leq y_{4} \\
      P_{\bar{x}} &,\text{ else} 
   \end{cases}\ ,
\]
we can simplify:
\begin{multline}
\epsilon_{J_1\,J_2\,J_3}\,\epsilon_{M_1\,M_2\,M_3}\,P_{\bar{y}}\of{x_{4},y_{4}}_{J_1\,M_1}\,P^{\dagger}_{\bar{y}}\of{y_{4},x_{4}}_{J_2\,M_2}\\
=\,\begin{cases} 
\trace_{c}\sof{P_{\bar{x}}}\,\delta_{J_3\,M_3}\,-\,P_{\bar{x},\,M_3\,J_3} &,\text{ if } x_{4} > y_{4} \\
\trace_{c}\sof{P^{\dagger}_{\bar{x}}}\,\delta_{J_3\,M_3}\,-\,P^{\dagger}_{\bar{x},\,M_3\,J_3} &,\text{ if } x_{4} < y_{4} \\
2\,\delta_{J_3\,M_3} &,\text{ if } x_{4}=y_{4}
\end{cases}\ ,
\end{multline}
and \eqref{eq:effsqbntex} becomes for $x_{4}>y_{4}$:
\begin{multline}
-\frac{a_{f}^{n_{t}}\of{\beta}}{2}\,\of{2\kappa\e^{\mu}}^{n_{t}}\of{\trace_{c}\sof{P_{\bar{y}}}\,\trace_{c}\sof{P_{\bar{y}}\sof{\id+\ssof{2\kappa\e^{\mu}}^{n_{t}}P_{\bar{y}}}^{-1}}-\trace_{c}\sof{P^{2}_{\bar{y}}\sof{\id+\ssof{2\kappa\e^{\mu}}^{n_{t}}P_{\bar{y}}}^{-1}}}\\
\of{\trace_{c}\sof{P^{\dagger}_{\bar{x}}}\,\trace_{c}\sof{\sof{\id+\ssof{2\kappa\e^{\mu}}^{n_{t}}P_{\bar{x}}}^{-1}}-\trace_{c}\sof{P^{\dagger}_{\bar{x}}\sof{\id+\ssof{2\kappa\e^{\mu}}^{n_{t}}P_{\bar{x}}}^{-1}}}\ ,\label{eq:effsqbntex1}
\end{multline}
and for $x_{4}<y_{4}$:
\begin{multline}
-\frac{a_{f}^{n_{t}}\of{\beta}}{2}\,\of{2\kappa\e^{\mu}}^{n_{t}}\of{\trace_{c}\sof{P^{\dagger}_{\bar{y}}}\,\trace_{c}\sof{\sof{\id+\ssof{2\kappa\e^{\mu}}^{n_{t}}P_{\bar{y}}}^{-1}}-\trace_{c}\sof{P^{\dagger}_{\bar{y}}\sof{\id+\ssof{2\kappa\e^{\mu}}^{n_{t}}P_{\bar{y}}}^{-1}}}\\
\of{\trace_{c}\sof{P_{\bar{x}}}\,\trace_{c}\sof{P_{\bar{x}}\sof{\id+\ssof{2\kappa\e^{\mu}}^{n_{t}}P_{\bar{x}}}^{-1}}-\trace_{c}\sof{P^{2}_{\bar{x}}\sof{\id+\ssof{2\kappa\e^{\mu}}^{n_{t}}P_{\bar{x}}}^{-1}}}\ .\label{eq:effsqbntex2}
\end{multline}
However, by using that for $P\in\SU{3}$, we have
\[
P^{\dagger}\,=\,P^{2}\,-\,\trace_{c}\of{P}\,P\,+\,\id\,\trace{P^{\dagger}}\ ,
\]
we see that \eqref{eq:effsqbntex1} and \eqref{eq:effsqbntex2} are actually the same, and can be written as, 
\begin{multline}
-\frac{a_{f}^{n_{t}}\of{\beta}}{2}\,\of{2\kappa\e^{\mu}}^{n_{t}}\of{\trace_{c}\sof{P_{\bar{y}}}\,\trace_{c}\sof{P_{\bar{y}}\sof{\id+\ssof{2\kappa\e^{\mu}}^{n_{t}}P_{\bar{y}}}^{-1}}-\trace_{c}\sof{P^{2}_{\bar{y}}\sof{\id+\ssof{2\kappa\e^{\mu}}^{n_{t}}P_{\bar{y}}}^{-1}}}\\
\of{\trace_{c}\sof{P_{\bar{x}}}\,\trace_{c}\sof{P_{\bar{x}}\sof{\id+\ssof{2\kappa\e^{\mu}}^{n_{t}}P_{\bar{x}}}^{-1}}-\trace_{c}\sof{P^{2}_{\bar{x}}\sof{\id+\ssof{2\kappa\e^{\mu}}^{n_{t}}P_{\bar{x}}}^{-1}}}\ .\label{eq:effsqbntex3}
\end{multline}
The terms corresponding to the remaining diagrams in the last row of Fig. \ref{fig:ksqbexmpl} can be computed in a completely analogous way. Collecting all the resulting terms and summing over $x_{4}\neq y_{4}$ then yields:
\begin{multline}
-\frac{a_{f}^{n_{t}}\of{\beta}\,n_{t}}{2}\,\bigg\{\bigg.\\
\of{n_{t}-1}\sof{2\kappa\e^{\mu}}^{n_{t}}\of{\trace_{c}\sof{P_{\bar{y}}}\trace_{c}\sof{P_{\bar{y}}\sof{\id+\of{2\,\kappa\e^{\mu}}^{n_{t}}P_{\bar{y}}}^{-1}}-\trace_{c}\sof{P^{2}_{\bar{y}}\sof{\id+\of{2\,\kappa\e^{\mu}}^{n_{t}}P_{\bar{y}}}^{-1}}}\\
\cdot\of{\trace_{c}\sof{P_{\bar{x}}}\trace_{c}\sof{P_{\bar{x}}\sof{\id+\of{2\,\kappa\e^{\mu}}^{n_{t}}P_{\bar{x}}}^{-1}}-\trace_{c}\sof{P^{2}_{\bar{x}}\sof{\id+\of{2\,\kappa\e^{\mu}}^{n_{t}}P_{\bar{x}}}^{-1}}}\\
+\frac{\sof{2\kappa}^{2}-\sof{2\kappa}^{2\,n_{t}}}{1-\sof{2\kappa}^{2}}\of{\trace_{c}\sof{P_{\bar{y}}}\trace_{c}\sof{P_{\bar{y}}\sof{\id+\of{2\,\kappa\e^{\mu}}^{n_{t}}P_{\bar{y}}}^{-1}}-\trace_{c}\sof{P^{2}_{\bar{y}}\sof{\id+\of{2\,\kappa\e^{\mu}}^{n_{t}}P_{\bar{y}}}^{-1}}}\\
\cdot\of{\trace_{c}\sof{P^{\dagger}_{\bar{x}}}\trace_{c}\sof{P^{\dagger}_{\bar{x}}\sof{\id+\of{2\,\kappa\e^{-\mu}}^{n_{t}}P^{\dagger}_{\bar{x}}}^{-1}}-\trace_{c}\sof{P^{\dagger\,2}_{\bar{x}}\sof{\id+\of{2\,\kappa\e^{-\mu}}^{n_{t}}P^{\dagger}_{\bar{x}}}^{-1}}}\\
+\frac{\sof{2\kappa}^{2}-\sof{2\kappa}^{2\,n_{t}}}{1-\sof{2\kappa}^{2}}\of{\trace_{c}\sof{P^{\dagger}_{\bar{y}}}\trace_{c}\sof{P^{\dagger}_{\bar{y}}\sof{\id+\of{2\,\kappa\e^{-\mu}}^{n_{t}}P^{\dagger}_{\bar{y}}}^{-1}}-\trace_{c}\sof{P^{\dagger\,2}_{\bar{y}}\sof{\id+\of{2\,\kappa\e^{-\mu}}^{n_{t}}P^{\dagger}_{\bar{y}}}^{-1}}}\\
\cdot\of{\trace_{c}\sof{P_{\bar{x}}}\trace_{c}\sof{P_{\bar{x}}\sof{\id+\of{2\,\kappa\e^{\mu}}^{n_{t}}P_{\bar{x}}}^{-1}}-\trace_{c}\sof{P^{2}_{\bar{x}}\sof{\id+\of{2\,\kappa\e^{\mu}}^{n_{t}}P_{\bar{x}}}^{-1}}}\\
+\of{n_{t}-1}\of{2\kappa\e^{-\mu}}^{n_{t}}\of{\trace_{c}\sof{P^{\dagger}_{\bar{y}}}\trace_{c}\sof{P^{\dagger}_{\bar{y}}\sof{\id+\of{2\,\kappa\e^{-\mu}}^{n_{t}}P^{\dagger}_{\bar{y}}}^{-1}}-\trace_{c}\sof{P^{\dagger\,2}_{\bar{y}}\sof{\id+\of{2\,\kappa\e^{-\mu}}^{n_{t}}P^{\dagger}_{\bar{y}}}^{-1}}}\\
\cdot\of{\trace_{c}\sof{P^{\dagger}_{\bar{x}}}\trace_{c}\sof{P^{\dagger}_{\bar{x}}\sof{\id+\of{2\,\kappa\e^{-\mu}}^{n_{t}}P^{\dagger}_{\bar{x}}}^{-1}}-\trace_{c}\sof{P^{\dagger\,2}_{\bar{x}}\sof{\id+\of{2\,\kappa\e^{-\mu}}^{n_{t}}P^{\dagger}_{\bar{x}}}^{-1}}}
\bigg.\bigg\}\ .
\end{multline}

Putting everything together, the effective nearest-neighbor Polyakov loop action, whose Boltzmann factor coincides with the full nearest-neighbor interaction term in \eqref{eq:nneffpartf} up to order $\order{\kappa_{s}^{2}\,a^{n_{t}}_{f}\of{\beta}}$, can be written as
\[
-S_{eff}\of{n_{t},\kappa,\beta,\mu}\,=\,-S_{g,eff}\of{n_{t},\beta}\,-\,S_{f,eff}\of{n_{t},\kappa,\beta,\mu}\ ,\label{eq:sefftotall}
\]
where
\[
-S_{g,eff}\of{n_{t},\beta}\,=\,a_{f}^{n_{t}}\of{\beta}\,\sum\limits_{\avof{\bar{x},\bar{y}}}\,\of{\trace_{c}\sof{P_{\bar{x}}}\,\trace_{c}\sof{P^{\dagger}_{\bar{y}}}+\trace_{c}\sof{P^{\dagger}_{\bar{x}}}\,\trace_{c}\sof{P_{\bar{y}}}}
\]
is the gauge part, already known from \eqref{eq:effplaction}, and
\begin{multline}
-\,S_{f,eff}\of{n_{t},\kappa,\beta,\mu}\,=\\
-\frac{2\,\kappa_{s}^{2}\,n_{t}}{3}\,\sum\limits_{\avof{\bar{x},\bar{y}}}\bigg\{\bigg.
\bof{\bigg.\trace_{c}\sof{\sof{\id+\of{2\,\kappa\e^{\mu}}^{n_{t}}P_{\bar{x}}}^{-1}}-\trace_{c}\sof{\sof{\id+\of{2\,\kappa\e^{-\mu}}^{n_{t}}P^{\dagger}_{\bar{x}}}^{-1}}\bigg.}\\
\qquad\cdot\bof{\bigg.\trace_{c}\sof{\big.\sof{\big.\id+\of{2\,\kappa\e^{\mu}}^{n_{t}}P_{\bar{y}}\big.}^{-1}\big.}-\trace_{c}\sof{\sof{\id+\of{2\,\kappa\e^{-\mu}}^{n_{t}}P^{\dagger}_{\bar{y}}}^{-1}}\bigg.}\\
-\frac{a_{f}\of{\beta}-a^{n_{t}}_{f}\of{\beta}}{1-a_{f}\of{\beta}}\bof{\bigg.\trace_{c}\sof{\sof{\id+\of{2\,\kappa\e^{\mu}}^{n_{t}}P_{\bar{y}}}^{-1}}\sof{3-\trace_{c}\sof{\sof{\id+\of{2\,\kappa\e^{\mu}}^{n_{t}}P_{\bar{x}}}^{-1}}}\\
+\sof{3-\trace_{c}\sof{\sof{\id+\of{2\,\kappa\e^{\mu}}^{n_{t}}P_{\bar{y}}}^{-1}}}\trace_{c}\sof{\sof{\id+\of{2\,\kappa\e^{\mu}}^{n_{t}}P_{\bar{x}}}^{-1}}\\
+\trace_{c}\sof{\sof{\id+\of{2\,\kappa\e^{-\mu}}^{n_{t}}P^{\dagger}_{\bar{y}}}^{-1}}\sof{3-\trace_{c}\sof{\sof{\id+\of{2\,\kappa\e^{-\mu}}^{n_{t}}P^{\dagger}_{\bar{x}}}^{-1}}}\\
+\sof{3-\trace_{c}\sof{\sof{\id+\of{2\,\kappa\e^{-\mu}}^{n_{t}}P^{\dagger}_{\bar{y}}}^{-1}}}\trace_{c}\sof{\sof{\id+\of{2\,\kappa\e^{-\mu}}^{n_{t}}P^{\dagger}_{\bar{x}}}^{-1}}\bigg.}\\
-\frac{\of{2\kappa}^{2}a_{f}\of{\beta}-\of{2\kappa}^{2\,n_{t}}a^{n_{t}}_{f}\of{\beta}}{1-\of{2\kappa}^{2}a_{f}\of{\beta}}\bof{\bigg.\trace_{c}\sof{\sof{\id+\of{2\,\kappa\e^{\mu}}^{n_{t}}P_{\bar{y}}}^{-1}}\trace_{c}\sof{\sof{\id+\of{2\,\kappa\e^{-\mu}}^{n_{t}}P^{\dagger}_{\bar{x}}}^{-1}}\\
+\trace_{c}\sof{\sof{\id+\of{2\,\kappa\e^{-\mu}}^{n_{t}}P^{\dagger}_{\bar{y}}}^{-1}}\trace_{c}\sof{\sof{\id+\of{2\,\kappa\e^{\mu}}^{n_{t}}P_{\bar{x}}}^{-1}}\bigg.}\\
-\frac{\of{2\kappa}^{2\,n_{t}}a_{f}\of{\beta}-\of{2\kappa}^{2}a^{n_{t}}_{f}\of{\beta}}{\of{2\kappa}^{2}-a_{f}\of{\beta}}\bof{\bigg.\trace_{c}\sof{P_{\bar{y}}\sof{\id+\of{2\,\kappa\e^{\mu}}^{n_{t}}P_{\bar{y}}}^{-1}}\trace_{c}\sof{P^{\dagger}_{\bar{x}}\sof{\id+\of{2\,\kappa\e^{-\mu}}^{n_{t}}P^{\dagger}_{\bar{x}}}^{-1}}\\
+\trace_{c}\sof{P^{\dagger}_{\bar{y}}\sof{\id+\of{2\,\kappa\e^{-\mu}}^{n_{t}}P^{\dagger}_{\bar{y}}}^{-1}}\trace_{c}\sof{P_{\bar{x}}\sof{\id+\of{2\,\kappa\e^{\mu}}^{n_{t}}P_{\bar{x}}}^{-1}}\bigg.}\\
-\frac{3\,\of{n_{t}-1}\,a_{f}^{n_{t}}\of{\beta}\,\sof{2\kappa\e^{\mu}}^{n_{t}}}{4}\of{\trace_{c}\sof{P_{\bar{y}}}\trace_{c}\sof{P_{\bar{y}}\sof{\id+\of{2\kappa\e^{\mu}}^{n_{t}}P_{\bar{y}}}^{-1}}-\trace_{c}\sof{P^{2}_{\bar{y}}\sof{\id+\of{2\,\kappa\e^{\mu}}^{n_{t}}P_{\bar{y}}}^{-1}}}\\
\cdot\of{\trace_{c}\sof{P_{\bar{x}}}\trace_{c}\sof{P_{\bar{x}}\sof{\id+\of{2\kappa\e^{\mu}}^{n_{t}}P_{\bar{x}}}^{-1}}-\trace_{c}\sof{P^{2}_{\bar{x}}\sof{\id+\of{2\kappa\e^{\mu}}^{n_{t}}P_{\bar{x}}}^{-1}}}\\
-\frac{3\,a_{f}^{n_{t}}\of{\beta}\,\sof{\sof{2\kappa}^{2}-\sof{2\kappa}^{2\,n_{t}}}}{4\,\sof{1-\sof{2\kappa}^{2}}}\of{\trace_{c}\sof{P_{\bar{y}}}\trace_{c}\sof{P_{\bar{y}}\sof{\id+\of{2\kappa\e^{\mu}}^{n_{t}}P_{\bar{y}}}^{-1}}-\trace_{c}\sof{P^{2}_{\bar{y}}\sof{\id+\of{2\kappa\e^{\mu}}^{n_{t}}P_{\bar{y}}}^{-1}}}\\
\cdot\of{\trace_{c}\sof{P^{\dagger}_{\bar{x}}}\trace_{c}\sof{P^{\dagger}_{\bar{x}}\sof{\id+\of{2\kappa\e^{-\mu}}^{n_{t}}P^{\dagger}_{\bar{x}}}^{-1}}-\trace_{c}\sof{P^{\dagger\,2}_{\bar{x}}\sof{\id+\of{2\kappa\e^{-\mu}}^{n_{t}}P^{\dagger}_{\bar{x}}}^{-1}}}\\
-\frac{3\,a_{f}^{n_{t}}\of{\beta}\,\sof{\sof{2\kappa}^{2}-\sof{2\kappa}^{2\,n_{t}}}}{4\,\sof{1-\sof{2\kappa}^{2}}}\of{\trace_{c}\sof{P^{\dagger}_{\bar{y}}}\trace_{c}\sof{P^{\dagger}_{\bar{y}}\sof{\id+\of{2\kappa\e^{-\mu}}^{n_{t}}P^{\dagger}_{\bar{y}}}^{-1}}-\trace_{c}\sof{P^{\dagger\,2}_{\bar{y}}\sof{\id+\of{2\kappa\e^{-\mu}}^{n_{t}}P^{\dagger}_{\bar{y}}}^{-1}}}\\
\cdot\of{\trace_{c}\sof{P_{\bar{x}}}\trace_{c}\sof{P_{\bar{x}}\sof{\id+\of{2\kappa\e^{\mu}}^{n_{t}}P_{\bar{x}}}^{-1}}-\trace_{c}\sof{P^{2}_{\bar{x}}\sof{\id+\of{2\kappa\e^{\mu}}^{n_{t}}P_{\bar{x}}}^{-1}}}\\
-\frac{3\,\of{n_{t}-1}\,a_{f}^{n_{t}}\of{\beta}\,\of{2\kappa\e^{-\mu}}^{n_{t}}}{4}\of{\trace_{c}\sof{P^{\dagger}_{\bar{y}}}\trace_{c}\sof{P^{\dagger}_{\bar{y}}\sof{\id+\of{2\kappa\e^{-\mu}}^{n_{t}}P^{\dagger}_{\bar{y}}}^{-1}}-\trace_{c}\sof{P^{\dagger\,2}_{\bar{y}}\sof{\id+\of{2\kappa\e^{-\mu}}^{n_{t}}P^{\dagger}_{\bar{y}}}^{-1}}}\\
\cdot\of{\trace_{c}\sof{P^{\dagger}_{\bar{x}}}\trace_{c}\sof{P^{\dagger}_{\bar{x}}\sof{\id+\of{2\kappa\e^{-\mu}}^{n_{t}}P^{\dagger}_{\bar{x}}}^{-1}}-\trace_{c}\sof{P^{\dagger\,2}_{\bar{x}}\sof{\id+\of{2\kappa\e^{-\mu}}^{n_{t}}P^{\dagger}_{\bar{x}}}^{-1}}}
\bigg.\bigg\}\label{eq:effplactionfull}
\end{multline}
is the fermionic part. Note that the $\kappa^{2} a_{f}^{n_t}\of{\beta}$ contribution and the $\kappa^4$ contributions are already contained in the general effective action derived in \cite{Langelage}, \cite{Langelage2} and \cite{Langelage3}.

By using
\[
\trace_{c}\sof{\sof{\id+\of{2\,\kappa\e^{\mu}}^{n_{t}}P}^{-1}}\,=\,\frac{3+2\of{2\kappa\e^{\mu}}^{n_{t}}\trace_{c}\sof{P}+\of{2\kappa\e^{\mu}}^{2\,n_{t}}\trace_{c}\sof{P^{\dagger}}}{1+\of{2\kappa\e^{\mu}}^{n_{t}}\trace_{c}\sof{P}+\of{2\kappa\e^{\mu}}^{2\,n_{t}}\trace_{c}\sof{P^{\dagger}}+\of{2\kappa\e^{\mu}}^{3\,n_{t}}}\label{eq:dl1}
\]
and
\[
\trace_{c}\sof{\sof{\id+\of{2\,\kappa\e^{-\mu}}^{n_{t}}P^{\dagger}}^{-1}}\,=\,\frac{3+2\of{2\kappa\e^{-\mu}}^{n_{t}}\trace_{c}\sof{P^{\dagger}}+\of{2\kappa\e^{-\mu}}^{2\,n_{t}}\trace_{c}\sof{P}}{1+\of{2\kappa\e^{-\mu}}^{n_{t}}\trace_{c}\sof{P^{\dagger}}+\of{2\kappa\e^{-\mu}}^{2\,n_{t}}\trace_{c}\sof{P}+\of{2\kappa\e^{-\mu}}^{3\,n_{t}}}\ ,\label{eq:dl2}
\]
as well as
\[
\trace_{c}\sof{P\sof{\id+\of{2\,\kappa\e^{\mu}}^{n_{t}}P}^{-1}}\,=\,\frac{3\of{2\kappa\e^{\mu}}^{2\,n_{t}}+\trace_{c}\sof{P}+2\of{2\kappa\e^{\mu}}^{n_{t}}\trace_{c}\sof{P^{\dagger}}}{1+\of{2\kappa\e^{\mu}}^{n_{t}}\trace_{c}\sof{P}+\of{2\kappa\e^{\mu}}^{2\,n_{t}}\trace_{c}\sof{P^{\dagger}}+\of{2\kappa\e^{\mu}}^{3\,n_{t}}}\label{eq:sl1}
\]
and
\[
\trace_{c}\sof{P^{\dagger}\sof{\id+\of{2\,\kappa\e^{-\mu}}^{n_{t}}P^{\dagger}}^{-1}}\,=\,\frac{3\of{2\kappa\e^{-\mu}}^{2\,n_{t}}+\trace_{c}\sof{P^{\dagger}}+2\of{2\kappa\e^{-\mu}}^{n_{t}}\trace_{c}\sof{P}}{1+\of{2\kappa\e^{-\mu}}^{n_{t}}\trace_{c}\sof{P^{\dagger}}+\of{2\kappa\e^{-\mu}}^{2\,n_{t}}\trace_{c}\sof{P}+\of{2\kappa\e^{-\mu}}^{3\,n_{t}}}\ ,\label{eq:sl2}
\]
and also
\begin{multline}
\sof{\trace_{c}\sof{P}\trace_{c}\sof{P\sof{\id+\of{2\kappa\e^{\mu}}^{n_{t}}P}^{-1}}-\trace_{c}\sof{P^{2}\sof{\id+\of{2\,\kappa\e^{\mu}}^{n_{t}}P}^{-1}}}\\
=\,\frac{2\,\trace_{c}\sof{P^{\dagger}}+\of{2\kappa\e^{\mu}}^{n_{t}}\sof{3+2\,\of{2\kappa\e^{\mu}}^{n_{t}}\trace_{c}\sof{P}+\trace_{c}\sof{P}\trace_{c}\sof{P^{\dagger}}}}{1+\of{2\kappa\e^{\mu}}^{n_{t}}\trace_{c}\sof{P}+\of{2\kappa\e^{\mu}}^{2\,n_{t}}\trace_{c}\sof{P^{\dagger}}+\of{2\kappa\e^{\mu}}^{3\,n_{t}}}\label{eq:ml1}
\end{multline}
and
\begin{multline}
\sof{\trace_{c}\sof{P^{\dagger}}\trace_{c}\sof{P^{\dagger}\sof{\id+\of{2\kappa\e^{-\mu}}^{n_{t}}P^{\dagger}}^{-1}}-\trace_{c}\sof{P^{\dagger\,2}\sof{\id+\of{2\,\kappa\e^{-\mu}}^{n_{t}}P^{\dagger}}^{-1}}}\\
=\,\frac{2\,\trace_{c}\sof{P}+\of{2\kappa\e^{-\mu}}^{n_{t}}\sof{3+2\,\of{2\kappa\e^{-\mu}}^{n_{t}}\trace_{c}\sof{P^{\dagger}}+\trace_{c}\sof{P^{\dagger}}\trace_{c}\sof{P}}}{1+\of{2\kappa\e^{-\mu}}^{n_{t}}\trace_{c}\sof{P^{\dagger}}+\of{2\kappa\e^{-\mu}}^{2\,n_{t}}\trace_{c}\sof{P}+\of{2\kappa\e^{-\mu}}^{3\,n_{t}}}\ ,\label{eq:ml2}
\end{multline}
equation \eqref{eq:effplactionfull} can be written completely in terms of Polyakov loops $L_{i}=\trace_{c}\sof{P_{\bar{x}_{i}}}$ and complex conjugate (or inverse) Polyakov loops $L_{j}^{*}=\trace_{c}\sof{P^{\dagger}_{\bar{x}_{j}}}$ at neighboring spatial sites $\bar{x}_{i}$, $\bar{x}_{j}$, $i\neq j$.\\

The partition function for the effective single flavor theory can now be written as
\begin{multline}
Z_{eff}\of{n_{t},\kappa,\beta,\mu}\,=\,\int\DD{P}\bcof{\bigg.\bof{\prod\limits_{\bar{\scriptstyle x}}\det_{c}^{2}\sof{\id+\of{2\kappa\e^{\mu}}^{n_{t}}P_{\bar{\scriptstyle x}}}\det_{c}^{2}\sof{\id+\of{2\kappa\e^{-\mu}}^{n_{t}}P^{\dagger}_{\bar{\scriptstyle x}}}}\\
\e^{-S_{f,eff}\of{n_{t},\kappa,\beta,\mu}}\,\e^{-S_{g,eff}\of{n_{t},\beta}}\bigg.}\ ,\label{eq:effnnpartf}
\end{multline}
where the correspondence between the terms in the integrand of \eqref{eq:effnnpartf} and the terms in the exact partition function \eqref{eq:sfpartf} is given by:
\[
\Det{D}\,\sim\,\bof{\prod\limits_{\bar{\scriptstyle x}}\det_{c}^{2}\sof{\id+\of{2\kappa\e^{\mu}}^{n_{t}}P_{\bar{\scriptstyle x}}}\det_{c}^{2}\sof{\id+\of{2\kappa\e^{-\mu}}^{n_{t}}P^{\dagger}_{\bar{\scriptstyle x}}}}\,\e^{-S_{f,eff}\of{n_{t},\kappa,\beta,\mu}}\label{eq:efftofullcorresp}
\]
and of course:
\[
\e^{-S_{g}}\,\sim\,\e^{-S_{g,eff}\of{n_{t},\beta}}\ .
\]
The generalization to $N_{f}$ flavors is, at the present expansion order, obtained by simply adding appropriate factors of \eqref{eq:efftofullcorresp} to the integrand of \eqref{eq:effnnpartf}.\\

The single site mean-field action corresponding to \eqref{eq:sefftotall}, which was used to generate the figures \ref{fig:polyakovloopfulleffcmp}, \ref{fig:polyakovloopfulleffcmpnq} and \ref{fig:effavsignvsmubeta} in Sec. \ref{ssec:fromhdtofulldet}, is obtained as usual, with the small complication that, as the action is no longer bi-linear in the Polyakov loop variables but a rational function, the usual approximation procedure:
\begin{enumerate}
\item write $L_{i}=\bar{L}+\delta L_{i}$, $L^{*}_{j}=\bar{L}^{*}+\delta L^{*}_{j}$ in $S_{eff}$,
\item keep terms only up to linear order in the perturbations $\delta L_{i}$, $\delta L^{*}_{j}$
\item substitute back $\delta L_{i}=L_{i}-\bar{L}$, $\delta L^{*}_{j}=L^{*}_{j}-\bar{L}^{*}$,
\item set $L_{i}=L_{a}$, $L^{*}_{i}=L^{*}_{a}$ for all $i$ (the subscript $a$ means "active") and divide by the system size,
\end{enumerate}
should be interpreted as coming from a leading order Taylor expansion of $S_{eff}$ around the state where the Polyakov loops and inverse Polyakov loops at each spatial site assume their mean-field values, i.e.:
\begin{multline}
S_{mf}\of{L^{\phantom{*}}_{a},L^{*}_{a},\bar{L}}\,=\,\frac{1}{V}\sum\limits_{i}^{V}\bcof{\bigg.\of{L_{a}-\bar{L}}\partd{S_{eff}\of{L_{1},\ldots,L_{V},L^{*}_{1},\ldots,L^{*}_{V}}}{L_{i}}\bigg|_{L_{i}=\bar{L}}\\
+\of{L^{*}_{a}-\bar{L}}\partd{S_{eff}\of{L_{1},\ldots,L_{V},L^{*}_{1},\ldots,L^{*}_{V}}}{L^{*}_{i}}\bigg|_{L^{*}_{i}=\bar{L}}\bigg.}\ ,\label{eq:meanfiledaction}
\end{multline}
where $V=n_{x}n_{y}n_{z}$ is the spatial volume and the leading constant piece has been dropped. Note also that the static-quark part of the fermion determinant, i.e. the product of single site static-quark fermion determinants in \eqref{eq:efftofullcorresp} is not part of the effective action and therefore not expanded during the mean-field approximation.\\

\end{document}